\documentclass[twocolumn,superscriptaddress,showpacs,nofootinbib]{revtex4}

\usepackage{bm,color}

\usepackage{graphicx}
\usepackage{amsmath}
\usepackage{amssymb}
\usepackage{amsfonts}
\usepackage{bm}
\usepackage{color}

\def\be{\begin{equation}}
\def\ee{\end{equation}}
\def\bea{\begin{eqnarray}}
\def\eea{\end{eqnarray}}

\begin{document}

\title{Jeans type  instability of a complex
self-interacting scalar field
 in general relativity}

\author{Abril Su\'arez}
\affiliation{Laboratoire de Physique Th\'eorique, Universit\'e Paul
Sabatier, 118 route de Narbonne 31062 Toulouse, France}
\affiliation{Departamento de Aeron\'autica, Universidad Polit\'ecnica
Metropolitana de Hidalgo, Ex-Hacienda San Javier, Tolcayuca, Hgo. C.P.
43860, Mexico}
\author{Pierre-Henri Chavanis}
\affiliation{Laboratoire de Physique Th\'eorique, Universit\'e Paul
Sabatier, 118 route de Narbonne 31062 Toulouse, France}

\begin{abstract}
We study the gravitational instability of a general relativistic complex
scalar field with a quartic self-interaction in an infinite homogeneous static
background. This
quantum relativistic Jeans problem provides a simplified framework to study
the formation of the large-scale structures of the Universe in the case where
dark matter is made of a scalar field. The scalar field may represent the
wave function of a relativistic self-gravitating Bose-Einstein condensate. Exact
analytical expressions for the
dispersion relation and Jeans length $\lambda_J$ are obtained
from a hydrodynamical
representation of the Klein-Gordon-Einstein equations  [Su\'arez and Chavanis,
Phys. Rev. D 92, 023510
(2015)]. We compare our results with those previously obtained with a simplified
relativistic model [Khlopov, Malomed and Y. B. Zeldovich, Mon. Not. R.
astr. Soc. {215}, 575 (1985)]. When relativistic effects are
fully accounted for, we find that the perturbations are stabilized at very large
scales of the order of the Hubble length $\lambda_H$. Numerical
applications are made for ultralight bosons without
self-interaction (fuzzy dark matter), for bosons with a repulsive
self-interaction, and for bosons with an attractive self-interaction
(QCD axions and ultralight axions). We show that the Jeans instability is
inhibited in the ultrarelativistic regime (early Universe and radiationlike era)
because the Jeans length is 
of the order of the Hubble length ($\lambda_J\sim\lambda_H$), 
except when the self-interaction of the bosons
is attractive.
By contrast, structure formation can take place in the nonrelativistic regime
(matterlike era) for $\lambda_J\le \lambda\le \lambda_H$. Since
the scalar field has a nonzero Jeans
length due to its quantum nature (Heisenberg uncertainty principle or quantum
pressure due to the self-interaction of the bosons), it appears that the wave
properties of the scalar field can
stabilize gravitational collapse at small scales, providing halo cores and
suppressing
small-scale linear power. This
may solve the CDM crisis such as the cusp problem and the missing satellite
problem. We also compare our results with those obtained in the case
where dark matter is made of fermions instead of bosons.

\end{abstract}



\maketitle


\section{Introduction: A brief review}

The formation of the large-scale structures of the Universe is an important
problem of
cosmology. In order to explain the evolution of cosmic structures, one must
understand the mechanism
that translates the almost homogeneous and isotropic early Universe
revealed by the cosmic microwave background (CMB) radiation to
the
greatly clustered Universe observed today.

The clumping of matter resulting from the gravitational attraction acting on an
initially uniform medium was first suggested by Newton in 1692 in a
famous letter to the Reverend Dr. Bentley \cite{newton}. He wrote: ``But if the
matter was evenly disposed throughout an infinite space, it could never convene
into one mass; but some of it would convene into one mass and some into another,
so as to make an infinite number of great masses, scattered at great distances
from one to another throughout all that infinite space. And thus might the sun
and fixt stars be formed, supposing the matter were of a lucid
nature.''

The
first serious theory of galaxy formation was proposed by Jeans in 1902 in a
paper entitled ``The Stability of a Spherical Nebula'' \cite{jeans1902}. He
supposed the
Universe to be filled with a fluid with mass density $\rho$, pressure $P$ and
velocity ${\bf v}$, and used classical hydrodynamic equations within the
framework of Newtonian gravity. In the last part of his paper,\footnote{As
indicated by the title of his paper, Jeans  \cite{jeans1902} was mainly
concerned
by the investigation of the gravitational stability of an
inhomogeneous
spherical
system. What has now become famous as the ``Jeans problem'', namely the
gravitational 
instability of an infinite uniform distribution of matter, represents only a
short section ({\it Infinite Space filled with Matter}) at the end of his
paper.} he studied the gravitational instability
of an
infinite and spatially uniform gas and treated the question of the formation of
self-gravitating objects by means of the interplay between the gravitational
attraction and the pressure forces acting on a mass of gas. By linearizing the
fluid equations about a static homogeneous state  and decomposing the
perturbations in Fourier modes of the form
$\delta\rho({\bf r},t)\sim {\rm exp}[i({\bf k}\cdot {\bf r}-\omega t)]$, he
obtained the dispersion relation 
\begin{eqnarray}
\omega^2=c_s^2k^2-4\pi G\rho,
\label{intro1}
\end{eqnarray}
where $c_s^2=P'(\rho)$ is the square of the speed
of sound in the medium. This equation  exhibits a characteristic  wavenumber
\begin{eqnarray}
k_J=\left (\frac{4\pi G\rho}{c_s^2}\right )^{1/2}
\label{intro2}
\end{eqnarray}
called the Jeans wavenumber. The
Jeans length $\lambda_J=2\pi/k_J$ gives an estimate of the minimum size of
the objects that can undergo gravitational collapse.
Perturbations for which the wavelength $\lambda$  is larger than the Jeans
length $\lambda_J$ (i.e. $\lambda>\lambda_J$ or $k<k_J$) can grow by
gravitational instability to form
the structures
observed in the Universe. According to Jeans' study, the perturbations grow
exponentially rapidly with a growth rate $\sigma(k)=\sqrt{-\omega(k)^2}$ where
$\omega(k)$ is given by Eq. (\ref{intro1}). The maximum growth
rate
\begin{eqnarray}
\sigma_{\rm max}=\sqrt{4\pi G\rho}
\label{intro3}
\end{eqnarray}
corresponds to a wavenumber
\begin{eqnarray}
k_*=0\qquad  {\rm i.e.}\qquad  \lambda_*\rightarrow +\infty.
\label{intro4}
\end{eqnarray}
On the other hand, perturbations for which the
wavelength $\lambda$ is
smaller than the Jeans length  $\lambda_J$ (i.e. $\lambda<\lambda_J$ or $k>k_J$)
oscillate with a
pulsation $\omega(k)$ like gravity-modified sound waves. Therefore, structure
formation is suppressed on scales smaller than the Jeans length, and allowed on
larger scales. The Jeans length $\lambda_J$ and the
corresponding Jeans mass 
\begin{eqnarray}
M_J=\frac{4}{3}\pi \rho \left (\frac{\lambda_J}{2}\right )^3
\label{intro4b}
\end{eqnarray}
determine the minimum size and the minimum mass above which a perturbation is
amplified by self-gravity (they correspond to the condition of
marginal
stability $\omega=0$).

The Jeans study \cite{jeans1902,jeansbook} has been extended in several
directions. It can be applied
to star formation in clouds of interstellar gas and to  galaxy
formation. The effect of a uniform rotation and a uniform magnetic field on the
Jeans criterion has been treated by Chandrasekhar
\cite{chandramagn,chandramagnbook}. The generalization of the
Jeans criterion to infinite uniform rotating collisionless stellar systems
described by the Vlasov equation has been treated by Lynden-Bell
\cite{lbjeans}. The computation of the Jeans length is the
standard starting point
for studying gravitational collapse. Even though today processes of stellar
formation and galaxy formation are known with precision, the Jeans theory
remains a good first approximation and has some pedagogical virtue.  We refer to
\cite{chavanisjeans} for a review of the Jeans instability problem for 
collisional gaseous
systems described by the Euler-Poisson equations and for  collisionless
stellar
systems described by the Vlasov-Poisson equations.

The Jeans stability analysis encounters several problems. Firstly, the
background state that is slightly perturbed in Jeans' analysis, namely an
infinite homogeneous distribution of matter, is not a steady state of the
hydrodynamic equations. Therefore, the problem is mathematically ill-posed at
the start even though the linearized  equations for the perturbations make
sense. This is what Binney and Tremaine \cite{bt} call the ``Jeans
swindle''.\footnote{See the Introduction of \cite{chavanisjeans} for a more
detailed discussion of the Jeans swindle and some possible justifications.}
Secondly, the
results of Eqs. (\ref{intro3}) and (\ref{intro4})
suggest that there is no upper limit to the  Jeans instability since the
maximum growth rate is
achieved for $\lambda_*\rightarrow +\infty$. Thirdly, as
noted
by Weinberg \cite{weinbergbook}, the Jeans analysis is not rigorously
applicable
to the formation of galaxies in an expanding Universe because Jeans assumed a
static medium whereas the rate of expansion of the Universe with a scale
factor $a(t)$ is given by the Hubble parameter
\begin{eqnarray}
H=\frac{\dot a}{a}=\left (\frac{8\pi G\rho}{3}\right )^{1/2}
\label{intro5}
\end{eqnarray}
which is of the same order as the maximum growth rate given
by Eq. (\ref{intro3}).
Therefore, we cannot assume that the Universe is static, or even quasistatic,
when doing the Jeans analysis.
A last limitation of the original Jeans stability analysis is that it applies to
nonrelativistic systems described by Newtonian gravity while the evolution of
the Universe is fundamentally relativistic even though the Newtonian
approximation is  relevant in most situations of astrophysical interest.

The first satisfactory theory of the instabilities in an expanding Universe was
given by Lifshitz \cite{lifshitz} in 1946 using general relativity. He showed
that disturbances at wavelengths above $\lambda_J$ grow, not exponentially, but
like a power of $t$. Taking into account the expansion of the
Universe is the correct manner to avoid the Jeans swindle. It leads, however, to
different results. A nonrelativistic
treatment of the instabilities in an
expanding Universe  was given by Bonnor \cite{bonnor} in 1957 using a Newtonian
world-model.\footnote{The Newtonian theory is applicable at the onset of the
matter-dominated
era, when the energy density of radiation drops below the rest mass density.
Therefore, the nonrelativistic analysis is adequate to study the
formation of structures in the matter-dominated era where $P\ll\epsilon$.
However, a relativistic treatment is needed to deal with the radiation era where
$P\sim\epsilon$.}  He obtained the equation
\begin{eqnarray}
\frac{d^2\delta_k}{dt^2}+2H\frac{d\delta_k}{dt}+\biggl
(\frac{c_s^2}{a^2}k^2
-{4\pi G\rho_b}
\biggr )\delta_k=0,
\label{intro6}
\end{eqnarray}
where $\delta=\delta\rho/\rho$ is the density contrast and $k$ denotes here the
comoving wavenumber (sometimes noted $k^c$). For a static Universe, one recovers
the Jeans dispersion relation
of Eq. (\ref{intro1}) by setting $a={\rm cte}$ and introducing the
physical wavenumber $k=k^c/a$. Equation (\ref{intro6}) is the fundamental
differential equation that
governs the growth or decay of gravitational condensations in an expanding
Universe. The case of collisionless stellar systems described by the Vlasov
equation in
an expanding Universe has been treated by Gilbert \cite{gilbert} in Newtonian
gravity. Further developments of these works for nonrelativistic and
relativistic fluids, and for collisionless stellar systems, can be found in the
books of Weinberg \cite{weinbergbook}, Peebles
\cite{peeblesbook}, and Padmanabhan \cite{paddy}.

In the previous studies, the Universe is treated  as a classical fluid
so that the Planck constant $\hbar$ does not appear in the equations. 
However, at very high energies, a standard hydrodynamical description of matter
is not
valid anymore. Rather, we expect that matter will be described in terms of
quantum
fields. In many particle physics models, scalar fields (SF) are introduced.
Their
role is to break high energy symmetries and to give mass to the particles by
the mechanism of spontaneous symmetry
breaking \cite{goldstone,higgs,weinbergmass}. These SFs are described by the
Klein-Gordon-Einstein (KGE) equations. A SF can be self-interacting, and
this self-interaction is described by a potential
$V(\varphi)$ \cite{coleman,weinberg}. It was soon
realized that due to the potential
$V(\varphi)$, SFs can play an important role in cosmology. 
In particular, if $\varphi$ is homogeneous in space, and trapped in a local
minimum of the potential, the scale factor of the Universe expands
exponentially rapidly. This leads to a primordial de Sitter
era\footnote{Previously, the
de Sitter solution (which is actually due to Lema\^itre) was introduced in
relation to the cosmological constant. A short account of the early development
of modern cosmology is given in the Introduction of \cite{cosmopoly1}.}
corresponding to what has been called inflation
\cite{starobinsky,guth,linde1,hm,as,astw,linde2,hawking,linde3}. A detailed
treatment of the evolution of a SF in the early Universe is performed in
\cite{belinsky,piran}. It is shown that a real SF
undergoes a stiff matter era followed by an inflation era which is an
attractor of the KGE equations. Finally, it oscillates about the minimum of
the potential and behaves in average as
radiation and, later, as pressureless matter.

After the discovery of the acceleration of the
Universe \cite{riess1,perlmutter,riess2}, SFs were introduced as dark energy
(DE) models.
This led to the concepts of quintessence \cite{quintessence},
Chaplygin gas model \cite{chaplygin}, tachyons \cite{tachyon}, 
phantom fields \cite{phantom}, Galileon model \cite{galileon}, polytropic
\cite{polytropic}, quadratic \cite{quadratic} and logotropic \cite{logotropic}
models etc.

SFs have also been introduced in the context of dark matter (DM) to solve the
cold dark matter (CDM) small-scale crisis such as the cusp problem
\cite{burkert,cusp}, missing satellite problem \cite{miss1,miss2}, and
too big to fail problem \cite{tbtf}.
Indeed, the wave properties of the SF can stabilize gravitational
collapse, providing halo cores and suppressing small-scale structures (or
linear power). Current works in particle physics and
cosmology have suggested different kinds of SF candidates for a solution to the
DM problem. Although these SFs have not yet been detected they are important DM
candidates. They could undergo Jeans instability and form boson stars and/or DM
halos as detailed below.

One possible DM candidate in
cosmology is the axion \cite{preskill,abbott,dine,turner1,davis}. Unlike the
Higgs
boson,
axions are sufficiently long-lived
to coalesce into DM halos which constitute the seeds of galaxy
formation. Axions are hypothetical pseudo-Nambu-Goldstone bosons of the
Peccei-Quinn \cite{pq} phase transition associated with a $U(1)$ symmetry that
solves the strong charge parity (CP) problem of quantum chromodynamics (QCD).
Axions also appear in string theory \cite{witten} leading to the notion of
string axiverse 
\cite{axiverse}. The QCD axion is a spin-$0$ particle
with a very small mass $m=10^{-4}\, {\rm eV}/c^2$ and an extremely weak
self-interaction $a_s=-5.8\times 10^{-53}\, {\rm m}$ (where $a_s$ is the
scattering length) arising from
nonperturbative effects in QCD. Axions are
extremely nonrelativistic and have
huge occupation numbers, so they can be described by a classical field. 
Axionic dark matter can form a Bose-Einstein condensate (BEC) during
the radiation-dominated era.  Axions can thus be described by a
relativistic quantum field theory with a real scalar field $\varphi$ whose
evolution is governed by the KGE equations.  In the nonrelativistic
limit, they can be described by an effective field theory with a complex scalar
field $\psi$ whose evolution is governed  by the Gross-Pitaevskii-Poisson (GPP)
equations.  In that case, the SF describes the wave function of the BEC. One
particularity of the axion is to have an attractive self-interaction ($a_s<0$).

The evolution of a Universe dominated by a cosmological SF
that could represent an axion field was studied by Turner
\cite{turner}. He considered a real self-interacting SF described by the
KGE equations with a potential of the form $V(\varphi)=a\varphi^{n}$ in an
isotropic and homogeneous cosmology. He showed that the SF experiences damped
oscillations but that, in average, it is equivalent to a perfect fluid with an
equation of state $P=[(n-2)/(n+2)]\epsilon$. For $n=2$, the SF behaves as
pressureless matter ($P=0$), and for $n=4$ it behaves as radiation
($P=\epsilon/3$) when the self-interaction is repulsive. Turner \cite{turner}
also mentioned the possibility of a stiff equation of state ($P=\epsilon$). The
formation of
structures in
an axion-dominated Universe was investigated by Hogan and Rees \cite{hr} and
Kolb and Tkachev \cite{kt,kt2}. They found very dense structures that they 
called
``axion miniclusters''  or ``axitons''.  These axitons have a mass $M_{\rm
axiton}\sim 10^{-12}\, M_{\odot}$ and a radius $R_{\rm
axiton}\sim 1\, R_{\odot}\sim 10^{9}\, {\rm m}$. 
These pseudosoliton
configurations  form in the very early Universe due to the
attractive self-interaction of the axions. In their work, self-gravity is
neglected. Kolb and Tkachev \cite{kt,kt2} mentioned the possibility to form
boson stars\footnote{Boson stars were introduced by Kaup  \cite{kaup} and
Ruffini and
Bonazzola \cite{rb} by coupling the KG equation to the Einstein equations.
Self-interacting boson stars with a repulsive self-interaction were considered
later by Colpi {\it et al.} \cite{colpi}, Tkachev \cite{tkachev} and Chavanis
and Harko \cite{chavharko}. These
authors showed that boson stars can exist only below a maximum mass due to
general relativistic effects (see Appendix \ref{sec_is}).} by Jeans instability
through Bose-Einstein
relaxation in the gravitationally bound clumps of axions. In other words,
axitons are expected to collapse and form
boson stars  when self-gravity becomes important.
Kolb and Tkachev \cite{kt,kt2} 
considered ordinary boson stars with a repulsive self-interaction such as those
sudied in \cite{tkachev}. However, since axions have an attractive
self-interaction, one cannot directly apply the standard results on boson stars
\cite{kaup,rb,colpi,tkachev,chavharko} to these objects. The case of
self-interacting
boson stars with an attractive
self-interaction, possibly representing axion stars, was
considered only
recently \cite{gu,bb,prd1,prd2,braaten,davidson,cotner,bectcoll,eby,
tkachevprl,helfer,dkr,phi6}. The analytical expression  of the maximum mass of
Newtonian self-gravitating BECs
with attractive self-interaction was first obtained in \cite{prd1,prd2} (see
Appendix \ref{sec_cis}). For QCD
axions, this leads to a maximum mass $M_{\rm max}=6.5\times
10^{-14}\, M_{\odot}$ and a radius $R_*=3.3\times 10^{-4}\, R_{\odot}=230\,
{\rm km}$ (the average density is $\rho_{\rm max}=2.54\, {\rm
kg/m^3}$). 
These are the maximum mass and minimum radius of
dilute miniaxion stars.
Coincidentally, these miniaxion stars have a mass comparable
to the mass of the axitons \cite{kt,kt2} but their mechanism of formation is
completely different since self-gravity plays a crucial role.  Obviously, QCD
axions cannot form giant BECs with the
dimension of DM halos. By contrast,
they may form mini axion stars (of the asteroid size) that could be
the constituents of DM halos under the form of mini massive compact halo objects
(mini MACHOs) \cite{bectcoll}. However, in that case, they would
essentially  behave as CDM and would not solve the small-scale
crisis of CDM.

Much bigger objects of the size of DM halos can be formed in
the case where the bosons have an ultrasmall mass of the order of $m\sim
10^{-22}\, {\rm eV}/c^2$ (ultralight bosons) and a very small
self-interaction. Their mass can be larger if they have a substantial
repulsive self-interaction.  At the
scale of DM halos, Newtonian gravity can be used and the KGE equations can be
replaced by the GPP equations. The
possibility that DM halos are made of a SF
representing a gigantic boson star described by the KGE or GPP equations has
been proposed by several authors
\cite{baldeschi,membrado,sin,jisin,leekoh,schunckpreprint,matosguzman,sahni,
guzmanmatos,fuzzy,peebles,goodman,mu,arbey1,silverman1,matosall,silverman,
lesgourgues,arbeycosmo,arbey} (see also
\cite{fm1,bohmer,fm2,bmn,fm3,sikivie,mvm,
lee09,ch1,leelim,prd1,prd2,prd3,briscese,
harkocosmo,harko,abrilMNRAS,aacosmo,velten,pires,park,rmbec,rindler,
lora,abrilJCAP,mhh,lensing,glgr1,robles,ch2,ch3,shapiro,bettoni,lora2,mlbec,
madarassy,
suarezchavanis1,suarezchavanis2,stiff,guthnew,souza,freitas,alexandre,schroven,
pop,martinez,ebydm,
cembranos,schwabe,fan,
calabrese,bectcoll,chavmatos,hui,zhang,total,suarezchavanis3,shapironew} 
for more recent
works). A
brief
history of the SFDM/BECDM scenario can be found in
the Introduction of
\cite{prd1} and in \cite{revueabril,revueshapiro,bookspringer,marshrevue}.

This brief review shows that SFs play a very important role in
particle physics, astrophysics and cosmology. When SFs are considered, quantum
mechanics arises in the problem and the Planck constant $\hbar$ enters
into the equations.

The gravitational  instability of a spatially homogeneous SF in a static
background in general relativity was first discussed by Khlopov {\it et al.}
\cite{khlopov}. They considered a real or complex SF and
demonstrated its instability to be analogous to the Jeans instability of a
classical self-gravitating gas in the sense that there is a critical wavelength
above which the system becomes unstable. For a noninteracting SF, Khlopov {\it
et al.} \cite{khlopov} obtained a quantum Jeans wavenumber given by\footnote{The
study of  Khlopov {\it et al.} \cite{khlopov} is expected to be valid for a
relativistic SF. However, we note that the quantum Jeans wavenumber of Eq.
(\ref{intro7})
is independent of $c$, meaning that it has seemingly the same expression in the
relativistic and nonrelativistic regimes. Actually, we shall argue that the
approach of Khlopov {\it et al.} \cite{khlopov} is not valid in the
relativistic regime because it neglects some terms of order $\Phi/c^2$ (this
limitation is
explicitly mentioned in \cite{khlopov}). We will show that Eq. (\ref{intro7})
is
only valid in the nonrelativistic regime while the general expression of the
quantum Jeans wavenumber is given by Eq. (\ref{em10}) below [see also Eqs.
(\ref{em16}) and
(\ref{em22})].}
\begin{eqnarray}
k_J=\left(\frac{16\pi G\rho m^2}{\hbar^2}\right)^{1/4}.
\label{intro7}
\end{eqnarray}
They also considered a self-interacting
SF with a quartic potential $V(|\varphi|^2)=(\lambda/4\hbar c)|\varphi|^4$.
They
treated the case of repulsive interactions ($\lambda>0$) but also the case of
attractive ones ($\lambda<0$) that arise in the Coleman-Weinberg 
model \cite{coleman} for not too large fields. This corresponds to a
$|\varphi|^4$ potential with the ``wrong'' sign. In
that case, they showed that the
quartic term develops an additional (besides gravitational) instability that
they called ``hydrodynamic'' because it corresponds to a fluid with a negative
pressure. This is a nongravitational instability of the SF.

Bianchi {\it et al.} \cite{bianchi} generalized the work of Khlopov {\it et
al.} \cite{khlopov} for a noninteracting SF to a
fully quantum context.  They used this system to make self-gravitating
structures
from a SF in a coherent state of a superfluid. They interpreted the density
perturbations of a
cosmological quantum real SF as excitations of a BEC at vanishing temperature.
They showed that
the dispersion
relation of the perturbations of the real SF obtained by Khlopov {\it et
al.} \cite{khlopov} matches the nonrelativistic Bogoliubov energy spectum
\cite{bogoliubov} of the
excitations of
bosonic ground states (valid for a general pair potential of interation
$V_k$) when the potential of interaction is the
gravitational one, $V_k=-4\pi Gm^2/k^2$. They mentioned that the physical
mechanism that
leads to a finite Jeans length has the same nature as that which accounts for
the equilibrium of the boson stars \cite{rb}.

Bianchi {\it et al.} \cite{bianchi} also considered the growth of structures
induced by a noninteracting SF in an expanding Universe. They first remarked
that, in the expression of the Jeans wavenumber (\ref{intro7}), the zero-point
pressure plays the same role which, in the traditional Jeans treatment, is
played by the pressure of a thermal gas in equilibrium against
gravitational attraction. Formally, this means that the quantum  Jeans
wavenumber (\ref{intro7}) can be obtained from the classical Jeans
wavenumber (\ref{intro2}) by making the
substitution 
\begin{eqnarray}
c_s^2=\frac{\hbar^2 k^2}{4m^2}.
\label{intro8}
\end{eqnarray}
Then,  applying the
linearized theory of perturbations for a classical fluid in general relativity
developed
by Lifshitz and Khalatnikov \cite{lk} and Weinberg \cite{weinbergbook}, and
making
the substitution from Eq. (\ref{intro8}), they obtained, in the nonrelativistic
limit (see Eq. (\ref{intro6})),
the following
equation\footnote{There is a factor $4$ missing in their expression.}
\begin{eqnarray}
\frac{d^2\delta_k}{dt^2}+2H\frac{d\delta_k}{dt}+\biggl (\frac{\hbar^2k^4}{
4m^2a^4 }-{4\pi G\rho_b}
\biggr )\delta_k=0
\label{intro9}
\end{eqnarray}
for the evolution of the density contrast. They contrasted this equation
from the one obtained by Ipser and Sikivie \cite{is} for a pressureless
axion field in which the speed of sound vanishes (obtained from Eq.
(\ref{intro6}) with $c_s=0$). They also considered the ultra-relativistic limit
of their model and showed that, in this case, the Jeans length is
larger than the Hubble length
(horizon) thereby preventing the growth of
structures.

The study of perturbations and the growth of structures for a relativistic SF
described by the KGE equations in an expanding background is complicated. It has
been treated in several
papers \cite{sasaki1,sasaki2,mukhanov,ratrapeebles,nambu,ratra,mukhanovrevue,
hwang,
jetzer,hu,joyce,ma,pb,brax,matos,hn1,hn2,malik,axiverse,mf,easther,phn,mmss,nph,
nhp,hlozek,alcubierre1,cembranos,ug,marshrevue,spintessence,jk,fm2,fm3}
in different contexts where
the SF can
represent the inflaton or can model DM or DE.  In these studies, the formation
of structures in an expanding Universe is studied directly from the field
equations for $\varphi$.

Instead of working in terms of field variables, a fluid
approach\footnote{The hydrodynamic representation of a SF is exact in the
case of a complex SF and approximate in the case of a real SF. In this paper, we
 consider the case of a complex SF. The hydrodynamic representation of a
real SF may give wrong results in the relativistic regime because of the
neglect of some oscillatory
terms (see Sec. II of \cite{phi6}). This is why the studies
of
\cite{sasaki1,sasaki2,mukhanov,ratrapeebles,nambu,ratra,mukhanovrevue,
hwang,
jetzer,hu,joyce,ma,pb,brax,matos,hn1,hn2,malik,axiverse,mf,easther,phn,mmss,nph,
nhp,hlozek,alcubierre1,cembranos,ug,marshrevue,spintessence,jk,fm2,fm3}
based on the field equations are necessary in that case. Note,
however, that the hydrodynamic representation of a real SF is valid in
the nonrelativistic regime (see Sec. II of \cite{phi6} for more details).} can
be
adopted (see a brief history of this approach in the Introduction
of \cite{chavmatos}). In the nonrelativistic case, this hydrodynamic approach
was introduced
by Madelung \cite{madelung} who showed that the Schr\"odinger equation is
equivalent to the
Euler equations for an irrotational fluid with an additional quantum potential
arising from the finite value of $\hbar$ accounting for
Heisenberg's uncertainty principle. This approach has been
generalized to the GPP equations in the context of DM halos by
\cite{fuzzy,bohmer,sikivie,prd1,prd2,prd3,aacosmo,rindler} among others. In
the relativistic case, de Broglie
\cite{broglie1927a,broglie1927b,broglie1927c} in his
so-called pilot wave theory, showed that the KG equations are equivalent to
hydrodynamic equations including a covariant quantum potential. This
approach has been generalized to the Klein-Gordon-Poisson (KGP) and KGE
equations in the context
of DM halos by \cite{abrilMNRAS,suarezchavanis1,suarezchavanis2,chavmatos} (see
also \cite{debbasch,chavharko,ns} for BEC stars). In
this hydrodynamic
representation, DM halos
result from the balance between the gravitational attraction and the quantum
pressure arising from the Heisenberg uncertainty principle or from the
self-interaction of the bosons. At small scales, quantum
effects are important
and can prevent the formation of singularities and solve the cusp problem and
the missing satellite problem. At large scales, quantum effects are
generally negligible (except in the early Universe) and one recovers the
hydrodynamic equations of the $\Lambda$CDM model.

It is possible to study the Jeans instability problem directly from these
hydrodynamic equations. This is closer in spirit to the original Jeans'
approach.  In the relativistic regime, valid during the radiation era or
earlier, one should use the fluid equations
based on the KGE equations. In the nonrelativistic regime, appropriate to study
structure formation in the matter era, the KGE equations reduce to the GPP
equations or to the Schr\"odinger-Poisson equations, and one can use the
corresponding
fluid equations.

The Jeans instability of a nonrelativistic SF  was qualitatively discussed by
Hu {\it et al.} \cite{fuzzy} 
who rederived
the quantum Jeans length of Eq. (\ref{intro7}). A more detailed analysis was
done by
Sikivie and Yang \cite{sikivie} who considered the growth of structures in
an expanding Universe and rederived Eq. (\ref{intro9}) directly from the
quantum hydrodynamic equations. In that case, the term $\hbar^2k^4/4m^2a^2$
arises directly from the quantum potential in the Euler
equation, and there is no need to make the formal (and mathematically illicit)
substitution from Eq. (\ref{intro8}). In these studies, the SF is assumed to be
noninteracting. The Jeans
instability of a  nonrelativistic SF with an arbitrary self-interaction
(repulsive or attractive) in a static Universe
has been treated in detail by \cite{prd1} who derived
the Jeans length
of Eq. (\ref{nr5}) below. The growth of structures of a self-interacting SF in
an
expanding Universe was treated independently by \cite{abrilMNRAS}
and \cite{aacosmo} 
who derived the equation 
\begin{eqnarray}
\frac{d^2\delta_k}{dt^2}+2H\frac{d\delta_k}{dt}+\biggl (\frac{\hbar^2k^4}{
4m^2a^4 }+\frac{c_s^2}{a^2}k^2
-{4\pi G\rho_b}
\biggr )\delta_k=0\nonumber\\
\label{intro10}
\end{eqnarray}
for the evolution of the density contrast. For $\hbar=0$, one recovers the
classical Bonnor equation (\ref{intro6}) and for $c_s=0$ one
recovers Eq. (\ref{intro9}) as  particular case.

The study of the Jeans instability of a relativistic complex SF with a
quartic potential based on the
hydrodynamic approach has been initiated in our previous paper
\cite{suarezchavanis1}. Preliminary results were given for a
static Universe. In particular, we obtained the dispersion relation of Eq.
(\ref{em1}) and the
Jeans length of Eq. (\ref{em10}). The case of an expanding
Universe was also considered in \cite{suarezchavanis1}. Using simplifying
approximations, which amount to neglecting some relativistic terms while
preserving the exact expression of the static Jeans length, we obtained the
equation 
\begin{eqnarray}
&&\frac{d^2\delta_k}{dt^2}+2H\frac{d\delta_k}{dt}+\biggl\lbrack
\frac{\hbar^2k^4}{
4m^2a^4 }+\frac{c_s^2}{a^2}k^2\nonumber\\
&&-\frac{4\pi G\rho_b}{1+\frac{3H^2a^2}{k^2c^2}}\left
(1+\frac{2c_s^2}{c^2}\right
)\left (
1+\frac{3c_s^2}{c^2}+\frac{\hbar^2k^2}{4m^2c^2a^2}
\right )\biggr\rbrack\delta_k=0\nonumber\\
\label{intro11}
\end{eqnarray} 
for the evolution of the density contrast. We
showed that general relativity prevents the growth of perturbations
whose wavelengths are larger than the Hubble length (horizon).
More precisely, the SF gives three distinct regions in Fourier space. On
scales smaller than the Jeans
length $\lambda_J$, the contrast in the energy density oscillates with constant
or growing amplitude (depending on the dominant regime, either noninteracting
or nonquantum). Above the Jeans length $\lambda_J$ but still below the
Hubble length $\lambda_H$, self-gravity prevails, the density contrast grows,
and the Jeans instability
comes into play. Lastly, at scales close to or larger than the Hubble
length  $\lambda_H$, the energy density freezes due to general relativistic
effects. These results are consistent with those obtained
in \cite{mukhanovbook} using SF equations.

In this paper, we continue the work initiated in \cite{suarezchavanis1}.
We consider a relativistic complex SF with a $|\varphi|^4$
self-interaction.  The cosmological evolution of a homogeneous SF with
a repulsive
self-interaction has been treated in \cite{shapiro} from
field equations (see also the previous works
of \cite{jetzer,sj,spintessence,arbeycosmo}) and in
\cite{suarezchavanis1,suarezchavanis3} from hydrodynamic equations. For weak
self-interactions (in a sense detailed in \cite{suarezchavanis3}), the SF
undergoes a stiff matter era followed by a matterlike
era. For stronger self-interactions, it undergoes a stiff matter era followed by
a radiationlike era, and finally a matterlike era. Phase diagrams are
provided in \cite{suarezchavanis3}. The case of a SF with an attractive
self-interaction has also been considered in \cite{suarezchavanis3} leading to
intriguing results. It appears that two evolutions, corresponding to two
different branches, are
possible. The SF behaves as DM on the normal branch and as DE on the peculiar
branch. In the latter, the SF maintains a constant energy density as
a result of spintessence \cite{spintessence}.

In the present paper, we restrict ourselves to a static background and study
the Jeans instability of a general relativistic complex SF with a $|\varphi|^4$
self-interaction. In Sec. \ref{sec_jia}, we expose our procedure to solve the
problem and refer to our previous work \cite{suarezchavanis1} for technical
details. In Sec. \ref{sec_nr}, we briefly recall the results of the Jeans
instability of a nonrelativistic SF \cite{prd1}. In Sec. \ref{sec_sm}, we
consider a simplified relativistic model, which turns out to be equivalent to
the one considered by Khlopov {\it et al.} \cite{khlopov}, where some terms are
neglected  in the dispersion relation. In
Sec. \ref{sec_em},
we consider the exact relativistic model where all the terms are retained in the
dispersion relation (the nongravitational limit is treated in Sec.
\ref{sec_nog}). When relativistic effects are
fully accounted for, we find that the perturbations are stabilized at very large
scales of the order of the Hubble length $\lambda_H$.
In the following sections, we consider
astrophysical
and cosmological applications of our theoretical results.  In Sec.
\ref{sec_anrad}, we show that the Jeans instability is inhibited in the
ultrarelativistic regime (early Universe and radiationlike era) because the
Jeans length is of the order
of the Hubble length ($\lambda_J\sim \lambda_H$), except
when the self-interaction
of the SF is attractive. In Sec. \ref{sec_app}, we show that the Jeans
instability can take place in the nonrelativistic regime (matterlike era)
for $\lambda_J\le\lambda\le \lambda_H$.
Numerical
applications are made for ultralight bosons without
self-interaction (fuzzy dark matter), for bosons with a repulsive
self-interaction, and for bosons with an attractive self-interaction
(QCD axions and ultralight axions). In Sec. \ref{sec_fermi}, we
compare our results with those obtained in the case where dark matter is made of
fermions instead of bosons.

\section{The Jeans instability of a relativistic complex SF}
\label{sec_jia}

We consider a relativistic complex SF, possibly representing the wave function
of a BEC
at $T=0$, with a quartic self-interaction potential of the form
\begin{equation}
V_{\rm tot}(|\varphi|^2)=\frac{m^2c^2}{2\hbar^2}|\varphi|^2+\frac{2\pi a_s
m}{\hbar^2}|\varphi|^4.
\label{jia1}
\end{equation}
The quadratic term is the rest-mass term and the quartic term
is the self-interaction term. Here, $m$ denotes the mass of the bosons and $a_s$
their scattering length. The self-interaction is repulsive when $a_s>0$ and
attractive when $a_s<0$
\cite{revuebec}. The evolution of the SF is described by the KG equation
\begin{equation}
\Box\varphi+2\frac{dV_{\rm tot}}{d|\varphi|^2}\varphi=0,
\label{jia2}
\end{equation}
where $\Box$ is the d'Alembertian operator in a curved spacetime:
\begin{equation}
\Box\equiv D_{\mu}(g^{\mu\nu}\partial_{\nu})=\frac{1}{\sqrt{-g}}
\partial_\mu(\sqrt{-g}\, g^{\mu\nu}\partial_\nu).
\label{jia3}
\end{equation}
For the specific SF potential (\ref{jia1}), the KG equation
takes the form
\begin{equation}
\square \varphi+\frac{m^2c^2}{\hbar^2}\varphi+\frac{8\pi a_s
m}{\hbar^2}|\varphi|^2\varphi=0.
\label{jia4}
\end{equation}
It is coupled to the Einstein equations
\begin{equation}
R_{\mu\nu}-\frac{1}{2}g_{\mu\nu}R=\frac{8\pi G}{c^4}T_{\mu\nu},
\label{jia5}
\end{equation}
where $R_{\mu\nu}$ is the Ricci tensor and $T^{\mu\nu}$ is the
energy-momentum tensor of the SF given by
\begin{eqnarray}
T_{\mu\nu}=\frac{1}{2}(\partial_{\mu}\varphi^*
\partial_{\nu}\varphi+\partial_{\nu}\varphi^* \partial_{\mu}\varphi)\nonumber\\
-g_{\mu\nu}\left
\lbrack\frac{1}{2}g^{\rho\sigma}\partial_{\rho}\varphi^*\partial_{\sigma}
\varphi-V_{\rm tot}(|\varphi|^2)\right \rbrack.
\label{jia6}
\end{eqnarray}

Our aim is to study the Jeans instability of an infinite homogeneous
relativistic complex SF in a
static background in relation to the formation of structures in cosmology.
Following our previous work \cite{suarezchavanis1}, we proceed as follows: 

(i) We first introduce the Newtonian gauge [see
Eq. (I-18)]\footnote{Here and in the following (I-x) refers to Eq. (x) of
\cite{suarezchavanis1} called Paper I.} to write the KGE equations in the weak
field
approximation [see
Eqs. (I-20) and (I-26)]. Since we consider the linear perturbation regime, there
is no limitation in using this gauge. However, we consider the simplest form of
the Newtonian gauge, only taking into account scalar perturbations which are the
ones that contribute to
the formation of structures in cosmology. Vector contributions (which are
supposed to be
always small) vanish during cosmic inflation and tensor contributions (which
account for gravitational waves) are neglected \cite{mukhanovrevue}. We
also neglect
anisotropic stress for simplicity. 

(ii) We make the Klein transformation  [see Eq. (I-34)], which amounts to
subtracting the rest mass energy, and obtain the general
relativistic GPE equations  [see
Eqs. (I-35) and (I-36)] from the KGE equations. The interest of this
transformation is that the GPE equations (unlike the KGE equations) have a well
defined nonrelativistic limit $c\rightarrow +\infty$, leading to the GPP
equations [see Eqs. (I-C1) and (I-C2)] that are relevant in the Newtonian
regime.

(iii) We use the Madelung transformation  [see
Eqs. (I-39) and (I-40)] to obtain a hydrodynamic representation
of the GPE equations. These hydrodynamic equations [see
Eqs. (I-41)-(I-44)] are equivalent to the GPE
equations (themselves equivalent to the KGE equations) and they put us in a
situation similar to the one investigated by Jeans \cite{jeans1902,jeansbook}
except
that we have
additional terms due to quantum mechanics ($\hbar$) and relativity ($c$).
These
hydrodynamic equations 
depend on a pseudo
rest-mass density defined by
\begin{eqnarray}
\rho=\frac{m^2}{\hbar^2} |\varphi|^2=|\psi|^2.
\label{jia7}
\end{eqnarray}
It is only in the nonrelativistic limit $c\rightarrow +\infty$ that $\rho$
corresponds to the mass density, but we can always introduce $\rho$ from Eq.
(\ref{jia7}) as a
convenient notation, even in the relativistic regime \cite{suarezchavanis1}.

(iv) The hydrodynamic
equations involve a
pseudo pressure given by the 
polytropic (quadratic) equation of
state
\begin{eqnarray}
p=\frac{2\pi a_s\hbar^2}{m^3}\rho^2.
\label{jia8}
\end{eqnarray}
This pressure is due to the self-interaction of the
bosons. This equation of state keeps  the same form in the
nonrelativistic and
relativistic regimes.\footnote{The equation of state $p(\rho)$ usually differs
from the true equation of state $P(\epsilon)$ that relates the pressure $P$
obtained from the energy-momentum tensor to the energy density $\epsilon$. In
specific situations, e.g. for a static background, the pressures $p$ and $P$
coincide (see below).} From this equation of state, we
can define the pseudo speed of sound
\begin{eqnarray}
c_s^2=p'(\rho)=\frac{4\pi a_s\hbar^2\rho}{m^3}.
\label{jia9}
\end{eqnarray}
We note that $c_s^2>0$ for a repulsive self-interaction and  $c_s^2<0$ for an
attractive self-interaction.

(v) We now consider the case of a static Universe corresponding to $a=1$
and $H=0$ in the equations of \cite{suarezchavanis1}. For the homogeneous
background, we have the exact relations  \cite{suarezchavanis1}:\footnote{We
note that these relations coincide with those obtained in
\cite{suarezchavanis3} for a homogeneous SF in an expanding Universe. In
that context, they apply to the fast oscillation
regime, equivalent to the Thomas-Fermi (TF) or semiclassical approximation,
in which the quantum
potential can be neglected. We note that the TF approximation is always valid
for a homogeneous SF in a static background because the
terms in $\hbar$ in the hydrodynamic equations
all involve a temporal derivative or a gradient (see
\cite{suarezchavanis1,suarezchavanis3} for details) and therefore vanish.
Note, however, that the TF approximation is not always valid at the level of the
perturbations.}
\begin{eqnarray}
\epsilon=\rho c^2\left (1+\frac{6\pi a_s\hbar^2}{m^3c^2}\rho\right ),
\label{jia9b}
\end{eqnarray}
\begin{eqnarray}
\rho=\frac{m^3c^2}{12\pi a_s\hbar^2}\left (\sqrt{1+\frac{24\pi
a_s\hbar^2}{m^3c^4}\epsilon}-1\right ),
\label{jia9c}
\end{eqnarray}
\begin{eqnarray}
P=p=\frac{m^3c^4}{72\pi a_s\hbar^2}\left (\sqrt{1+\frac{24\pi
a_s\hbar^2}{m^3c^4}\epsilon}-1\right )^2
\label{jia9d}
\end{eqnarray}
between the (uniform) pseudo rest mass density $\rho$, the (uniform) energy
density $\epsilon$, and the (uniform) pressure $P$.  For a
noninteracting SF ($a_s=0$), these relations reduce to
\begin{eqnarray}
\epsilon=\rho c^2,\qquad P=0, 
\label{dim1}
\end{eqnarray}
in all the regimes (relativistic and nonrelativistic). We now have to
distinguish attractive and
repulsive self-interactions. We first
consider repulsive self-interactions ($a_s>0$).  In the ultrarelativistic limit
$\rho\gg \rho_R=m^3c^2/6\pi
a_s\hbar^2$ (see Appendix \ref{sec_is}), we get 
\begin{eqnarray}
\epsilon\sim \frac{6\pi a_s\hbar^2}{m^3}\rho^2,\qquad P\sim
\frac{1}{3}\epsilon. 
\label{dim2}
\end{eqnarray}
This corresponds to the radiationlike era in \cite{shapiro,suarezchavanis3}.
Indeed, in the ultrarelativistic limit, the SF behaves like radiation at the
background level. In the
nonrelativistic limit  $\rho\ll \rho_R$, we get 
\begin{eqnarray}
\epsilon\sim \rho c^2,\qquad P\sim \frac{2\pi
a_s\hbar^2}{m^3c^4}\epsilon^2.
\label{dim3}
\end{eqnarray}
This corresponds to the pressureless matterlike era in
\cite{shapiro,suarezchavanis3}. Indeed, in the nonrelativistic limit, the SF
behaves as pressureless matter at the bakground level. We now consider
attractive self-interactions
($a_s<0$). In the ultrarelativistic limit, coming back temporarily to the more
realistic case of an expanding Universe, we know from the results of
\cite{suarezchavanis3} that the pseudo rest-mass density tends to
the value (see Appendix \ref{sec_ngsi}):
\begin{eqnarray}
\rho_i=\frac{m^3c^2}{12\pi |a_s|\hbar^2}.
\label{badlu}
\end{eqnarray}
We also have
\begin{eqnarray}
\epsilon_i=\frac{m^3c^4}{24\pi
|a_s|\hbar^2},\qquad P_i=-\frac{1}{3}\epsilon_i.
\label{dim4}
\end{eqnarray}
This corresponds to the cosmic stringlike era in \cite{suarezchavanis3}. In
the
nonrelativistic limit $\rho\ll\rho_i$, we recover Eq. (\ref{dim3}).

(vi) We linearize the hydrodynamic equations about a spatially homogeneous
state (making the Jeans swindle)
and decompose the perturbation in Fourier modes of the form
${\rm exp}[i({\bf k}\cdot {\bf r}-\omega t)]$. In this way, we obtain the
dispersion relation $\omega(k)$ that relates the complex pulsation $\omega$ to
the wave number $k=2\pi/\lambda$. From this dispersion
relation,\footnote{Although we work in terms of {\it pseudo} hydrodynamic
variables such as $\rho$, $p$ etc., the dispersion relation
$\omega(k)$ is independent of our choice of variables. In this sense, our study
is general.} we can determine the oscillatory modes ($\omega^2>0$), the
growing
modes ($\omega^2<0$) giving rise to a linear instability,
the Jeans length $\lambda_J$ (corresponding to the neutral mode $\omega=0$)
separting stable and
unstable modes, and the optimal length $\lambda_*$ having the maximum growth
rate
$\sigma_{\rm max}$.

The general dispersion relation corresponding to the full KGE equations without
approximation  has been determined in our previous paper
\cite{suarezchavanis1}, see Eq. (I-69),  but
only a preliminary analysis of its solutions was given. In the present
paper, we study it in greater detail. To increase the complexity of the
problem progressively, we first consider the nonrelativistic model (Sec.
\ref{sec_nr}), then a
simplified relativistic model where some terms are neglected in the
KGE equations (Sec. \ref{sec_sm}), and finally the exact relativistic model
(Sec. \ref{sec_em}). In each section, we provide general equations valid for
repulsive ($c_s^2>0$) and attractive ($c_s^2<0$) self-interactions but in the
discussion, for simplicity, we restrict ourselves to repulsive
self-interactions. The case of attractive self-interactions is considered in
greater detail in the nongravitational limit (Sec. \ref{sec_nog}).

\section{The nonrelativistic model}
\label{sec_nr}

In this section, we study the dispersion relation of a self-gravitating BEC in
the nonrelativistic limit $c\rightarrow +\infty$. In that limit, the KGE
equations reduce to the GPP equations [see Eqs. (I-C1) and (I-C2)]. The Jeans
problem based on the
hydrodynamic equations derived from the GPP equations [see Eqs.
(I-C3)-(I-C6)] has been studied
in detail in \cite{prd1} for both repulsive ($a_s>0$) and  attractive
($a_s<0$)
self-interactions between the bosons. In this section, we briefly recall the
main
results of this study (restricting ourselves to the case $a_s>0$) in
order to facilitate the comparison with the relativistic results discussed in
the following sections.

\subsection{The general case}

In the  nonrelativistic limit, the dispersion relation characterizing the
small perturbations of a spatially homogeneous self-gravitating BEC is given by
(see Sec. V of \cite{prd1}):
\begin{eqnarray}
\omega^2=\frac{\hbar^2k^4}{4m^2}+c_s^2k^2-4\pi G\rho.
\label{nr1}
\end{eqnarray}
Equation  (\ref{nr1}) corresponds to the
Bogoliubov energy spectrum of the excitations of a self-gravitating and weakly
self-interacting 
BEC \cite{bogoliubov}. The function $\omega^2(k)$ is
plotted in Fig. \ref{dispnewton} using the normalization of Appendix
\ref{sec_dv1} with $\chi\propto 2/c^2=0$ (nonrelativistic limit).
It starts from $\omega^2(0)=-4\pi G\rho$ at $k=0$ and
increases monotonically with $k$. For $k\rightarrow 0$, we obtain
\begin{eqnarray}
\omega^2\simeq -4\pi G\rho+c_s^2k^2,
\label{nr2}
\end{eqnarray}
corresponding to the classical regime. For $k\rightarrow +\infty$, we obtain
\begin{eqnarray}
\omega^2\sim \frac{\hbar^2k^4}{4m^2},
\label{nr3}
\end{eqnarray}
corresponding to the strongly quantum regime.

\begin{figure}[h]
\scalebox{0.33}{\includegraphics{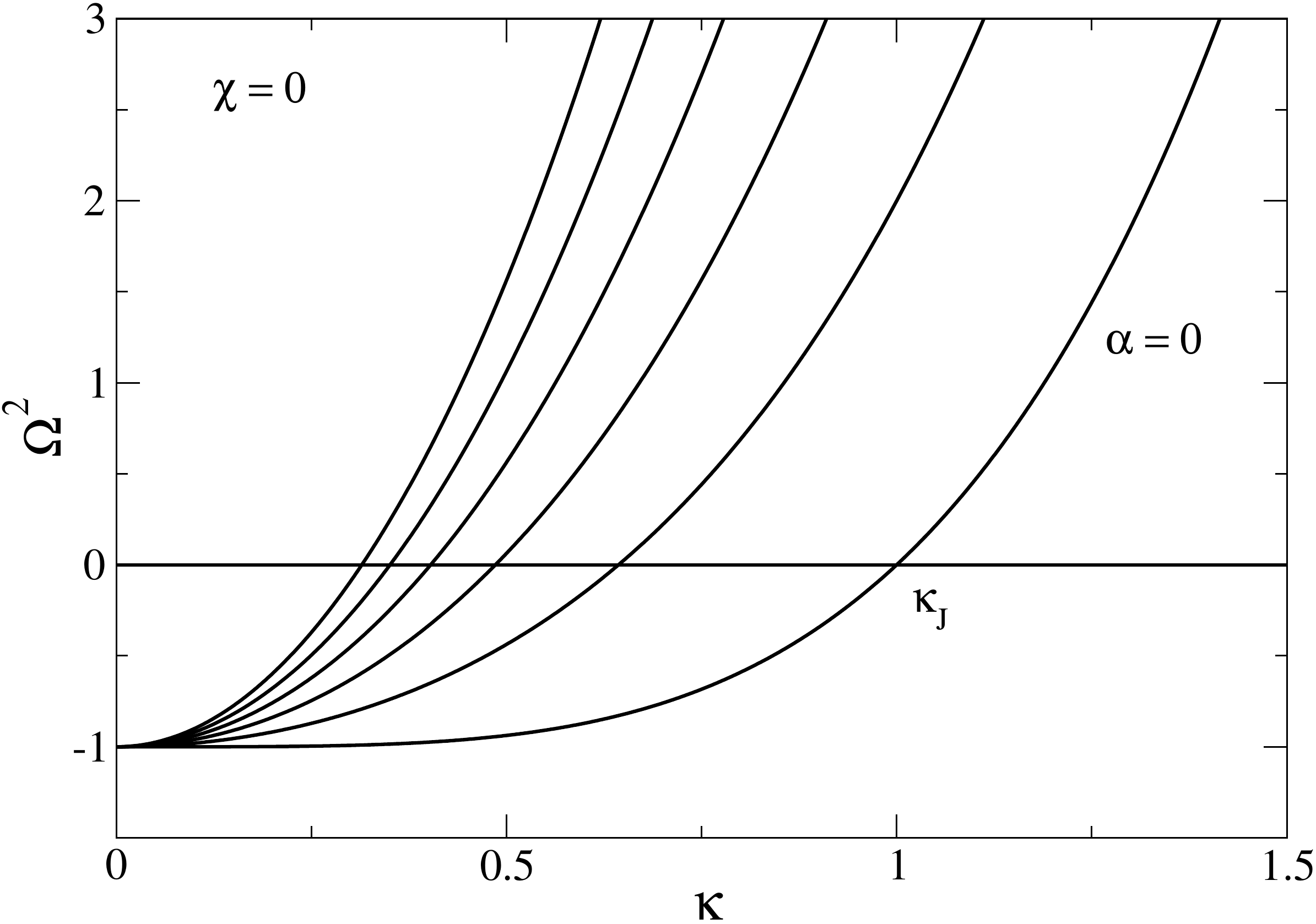}} 
\caption{Dispersion relation $\Omega^2(\kappa)$ of the nonrelativistic model
($\chi=0$) for
different values of the speed of sound (self-interaction) $\alpha=0,1,2,3,4,5$.}
\label{dispnewton}
\end{figure}

The Jeans wavenumber $k_J$, corresponding to $\omega=0$, is
determined by the equation
\begin{eqnarray}
\frac{\hbar^2k_J^4}{4m^2}+c_s^2k_J^2-4\pi
G\rho=0.
\label{nr4}
\end{eqnarray}
This is a second degree equation in $k_J^2$. Its physical solution
(corresponding to a real Jeans wavenumber) is \cite{prd1}:
\begin{eqnarray}
k_J^2=\frac{2m^2}{\hbar^2}\left ( -c_s^2+\sqrt{c_s^4
+\frac{4\pi G\rho\hbar^2}{m^2}}\right ).
\label{nr5}
\end{eqnarray}
The Jeans wavenumber $k_J^2$ is plotted as a function of the speed
of sound (self-interaction)  $\alpha\propto c_s^2/2$ in Fig. \ref{kjnewton}
using the normalization
of
Appendix \ref{sec_dv1}. The system is stable
($\omega^2>0$) for $k>k_J$ and
unstable ($\omega^2<0$) for $k<k_J$ (see Fig. \ref{dispnewton}). In the first
case, the perturbations
oscillate with a
pulsation $\omega=\sqrt{\omega^2}$. In the second case, the perturbations grow
exponentially rapidly with a growth rate $\sigma=\sqrt{-\omega^2}$. The
maximum growth rate
corresponds to $k_*=0$ (infinitely large
scales) and is equal to 
\begin{eqnarray}
\sigma_{\rm max}=(4\pi G\rho)^{1/2}.
\label{nr5b}
\end{eqnarray}
It is of the order of the inverse of the dynamical time
$t_{D}=(4\pi G\rho)^{-1/2}$ \cite{bt}. We
note from Figs. \ref{dispnewton} and \ref{kjnewton}
that the Jeans length increases with the speed of sound (self-interaction) while
the maximum growth
rate remains constant. Therefore, the self-interaction increases the Jeans
length
but does not change the maximum growth rate.

\begin{figure}[h]
\scalebox{0.33}{\includegraphics{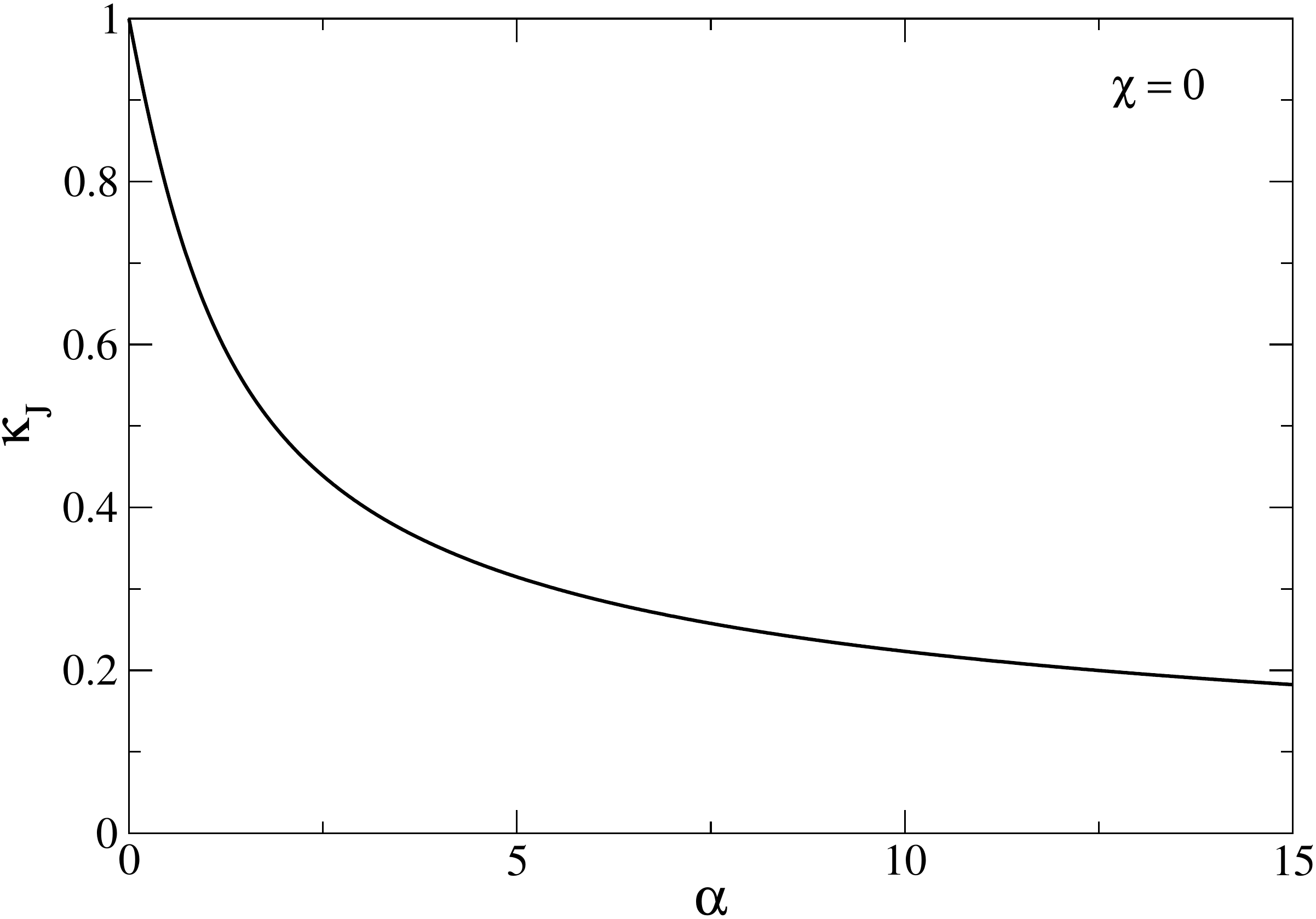}} 
\caption{Jeans wavenumber $\kappa_J$ of the nonrelativistic model ($\chi=0$) as
a function of the speed of sound 
$\alpha$ (self-interaction).}
\label{kjnewton}
\end{figure}

\subsection{The noninteracting limit}

In the noninteracting limit ($c_s=0$), the dispersion relation of Eq.
(\ref{nr1})
reduces to
\begin{eqnarray}
\omega^2=\frac{\hbar^2k^4}{4m^2}-4\pi
G\rho.
\label{nr6}
\end{eqnarray}
The Jeans wavenumber  is given by
\begin{eqnarray}
k_J^2=\left(\frac{16\pi G\rho m^2}{\hbar^2}\right)^{1/2}.
\label{nr7}
\end{eqnarray}
We  call it the quantum Jeans wavenumber because it is due to the competition
between the gravitational attraction and the repulsion due to the quantum
pressure arising from the Heisenberg uncertainty principle.

\subsection{The TF limit}

In the TF limit in which the quantum potential can be neglected
($\hbar=0$), the dispersion relation of Eq. (\ref{nr1}) reduces to
\begin{eqnarray}
\omega^2=c_s^2k^2-4\pi
G\rho.
\label{nr8}
\end{eqnarray}
This is the classical Jeans dispersion relation except that, in the present
context, the pressure term is due to the
self-interaction of the bosons instead of thermal motion (we recall that $T=0$
for our system). For $k\rightarrow
+\infty$:
\begin{eqnarray}
\omega^2(k)\sim c_s^2k^2.
\label{nr9}
\end{eqnarray}
For large wavenumbers (short wavelengths), the pressure term dominates the
gravitational term and the waves behave as sound waves.

The Jeans wavenumber is given by 
\begin{eqnarray}
k_J^2=\frac{4\pi G\rho}{c_s^2}.
\label{nr10}
\end{eqnarray}
It results from the competition between the gravitational attraction and the
repulsion due to the pressure arising from the self-interaction of the bosons. 
We call it the classical Jeans wavenumber because it is similar to
the ordinary Jeans wavenumber obtained from a classical hydrodynamic approach.
We
recall, however, that, in the present context, the pressure due to the
self-interaction of the bosons has an intrinsic quantum nature.

{\it Remark:} In the case $\hbar=c_s=0$, the dispersion relation of Eq.
(\ref{nr1}) reduces to $\omega^2=-4\pi G\rho$. The system is unstable at all
wavelengths and the growth rate has a constant value $\sigma=(4\pi
G\rho)^{1/2}$.

\section{The simplified relativistic model}
\label{sec_sm}

In this section, we consider a simplified relativistic model which corresponds
to the weak
field limit $\Phi/c^2\rightarrow 0$ of the KGE equations (see Appendix D
of \cite{suarezchavanis1} and Sec. 11 of \cite{suarezchavanis2}). This is
different from the nonrelativistic limit
$c\rightarrow +\infty$ considered in the previous section. This model coincides
with the one 
studied by  Khlopov {\it et al.} \cite{khlopov}. However, this model is not
fully accurate
because it neglects some terms in the
linearized KGE equations (the terms  $\Phi/c^2$ that are not
multiplied by $c^2$). Actually, in the relativistic regime, one must
take into account all the terms of order  $\Phi/c^2$ in the linearized
KGE equations. The exact relativistic model will be considered in Sec.
\ref{sec_em}.
However,
we first consider this simplified model in order to compare it later with the
exact one.

\subsection{The general case}

In the simplified  relativistic model, the dispersion relation is given by (see
Appendix D of \cite{suarezchavanis1}):
\begin{eqnarray}
\frac{\hbar^2}{4m^2c^4}\omega^4
&-&\biggl(1+\frac{3c_s^2}{c^2}+\frac{\hbar^2k^2}{2m^2c^2}\biggr)\omega^2+\frac{\hbar^2k^4}{4m^2}+c_s^2k^2\nonumber\\
&-&4\pi G\rho\left(1+\frac{2c_s^2}{c^2}\right)=0.
\label{sm1}
\end{eqnarray}
This is a second degree equation in $\omega^2$. The
discriminant of this equation
\begin{eqnarray}
\Delta=\left (1+\frac{3c_s^2}{c^2}\right )^2+\left (1+\frac{2c_s^2}{c^2}\right
)\frac{\hbar^2 k^2}{m^2c^2}\nonumber\\
+\frac{\hbar^2}{m^2c^4}4\pi G\rho \left
(1+\frac{2c_s^2}{c^2}\right )
\label{sm2}
\end{eqnarray}
is manifestly positive. Therefore, the dispersion
relation has two real
branches $\omega_{\pm}^2(k)$.  For $k=0$:
\begin{eqnarray}
\omega_{\pm}^2(0)=\frac{2m^2c^4}{\hbar^2}\Biggl\lbrace
1+\frac{3c_s^2}{c^2}\pm\Biggl\lbrack\left (1+\frac{3c_s^2}{c^2}\right
)^2\nonumber\\
+\frac{4\pi G\rho\hbar^2}{m^2c^4}\left
(1+\frac{2c_s^2}{c^2}\right )\Biggr\rbrack^{1/2}\Biggr \rbrace.
\label{sm3}
\end{eqnarray}
We note that $\omega_{+}^2(0)>0$ and $\omega_{-}^2(0)<0$. For $k\rightarrow 0$:
\begin{eqnarray}
\omega_{\pm}^2(k)\simeq \omega_{\pm}^2(0)+c_{\rm eff}^2k^2.
\label{sm4}
\end{eqnarray}
The coefficient in front of $k^2$ can be interpreted as an effective speed of
sound (compare Eq. (\ref{sm4}) with Eq. (\ref{nr2})). It is given by
\begin{eqnarray}
\frac{c_{\rm
eff}^2}{c^2}=1\pm\frac{1+\frac{2c_s^2}{c^2}}{\sqrt{\left
(1+\frac{3c_s^2}{c^2}\right )^2+\frac{4\pi G\rho\hbar^2}{m^2 c^4}\left
(1+\frac{2c_s^2}{c^2}\right )}}.
\label{sm5}
\end{eqnarray}
For
$k\rightarrow +\infty$:
\begin{eqnarray}
\omega_{\pm}^2(k)\simeq c^2k^2\left (1\pm\frac{2mc}{\hbar
k}\sqrt{1+\frac{2c_s^2}{c^2}}\right ).
\label{sm6}
\end{eqnarray}
In that limit, the effective speed of sound is equal to the speed of light
($c_{\rm eff}=c$).

The square pulsation $\omega_{+}^2(k)$ starts from a positive value, increases,
and tends to $+\infty$ as $k\rightarrow +\infty$. Therefore, the branch $(+)$
corresponds to stable modes that oscillate
with a pulsation $\omega_+=(\omega_{+}^2)^{1/2}$. The minimum
pulsation corresponds to $k=0$ (infinitely large scales) and is given by
$(\omega_+)_{\rm
min}=\omega_{+}(0)$ where $\omega_{+}^2(0)$ is given by
Eq. (\ref{sm3}). In the nonrelativistic limit $c\rightarrow +\infty$, the
minimum pulsation  
$(\omega_+)_{\rm min}$ tends to infinity.
This is the reason why the branch $(+)$ does not appear
in the nonrelativistic analysis of Sec. \ref{sec_nr} (it is rejected at
infinity).

The  square pulsation $\omega_{-}^2(k)$ starts from a
negative value 
$\omega_{-}^2(0)<0$, increases, vanishes at $k=k_J$, and tends to $+\infty$ as
$k\rightarrow +\infty$.
Therefore, there exist stable and unstable modes. 
The Jeans wavenumber $k_J$, corresponding to 
$\omega=0$, is determined by the equation
\begin{eqnarray}
\frac{\hbar^2k_J^4}{4m^2}+c_s^2k_J^2-4\pi
G\rho\left(1+\frac{2c_s^2}{c^2}\right)=0.
\label{sm7}
\end{eqnarray}
This is a second degree equation in $k_J^2$. Its physical
solution is
\begin{equation}
k_J^2=\frac{2m^2}{\hbar^2}\Biggl\lbrace -c_s^2+\Biggl\lbrack c_s^4
+\frac{4\pi G\rho\hbar^2}{m^2}\left
(1+\frac{2c_s^2}{c^2}\right )\Biggr\rbrack^{1/2}\Biggr \rbrace.
\label{sm8}
\end{equation}
The system is stable ($\omega^2>0$) for $k>k_J$ and
unstable ($\omega^2<0$) for $k<k_J$. In the first case, the perturbations
oscillate with a
pulsation $\omega_-=(\omega_-^2)^{1/2}$. In the second case, the perturbations
grow exponentially rapidly with a growth rate $\sigma=(-\omega_-^2)^{1/2}$. The
maximum growth rate corresponds to $k=0$ (infinitely large scales) and is given
by $\sigma_{\rm max}=(-\omega_{-}^2(0))^{1/2}$ where $\omega_{-}^2(0)$ is given
by Eq. (\ref{sm3}). In the nonrelativistic limit $c\rightarrow +\infty$, the
maximum growth rate $\sigma_{\rm max}$ tends to the value $(4\pi
G\rho)^{1/2}$ of Sec. \ref{sec_nr}.

\subsection{The noninteracting limit}

\begin{figure}[h]
\scalebox{0.33}{\includegraphics{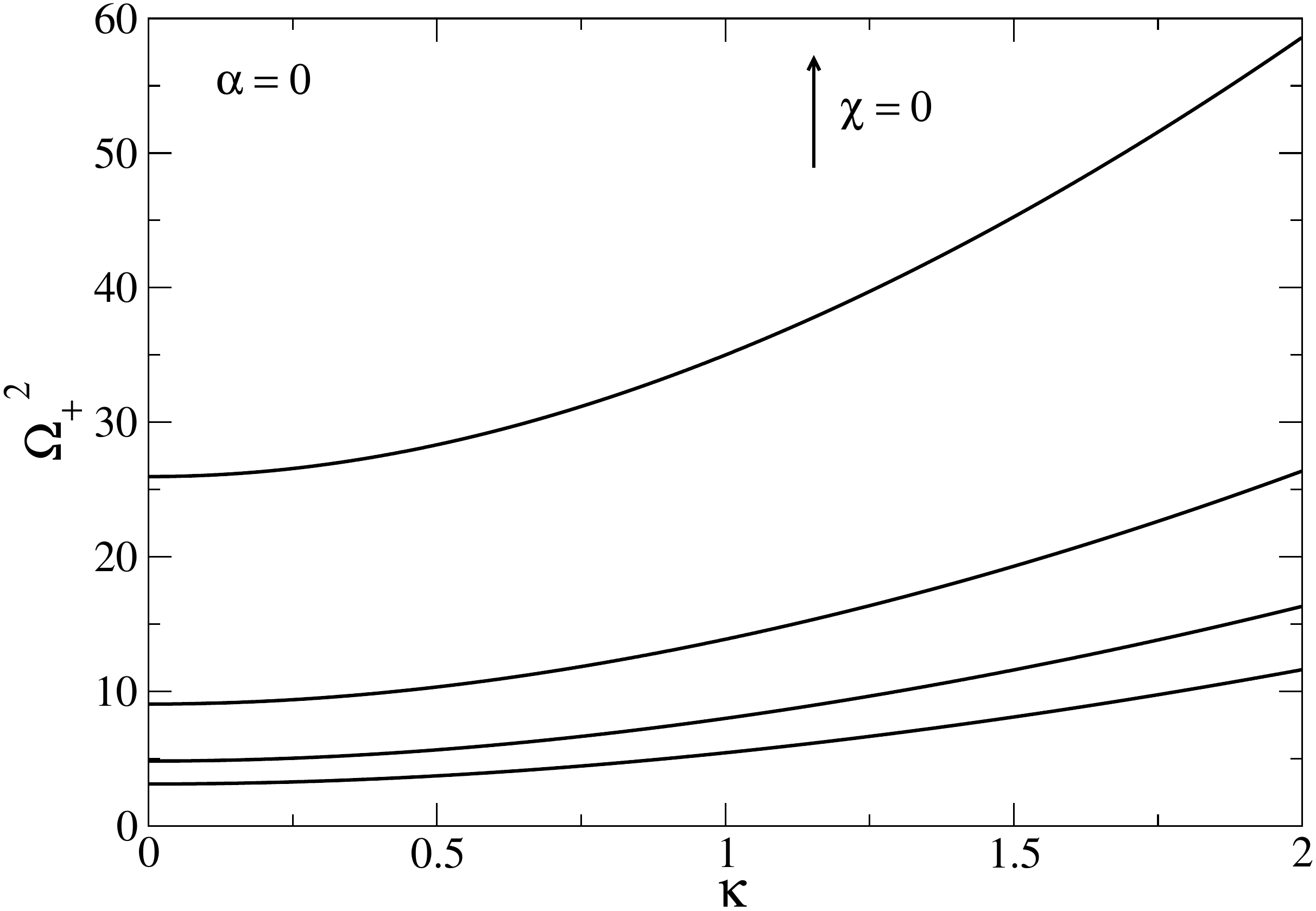}} 
\caption{Dispersion relation
$\Omega_+^2(\kappa)$ of the simplified relativistic model in the noninteracting
limit ($\alpha=0$) for different values of the relativistic parameter
$\chi=0.4,0.7,1,1.3$.}
\label{dispsimple}
\end{figure}

\begin{figure}[h]
\scalebox{0.33}{\includegraphics{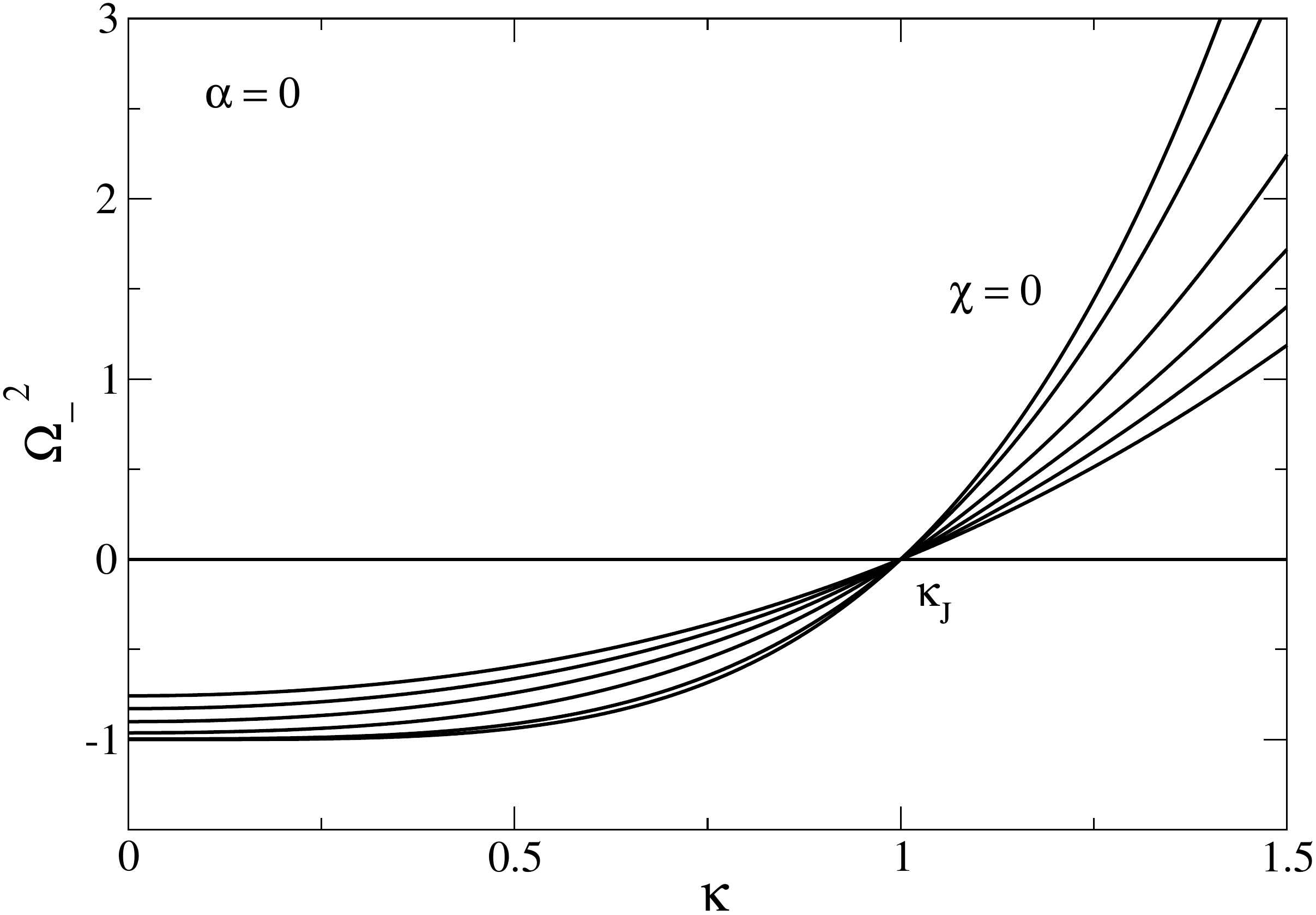}} 
\caption{Dispersion
relation $\Omega_-^2(\kappa)$ of the simplified relativistic model in the
noninteracting limit ($\alpha=0$) for different values of the relativistic
parameter $\chi=0,0.1,0.4,0.7,1,1.3$. We note that
the Jeans wavenumber $\kappa_J=1$ is independent of
the relativistic parameter.}
\label{dispsimpleMOINS}
\end{figure}

In the noninteracting limit ($c_s=0$), the dispersion relation of Eq.
(\ref{sm1}) reduces to
\begin{eqnarray}
\frac{\hbar^2}{4m^2c^4}\omega^4-\biggl (
1+\frac{\hbar^2k^2}{2m^2c^2}\biggr )\omega^2
+\frac{\hbar^2k^4}{4m^2}-4\pi
G\rho=0.
\label{sm9}
\end{eqnarray}
The functions $\omega_{\pm}^2(k)$ are
represented in Figs. \ref{dispsimple} and \ref{dispsimpleMOINS} for different
values of the relativistic
parameter $\chi\propto 2/c^2$ using the normalization of Appendix
\ref{sec_dv1}. For $k\rightarrow 0$, the solution of Eq.
(\ref{sm9}) can be written as Eq. (\ref{sm4}) with
\begin{eqnarray}
\omega_{\pm}^2(0)=\frac{2m^2c^4}{\hbar^2}\left (
1\pm\sqrt{1+\frac{4\pi G\rho\hbar^2}{m^2c^4}}\right )
\label{sm10}
\end{eqnarray}
and
\begin{eqnarray}
\frac{c_{\rm
eff}^2}{c^2}=1\pm\frac{1}{\sqrt{1+\frac{4\pi G\rho\hbar^2}{m^2 c^4}}}.
\label{sm11}
\end{eqnarray}
For $k\rightarrow +\infty$:
\begin{eqnarray}
\omega_{\pm}^2(k)\simeq c^2k^2\left (1\pm\frac{2mc}{\hbar
k}\right ).
\label{sm12}
\end{eqnarray}

Concerning the branch (+), the minimum pulsation corresponds to $k=0$ and is
given by
$(\omega_+)_{\rm min}=(\omega_{+}^2(0))^{1/2}$  where $\omega_{+}^2(0)$ is given
by Eq. (\ref{sm10}). It is plotted as a function of the relativistic parameter 
$\chi\propto 2/c^2$ in Fig. \ref{Wplus} of the next section  using the
normalization of
Appendix
\ref{sec_dv1}. Its asymptotic behaviors are given in Appendix \ref{sec_mgrnis}.
 We note that the minimum pulsation
$(\omega_+)_{\rm min}$ decreases as relativistic effects
increase.

Concerning the branch (-), the
Jeans wavenumber is given by
\begin{eqnarray}
k_J^2=\left(\frac{16\pi G\rho m^2}{\hbar^2}\right)^{1/2}.
\label{sm13}
\end{eqnarray}
It has the same expression as in the nonrelativistic model [see Eq.
(\ref{nr7})]. The maximum growth rate corresponds to $k_*=0$ (infinitely large
scales) and
is given by $\sigma_{\rm max}=(-\omega_{-}^2(0))^{1/2}$  where $\omega_{-}^2(0)$
is given by Eq. (\ref{sm10}). It is plotted as a function of  the relativistic
parameter $\chi\propto 2/c^2$ in Fig. \ref{maxgrowthrateOm} of the
next section  using the
normalization of Appendix
\ref{sec_dv1}. Its asymptotic behaviors are given in Appendix
\ref{sec_mgrnis}. We note that the  maximum growth rate $\sigma_{\rm max}$
decreases as
relativistic effects increase and tends to zero in the ultrarelativistic limit
($c\rightarrow 0$).

\subsection{The TF limit}

\begin{figure}[h]
\scalebox{0.33}{\includegraphics{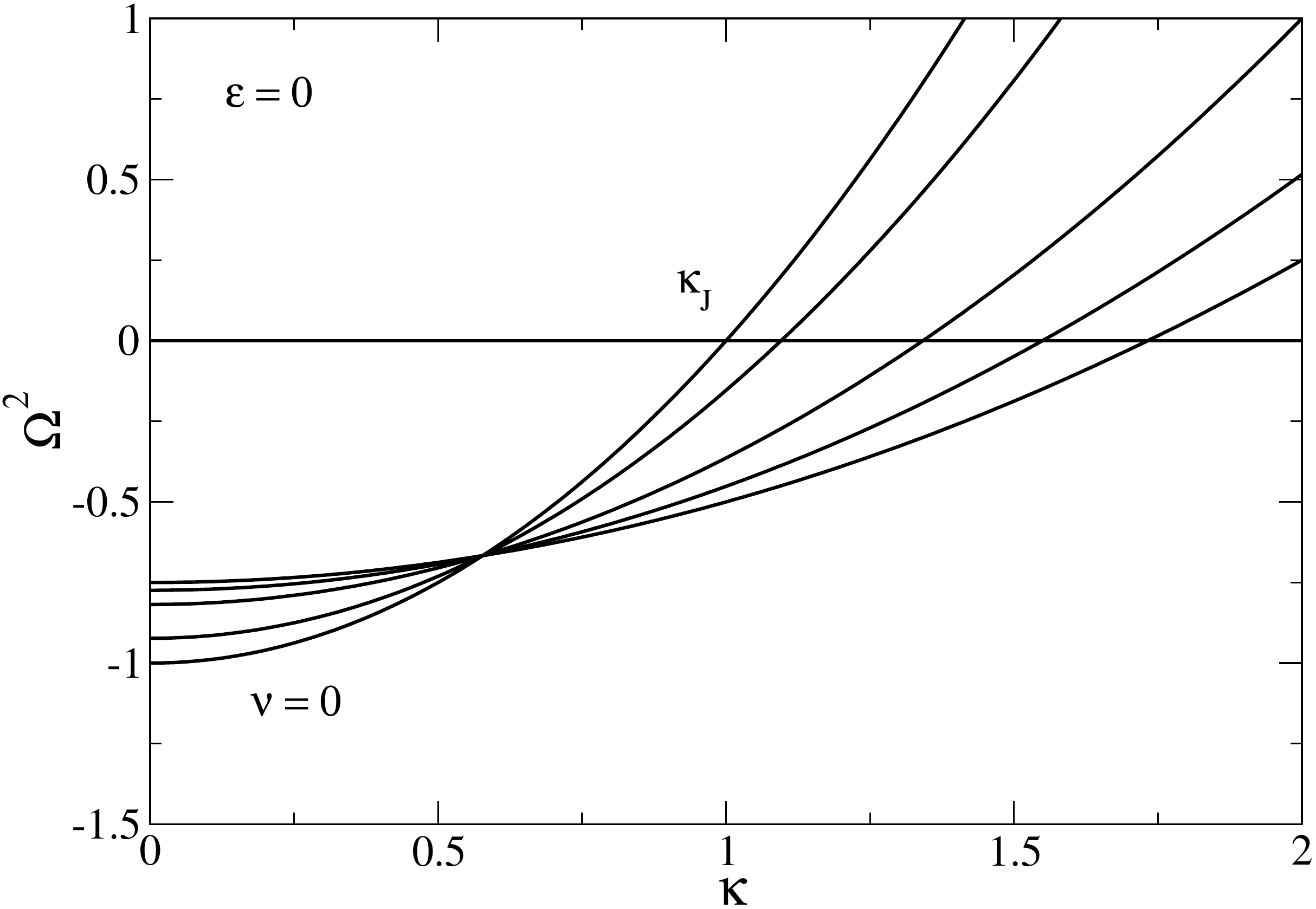}} 
\caption{Dispersion relation $\Omega^2(\kappa)$ of the simplified
relativistic model in the TF limit ($\epsilon=0$)  for different values of the
relativistic parameter $\nu=0,0.1,0.4,0.7,1$. We note
the remarkable fact that all the curves cross each
other at the point  $(\kappa,\Omega^2)=(1/\sqrt{3},-2/3)$. In other words, for
$\kappa=-1/\sqrt{3}$ the pulsation is $\Omega^2=-2/3$ independently of the
value of the relativistic parameter.}
\label{dispsimpleTF}
\end{figure}

In the TF limit ($\hbar=0$), the dispersion relation of Eq.
(\ref{sm1}) reduces to
\begin{eqnarray}
\omega^2=\frac{c_s^2k^2-4\pi
G\rho\left(1+\frac{2c_s^2}{c^2}\right)}{1+\frac{3c_s^2}{c^2}}.
\label{sm14}
\end{eqnarray}
The function $\omega^2(k)$ is
represented in Fig. \ref{dispsimpleTF} for different values of the relativistic
parameter
$\nu\propto 1/c^2$ using the normalization of Appendix \ref{sec_dv2}. It
corresponds to the limit form of the branch $(-)$. The
branch $(+)$ is
rejected at infinity. Equation
(\ref{sm14}) can be written as Eq.
(\ref{sm4}) with
\begin{eqnarray}
\omega^2(0)=-4\pi G\rho\frac{1+\frac{2c_s^2}{c^2}}{1+\frac{3c_s^2}{c^2}}
\label{sm15}
\end{eqnarray}
and
\begin{eqnarray}
c_{\rm eff}^2=\frac{c_s^2}{1+\frac{3c_s^2}{c^2}}.
\label{sm16}
\end{eqnarray}

The Jeans wavenumber is given by
\begin{eqnarray}
k_J^2=\frac{4\pi G\rho}{c_s^2}\left (1+\frac{2c_s^2}{c^2}\right ).
\label{sm17}
\end{eqnarray}
It is plotted as a function of the relativistic parameter 
$\nu\propto 1/c^2$ in Fig. \ref{kjTF} of the next section  using the
normalization of
Appendix
\ref{sec_dv2}. Its asymptotic behaviors are given in Appendix
\ref{sec_mgrtfsm}.   We
note that the Jeans length
decreases as relativistic effects increase. The
maximum growth
rate corresponds to $k_*=0$ (infinitely large scales) and is given by
$\sigma_{\rm max}=(-\omega_{-}^2(0))^{1/2}$  where $\omega_{-}^2(0)$ is given by
Eq. (\ref{sm15}). It is plotted as a function of  the relativistic
parameter $\nu\propto 1/c^2$ in Fig. \ref{nuOmegaexact} of the next section 
using the
normalization of Appendix
\ref{sec_dv2}. Its asymptotic behaviors are given in Appendix
\ref{sec_mgrtfsm}.  We
note that the maximum growth rate  $\sigma_{\rm max}$ decreases as
relativistic effects increase and tends to a constant in the 
ultrarelativistic limit ($c\rightarrow 0$).

{\it Remark:} In the case $\hbar=c_s=0$, the dispersion relation of Eq.
(\ref{sm1}) reduces to $\omega^2=-4\pi G\rho$ like in the nonrelativistic model.

\section{The exact relativistic model}
\label{sec_em}

In this section, we consider the exact relativistic model of
\cite{suarezchavanis1} which is based on the
exact linearized KGE equations.

\subsection{The general case}
\label{sec_empa}

In the exact relativistic model, the dispersion relation is given by (see Sec.
IV.D of \cite{suarezchavanis1}):
\begin{eqnarray}
\frac{\hbar^2}{4m^2c^4}\omega^4-\left (
\frac{1+\gamma}{3\gamma+1}\frac{\hbar^2k^2}{2
m^2c^2}+1+\frac{3c_s^2}{c^2}\right )\omega^2\nonumber\\
+\frac{1}{1+3\gamma}\Biggl\lbrack
(1-\gamma)\frac{\hbar^2k^4}{4m^2}+(1-3\gamma) k^2 c_s^2\nonumber\\
-4\pi G\rho
\left (1+\frac{2c_s^2}{c^2}\right )
\Biggr\rbrack=0,
\label{em1}
\end{eqnarray}
where we have introduced the abbreviation
\begin{eqnarray}
\gamma=\frac{4\pi G\rho}{k^2c^2}\left (1+\frac{2c_s^2}{c^2}\right ).
\label{em2}
\end{eqnarray}
This is a second degree equation in
$\omega^2$. The
discriminant of this equation
\begin{eqnarray}
\Delta=\frac{\hbar^4k^4}{m^4c^4}\frac{\gamma^2}{(3\gamma+1)^2}
+\frac{
\hbar^2k^2 } {m^2c^2}\frac{1}{3\gamma+1}\nonumber\\
\times\left\lbrack
1+2\gamma+(6\gamma+2)\frac{c_s^2}{c^2}\right\rbrack
+\left
(1+\frac{3c_s^2}{c^2}\right )^2
\label{em3}
\end{eqnarray}
is manifestly positive. Therefore, the dispersion relation has two real
branches $\omega_{\pm}^2(k)$. For $k\rightarrow 0$:
\begin{eqnarray}
\omega_+^2(k)\simeq \frac{4m^2c^4}{\hbar^2}\left
(1+\frac{3c_s^2}{c^2}\right )+k^2c^2+...,
\label{em6}
\end{eqnarray}
\begin{eqnarray}
\omega_-^2(k)\sim -\frac{1}{3}k^2c^2.
\label{em7}
\end{eqnarray}
For $k\rightarrow +\infty$:
\begin{eqnarray}
\omega_{\pm}^2(k)\simeq c^2k^2\left (1\pm\frac{2mc}{\hbar
k}\sqrt{1+\frac{2c_s^2}{c^2}}\right ).
\label{em8}
\end{eqnarray}
The effective speed of sound on the
branch ($+$) is equal to the speed of light ($c_{\rm eff}=c$) for both small
and large $k$. The effective speed of sound on the
branch ($-$) is  imaginary ($c_{\rm eff}^2=-c^2/3$) for small $k$ and is equal
to the speed of light ($c_{\rm eff}=c$) for large
$k$.

The square pulsation $\omega_{+}^2(k)$ starts from a positive
value $\omega_{+}^2(0)>0$, increases,
and tends to $+\infty$ as $k\rightarrow +\infty$. Therefore, the branch $(+)$
corresponds to stable modes that oscillate
with a pulsation $\omega_+=(\omega_{+}^2)^{1/2}$. The minimum
pulsation corresponds to $k=0$ (infinitely large scales) and is given by
$(\omega_+)_{\rm
min}=\omega_{+}(0)$ where $\omega_{+}^2(0)$ is given by the first term in 
Eq. (\ref{em6}). In the nonrelativistic limit $c\rightarrow +\infty$, the
minimum pulsation  
$(\omega_+)_{\rm min}$ tends to infinity. This is the reason why the branch
$(+)$ does not appear
in the nonrelativistic analysis of Sec. \ref{sec_nr} (it is rejected at
infinity).

The square pulsation $\omega_-^2(k)$ starts from zero, decreases, reaches a
minimum value $\omega_-^2(k_*)$  at $k_*$, increases,
vanishes at $k=k_J$, and
tends to
$+\infty$ as $k\rightarrow +\infty$.  Therefore, the branch ($-$) has 
stable (oscillating) and unstable (growing) modes. The Jeans wavenumber, 
corresponding to $\omega=0$, is determined by the equation 
\begin{eqnarray}
\frac{\hbar^2k_J^4}{4m^2}+\left\lbrack c_s^2-\frac{\pi 
G\rho\hbar^2}{m^2c^2}\left (1+\frac{2c_s^2}{c^2}\right )\right\rbrack
k_J^2\nonumber\\
-4\pi G\rho \left (1+\frac{2c_s^2}{c^2}\right ) \left
(1+\frac{3c_s^2}{c^2}\right )=0.
\label{em9}
\end{eqnarray}
This is a second degree equation in $k_J^2$. Its physical
solution is
\begin{eqnarray}
k_J^2=\frac{2m^2}{\hbar^2}\Biggl (-c_s^2+\frac{\pi
G\rho \hbar^2}{m^2c^2}\left (1+\frac{2c_s^2}{c^2}\right )\nonumber\\
+\Biggl\lbrace
\Biggl \lbrack c_s^2-\frac{\pi
G\rho \hbar^2}{m^2c^2}\left (1+\frac{2c_s^2}{c^2}\right
)\Biggr\rbrack^2\nonumber\\
+\frac{4\pi G\rho\hbar^2}{m^2}\left
(1+\frac{2c_s^2}{c^2}\right )\left (1+\frac{3c_s^2}{c^2}\right
)\Biggr\rbrace^{1/2}\Biggr ).
\label{em10}
\end{eqnarray}
The system is stable ($\omega_-^2>0$) for $k>k_J$ and
unstable ($\omega_-^2<0$) for $k<k_J$. In the first case, the perturbations
oscillate with a
pulsation  $\omega_-=(\omega_-^2)^{1/2}$. In the second case, the
perturbations grow
exponentially rapidly with a growth rate
$\sigma=(-\omega_-^2)^{1/2}$. The maximum growth rate
corresponds to the optimal wavenumber $k_*$ and is
given by $\sigma_{\rm max}=(-\omega_{-}^2(k_*))^{1/2}$. In the
nonrelativistic limit $c\rightarrow +\infty$, $k_*\rightarrow 0$ and
$\sigma_{\rm max}\rightarrow (4\pi
G\rho)^{1/2}$ as in Sec. \ref{sec_nr}.

{\it Remark:} In the exact
relativistic model, the system tends to be stabilized at
very large scales since the growth rate $\sigma(k)$ vanishes at $k=0$ while
in the nonrelativistic model and
in the simplified relativistic model the growth rate $\sigma(k)$ is maximum at
$k=0$. Therefore, the exact model turns out to be very different from
the nonrelativistic model and from the simplified model at large scales.
General relativistic effects, when they are fully taken into account, tend to
stabilize
the system at large scales. For $c\rightarrow +\infty$, using
Eq. (\ref{em7}), we find that this
stabilization occurs when $-\omega_-^2\sim
({1}/{3})k^2c^2$ is of the order of $\sigma_{\rm
max,0}^2=4\pi G\rho$, corresponsing to a lengthscale of the order of
the
Hubble length $\lambda_H=c/H\sim c/\sqrt{G\rho}$ (see Appendix
\ref{sec_hs}). Above the Hubble length $\lambda_H$
(horizon), the growth rate decreases towards zero. A similar
stabilization at large scales, above the Hubble length, due to
general relativity, was found in \cite{suarezchavanis1} when
considering the growth of structures in an expanding Universe.

\subsection{The noninteracting limit}
\label{sec_emni}

\begin{figure}[h]
\scalebox{0.33}{\includegraphics{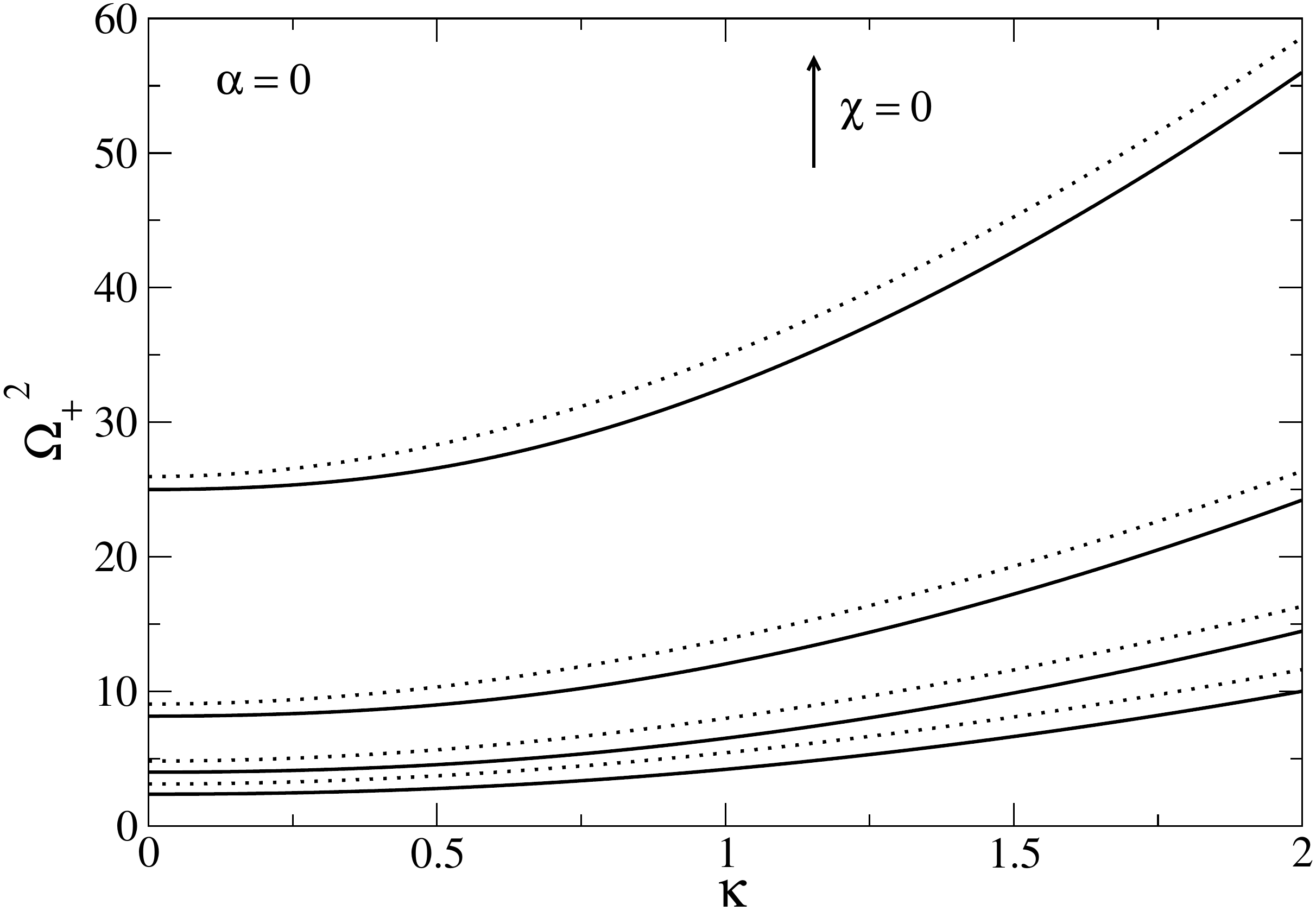}} 
\caption{Dispersion relation $\Omega_+^2(\kappa)$ of the relativistic model in
the noninteracting limit ($\alpha=0$) for different values of the
relativistic parameter $\chi=0.4,0.7,1,1.3$ (solid
lines: exact relativistic model; dotted lines:
simplified relativistic model).}
\label{dispexact}
\end{figure}

\begin{figure}[h]
\scalebox{0.33}{\includegraphics{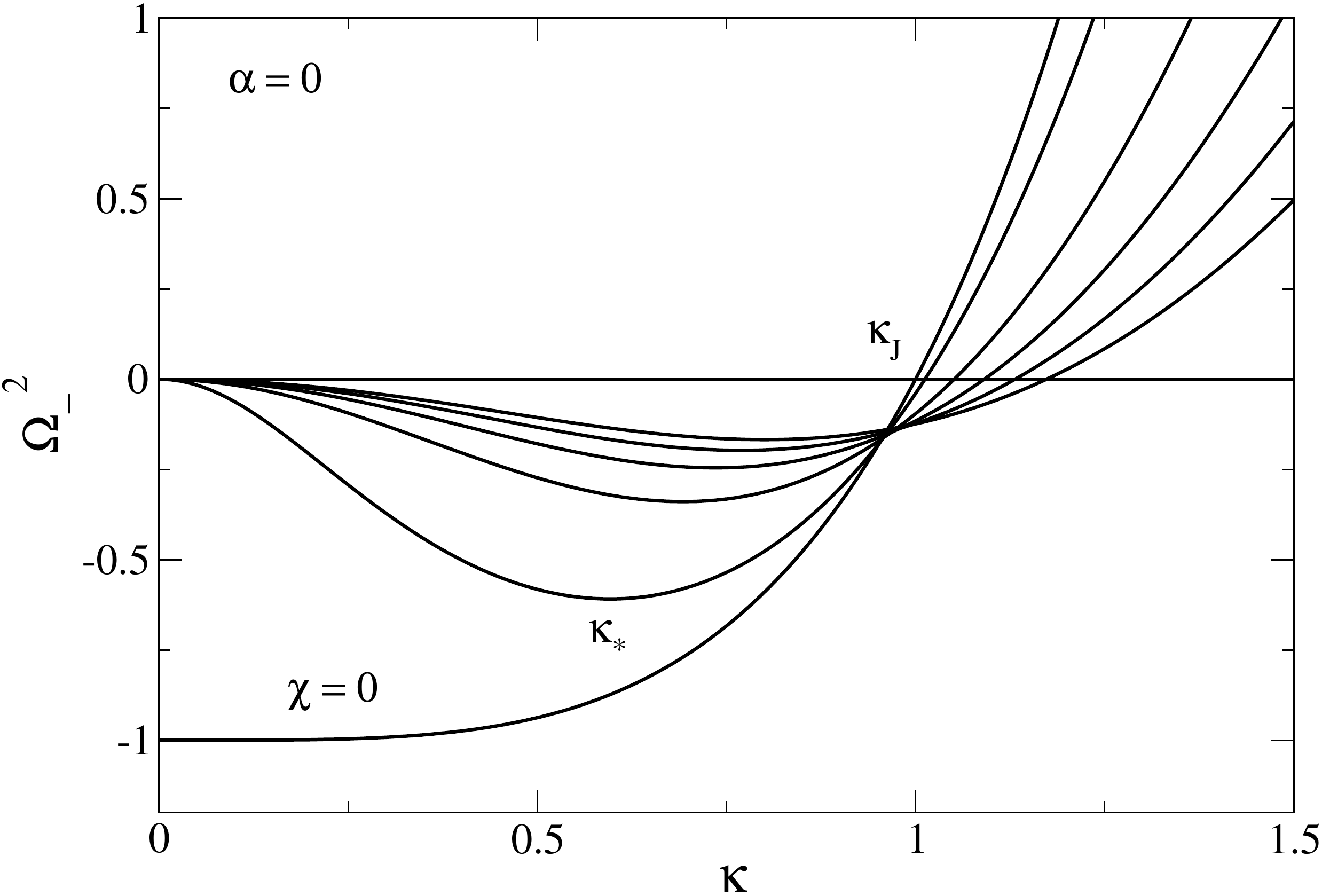}} 
\caption{Dispersion
relation $\Omega_-^2(\kappa)$ of the exact relativistic model in the
noninteracting limit ($\alpha=0$) for different values of the relativistic
parameter $\chi=0,0.1,0.4,0.7,1,1.3$. The growth
rate $\Sigma(\kappa)=(-\Omega_-^2(\kappa))^{1/2}$ reaches a maximum value
at a nonzero wavenumber $\kappa_*$. We note that the curves
do not exactly cross each other at the same point  contrary to what might be
thought
from the figure.}
\label{dispexactMOINS}
\end{figure}

In the noninteracting limit ($c_s=0$), the dispersion relation of Eq.
(\ref{em1})
reduces to
\begin{eqnarray}
\frac{\hbar^2}{4m^2c^4}\omega^4-\left (
\frac{1+\gamma}{3\gamma+1}\frac{\hbar^2k^2}{2
m^2c^2}+1\right )\omega^2\nonumber\\
+\frac{1}{1+3\gamma}\Biggl\lbrack
(1-\gamma)\frac{\hbar^2k^4}{4m^2}
-4\pi G\rho
\Biggr\rbrack=0,
\label{em11}
\end{eqnarray}
where 
\begin{eqnarray}
\gamma=\frac{4\pi G\rho}{k^2c^2}.
\label{em12}
\end{eqnarray}
The functions $\omega_{\pm}^2(k)$ are
represented in Figs. \ref{dispexact} and \ref{dispexactMOINS} for different
values of the relativistic
parameter $\chi\propto 2/c^2$ using the normalization of Appendix
\ref{sec_dv1}. For $k\rightarrow 0$:
\begin{eqnarray}
\omega_+^2(k)\simeq \frac{4m^2c^4}{\hbar^2}+k^2c^2+...,
\label{em13}
\end{eqnarray}
\begin{eqnarray}
\omega_-^2(k)\sim -\frac{1}{3}k^2c^2.
\label{em14}
\end{eqnarray}
For $k\rightarrow +\infty$:
\begin{eqnarray}
\omega_{\pm}^2(k)\sim k^2c^2\left(1\pm\frac{2mc}{\hbar k}\right).
\label{em15}
\end{eqnarray}

Concerning the branch (+), the minimum pulsation corresponds to $k=0$ and is
given by $(\omega_+)_{\rm min}=2mc^2/\hbar$. It is plotted as a function of the
relativistic parameter $\chi\propto 2/c^2$ in Fig.
\ref{Wplus} using the normalization of Appendix
\ref{sec_dv1} (see also Appendix
\ref{sec_mgrne}). We note that the minimum pulsation decreases
as relativistic effects increase. In the nonrelativistic limit
($\chi\rightarrow 0$), we find that $\omega_+(0)\rightarrow +\infty$. In the
ultrarelativistic limit
($\chi\rightarrow +\infty$), we find
that  $\omega_+(0)\rightarrow 0$.

\begin{figure}[h]
\scalebox{0.33}{\includegraphics{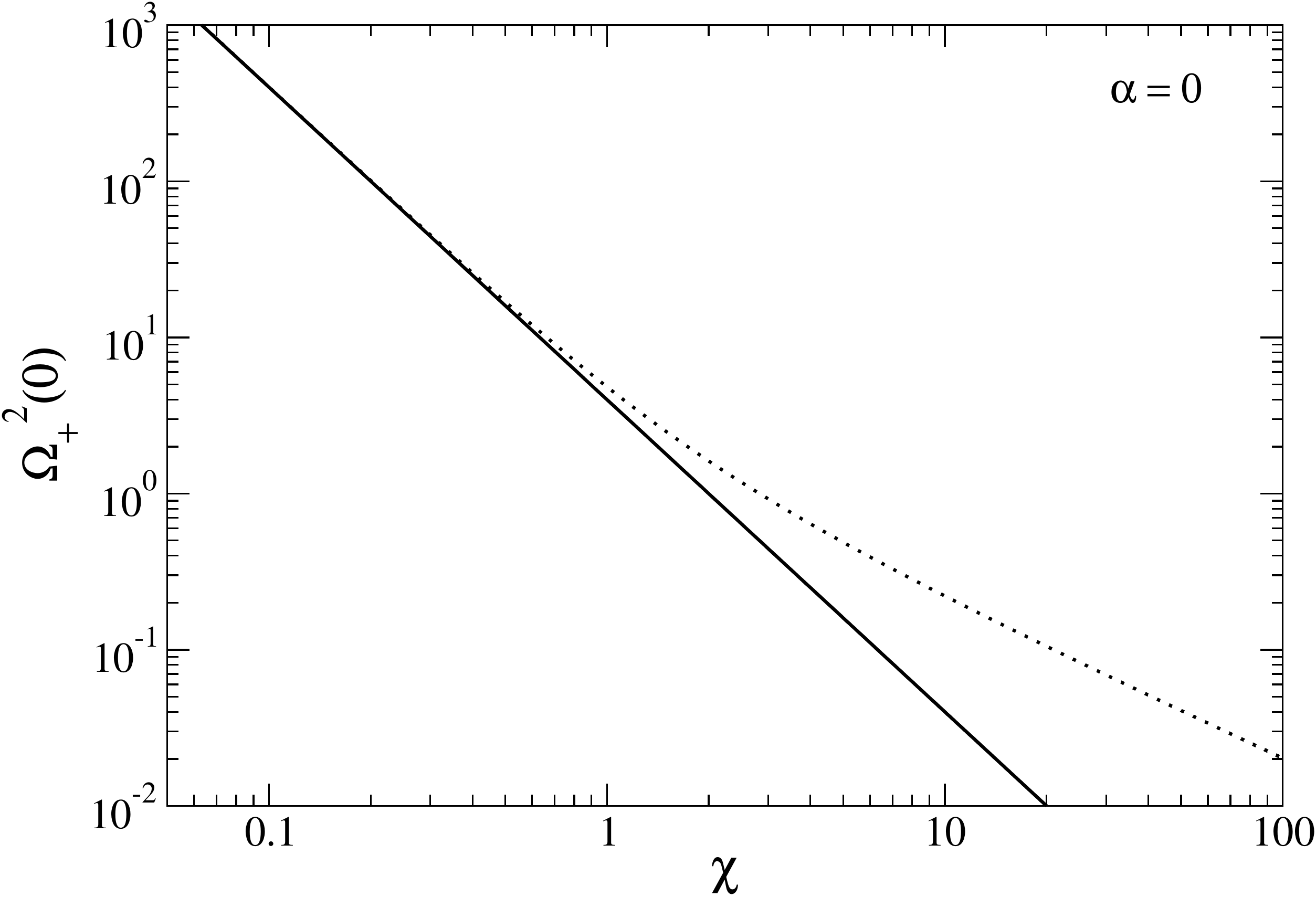}} 
\caption{Minimum pulsation $\Omega_+^2(0)$ of the branch
(+) as a function of the
relativistic parameter $\chi$ (solid
line: exact relativistic model; dotted line:
simplified relativistic model). }
\label{Wplus}
\end{figure}

Concerning the branch (-), the Jeans wavenumber is given by
\begin{eqnarray}
k_J^2=\frac{2\pi G\rho}{c^2}\left (1+\sqrt{1+\frac{4m^2c^4}{\pi
G\rho\hbar^2}}\right ).
\label{em16}
\end{eqnarray}
We note that the exact expression (\ref{em16}) of the Jeans wavenumber depends
on $c$ contrary to the expression (\ref{sm13}) obtained from the simplified
model.
The Jeans wavenumber $k_J$ is plotted as a function
of the relativistic parameter $\chi\propto 2/c^2$ in Fig. \ref{kj} using the
normalization of Appendix
\ref{sec_dv1}.  Its asymptotic behaviors are given in Appendix
\ref{sec_mgrne}. We note that the Jeans length decreases as relativistic
effects increase.   The optimal wavenumber
$k_*$ and
the maximum growth rate $\sigma_{\rm max}$ are plotted as a function
of the relativistic parameter $\chi\propto 2/c^2$ in Figs.
\ref{maxgrowthrateK} and \ref{maxgrowthrateOm} using the
normalization of Appendix
\ref{sec_dv1}. Their asymptotic behaviors are given in Appendix \ref{sec_mgrne}.
We note that the optimal wavelength  and the maximum growth rate both decrease
as relativistic effects increase. In the nonrelativistic limit
($\chi\rightarrow 0$), we find that $\lambda_J\rightarrow 2\pi(\hbar^2/16\pi
G\rho m^2)^{1/4}$, $\lambda_*\rightarrow +\infty$ and
$\sigma_{\rm max}\rightarrow (4\pi G\rho)^{1/2}$. In the ultrarelativistic limit
($\chi\rightarrow +\infty$), we find
that $\lambda_*\sim 1.47\lambda_J\rightarrow 0$ and $\sigma_{\rm max}\rightarrow
0.268 (4\pi
G\rho)^{1/2}$.

\begin{figure}[h]
\scalebox{0.33}{\includegraphics{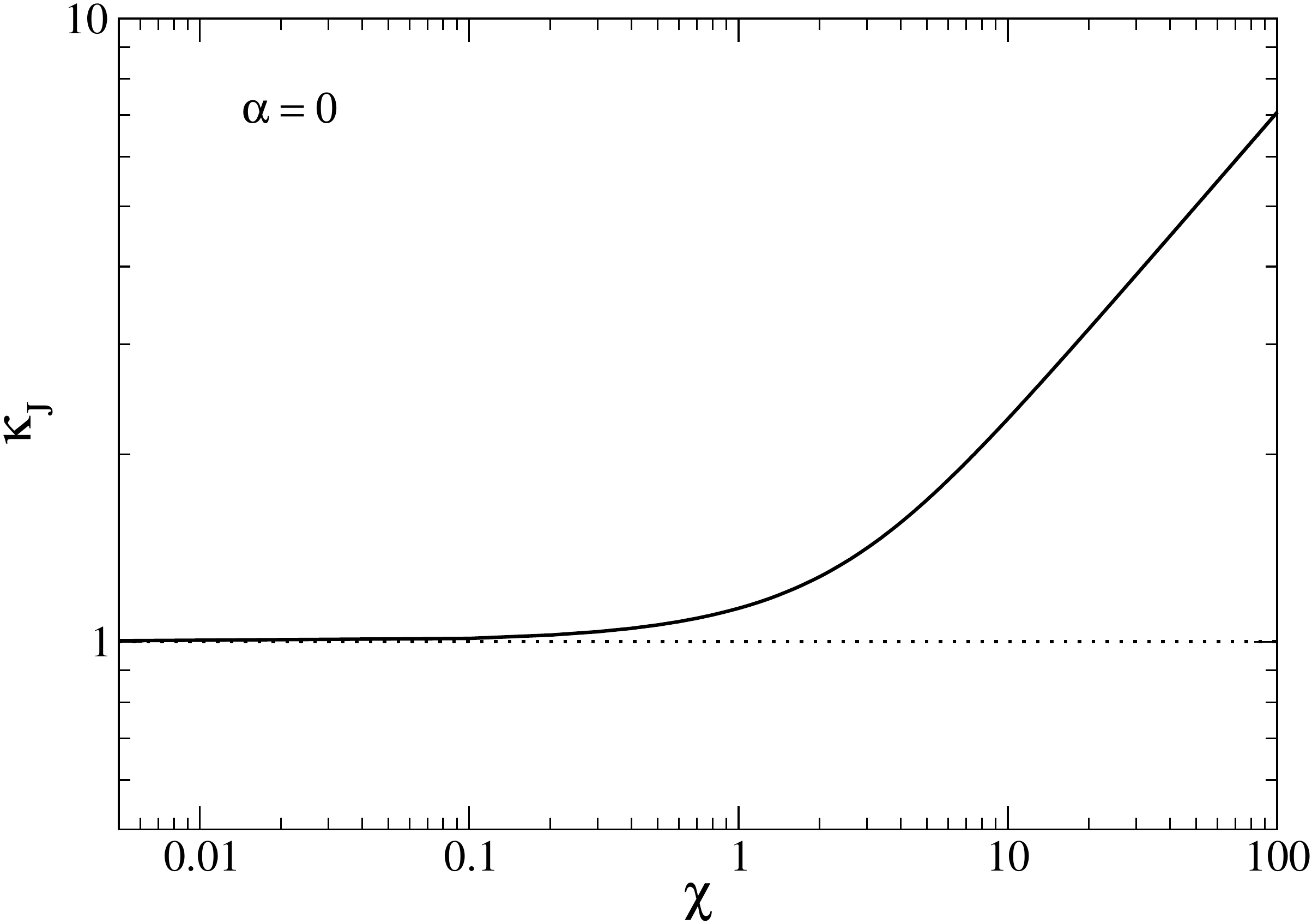}} 
\caption{Jeans wavenumber $\kappa_J$ of the relativistic model in the
noninteracting limit ($\alpha=0$) as a function of the relativistic
parameter $\chi$ (solid line: exact relativistic model; dotted line:
simplified relativistic model).}
\label{kj}
\end{figure}

\begin{figure}[h]
\scalebox{0.33}{\includegraphics{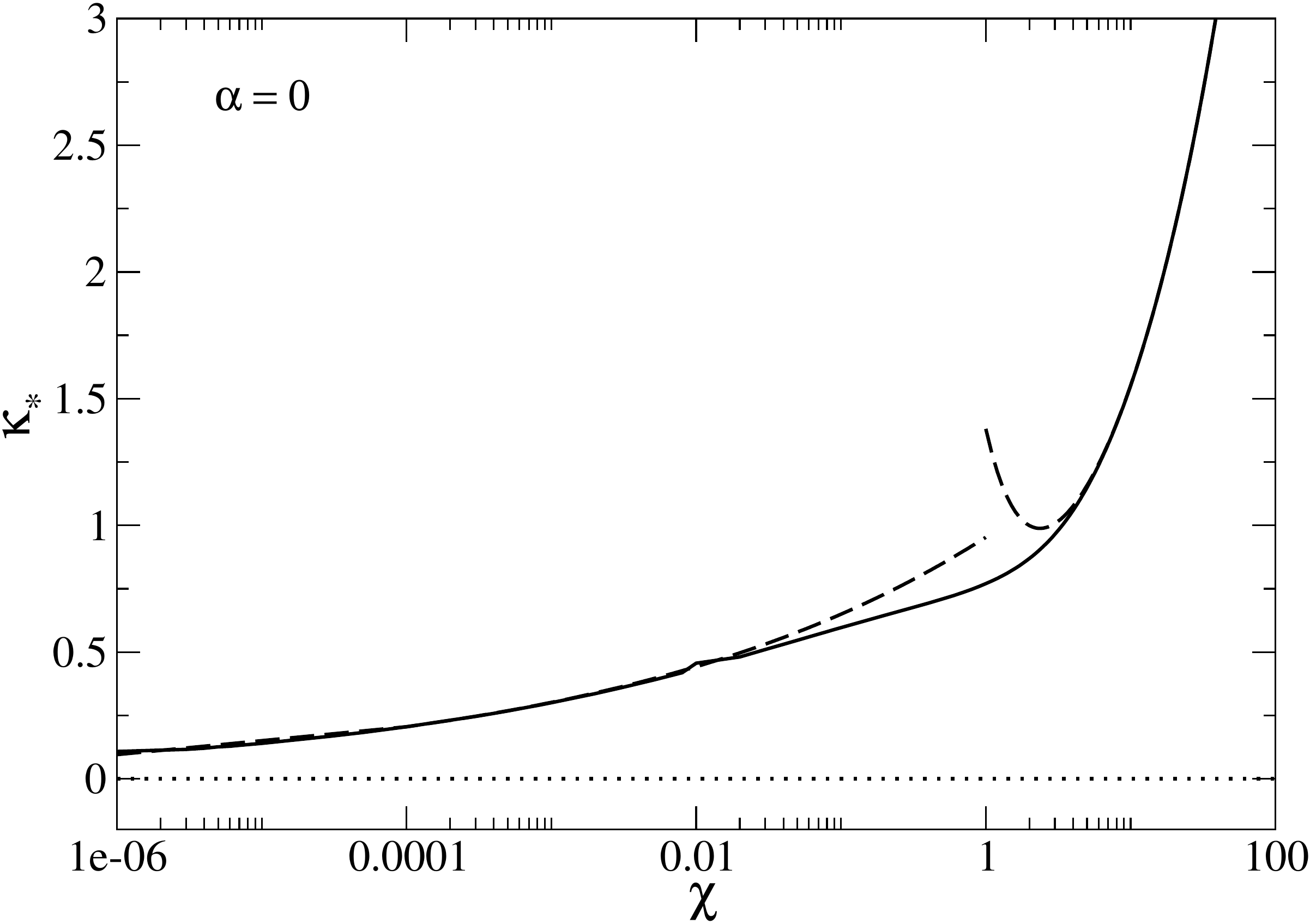}} 
\caption{Most unstable wavenumber $\kappa_*$ of the relativistic model in the
noninteracting limit ($\alpha=0$) as a function of the
relativistic parameter  $\chi$ (solid line: exact relativistic model;
dashed lines: asymptotic behaviors of the exact relativistic model; dotted line:
simplified relativistic model).}
\label{maxgrowthrateK}
\end{figure}

\begin{figure}[h]
\scalebox{0.33}{\includegraphics{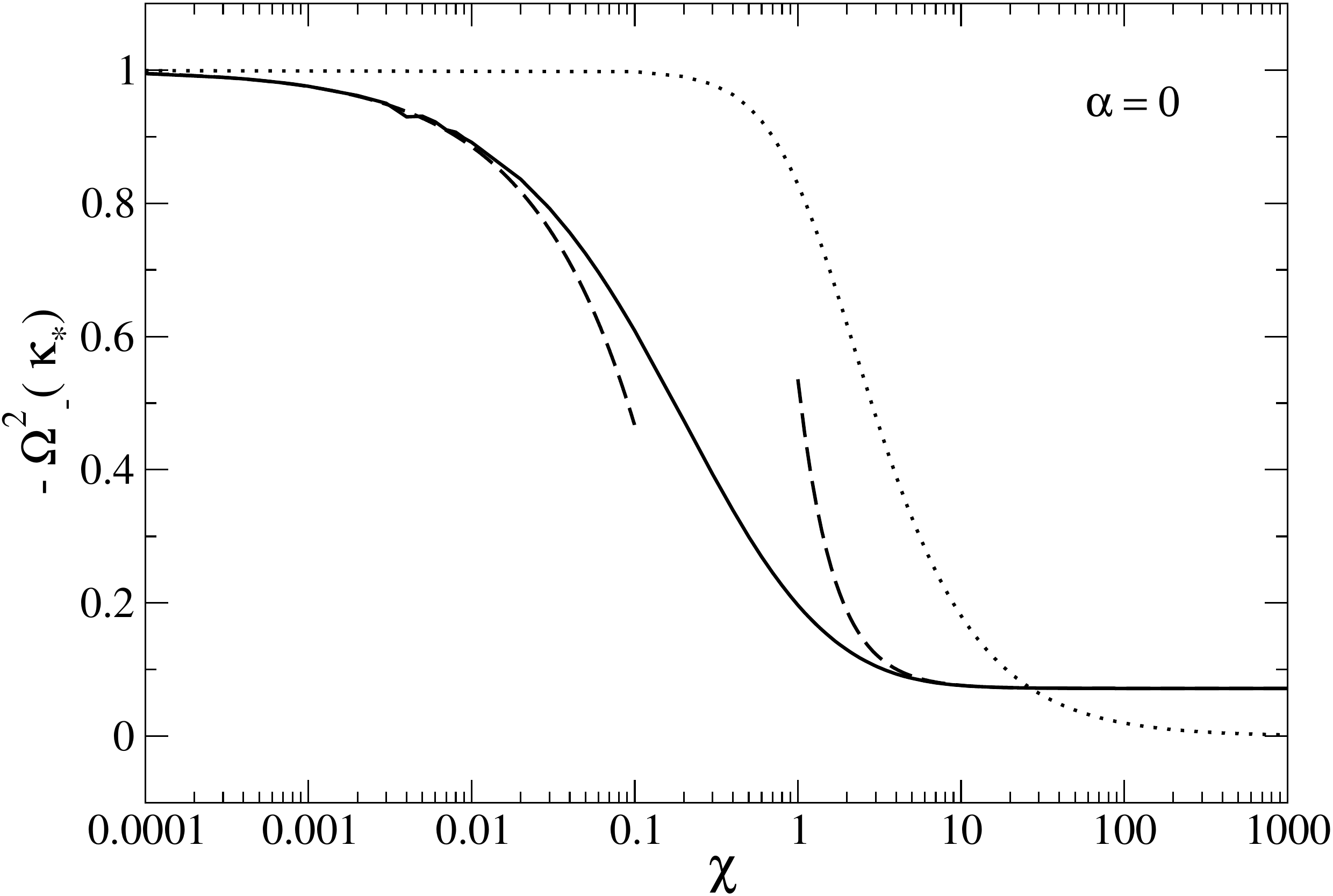}} 
\caption{Maximum growth
rate $-\Omega_-^2(\kappa_*)$ of the relativistic model in
the noninteracting limit ($\alpha=0$) as a function of the
relativistic parameter $\chi$ (solid line: exact relativistic model;
dashed lines: asymptotic behaviors of the exact relativistic model; dotted line:
simplified relativistic model).}
\label{maxgrowthrateOm}
\end{figure}

\subsection{The TF limit}
\label{sec_asc}

\begin{figure}[h]
\scalebox{0.33}{\includegraphics{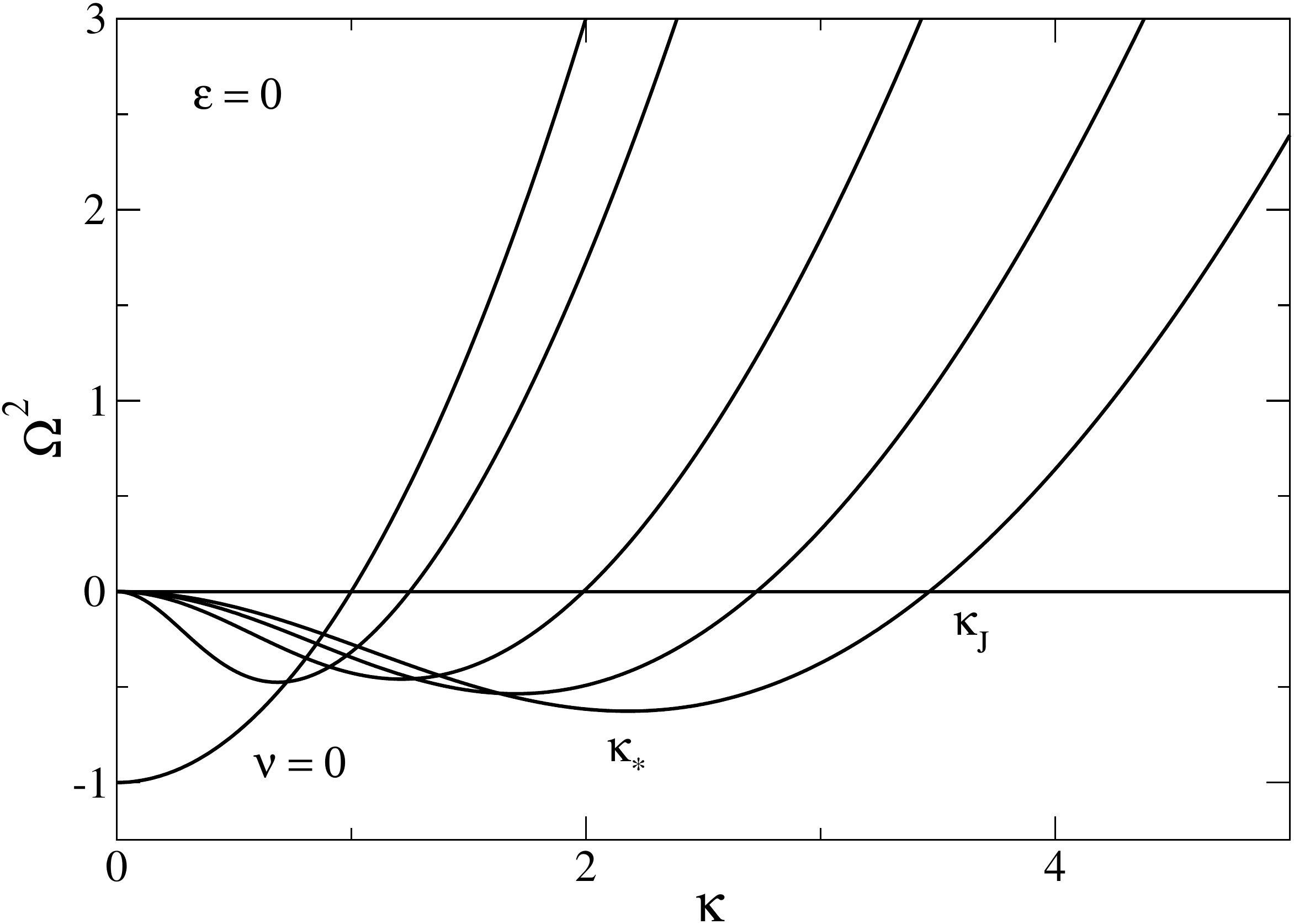}} 
\caption{Dispersion relation $\Omega^2(\kappa)$ of the exact relativistic
model in the TF limit ($\epsilon=0$) for different
values of the relativistic parameter $\nu=0,0.1,0.4,0.7,1$.}
\label{dispexactTF}
\end{figure}

In the TF approximation ($\hbar=0$), the
dispersion relation of Eq. (\ref{em1}) reduces to
\begin{eqnarray}
\omega^2=\frac{1}{1+3\gamma}\frac{
(1-3\gamma) k^2 c_s^2
-4\pi G\rho
\left (1+\frac{2c_s^2}{c^2}\right )}{1+\frac{3c_s^2}{c^2}},
\label{em17}
\end{eqnarray}
where
\begin{eqnarray}
\gamma=\frac{4\pi G\rho}{k^2c^2}\left (1+\frac{2c_s^2}{c^2}\right ).
\label{em18}
\end{eqnarray}
The function $\omega^2(k)$ is
represented in Fig. \ref{dispexactTF} for
different values of the relativistic parameter
$\nu\propto 1/c^2$ using the normalization of Appendix \ref{sec_dv2}. It
corresponds to the limit form of the branch $(-)$. The
branch $(+)$ is rejected at infinity. For $k\rightarrow 0$:
\begin{eqnarray}
\omega^2\sim -\frac{1}{3}k^2c^2.
\label{em19}
\end{eqnarray}
The effective speed of sound is  imaginary ($c_{\rm eff}^2=-c^2/3$). For
$k\rightarrow +\infty$: 
\begin{eqnarray}
\omega^2\sim \frac{c_s^2}{1+\frac{3c_s^2}{c^2}}k^2.
\label{em20}
\end{eqnarray}
In that case, the effective speed of sound is
\begin{eqnarray}
c_{\rm eff}^2=\frac{c_s^2}{1+\frac{3c_s^2}{c^2}}.
\label{em21}
\end{eqnarray}
The Jeans
wavenumber is given by
\begin{eqnarray}
k_J^2=\frac{4\pi G\rho}{c_s^2}\left (1+\frac{3c_s^2}{c^2}\right )\left
(1+\frac{2c_s^2}{c^2}\right ).
\label{em22}
\end{eqnarray}
It is plotted as a function of the relativistic parameter
$\nu\propto 1/c^2$ in Fig. \ref{kjTF} using the normalization of Appendix
\ref{sec_dv2}. Its asymptotic behaviors are given in Appendix
\ref{sec_mgrtfe}. We note that the Jeans length decreases as
relativistic effects increase. The optimal wavenumber $k_*$ and the maximum
growth rate $\sigma_{\rm max}$  are plotted  as a function of the
relativistic parameter
$\nu\propto 1/c^2$ in Figs. \ref{nukappaexact} and \ref{nuOmegaexact} using the
normalization of Appendix
\ref{sec_dv2}. Their asymptotic behaviors are given in Appendix
\ref{sec_mgrtfe}. We note that the optimal wavelength
decreases as relativistic effects increase. The maximum growth rate first
decreases, reaches a minimum and finally increases as relativistic effects
increase (see Appendix \ref{sec_mgrtfe}). In the nonrelativistic
limit
($\nu\rightarrow 0$), we find that $\lambda_J\rightarrow
2\pi c_s/(4\pi G\rho)^{1/2}$, $\lambda_*\rightarrow +\infty$ and
$\sigma_{\rm max}\rightarrow (4\pi G\rho)^{1/2}$. In the ultrarelativistic limit
($\nu\rightarrow +\infty$), we find
that $\lambda_*\sim 1.55\lambda_J\rightarrow 0$ and $\sigma_{\rm
max}\rightarrow
+\infty$. The maximum growth rate reaches its minimum value $(\sigma_{\rm
max})_{\rm min}=0.662(4\pi G\rho)^{1/2}$ for $\nu=0.23$.

\begin{figure}[h]
\scalebox{0.33}{\includegraphics{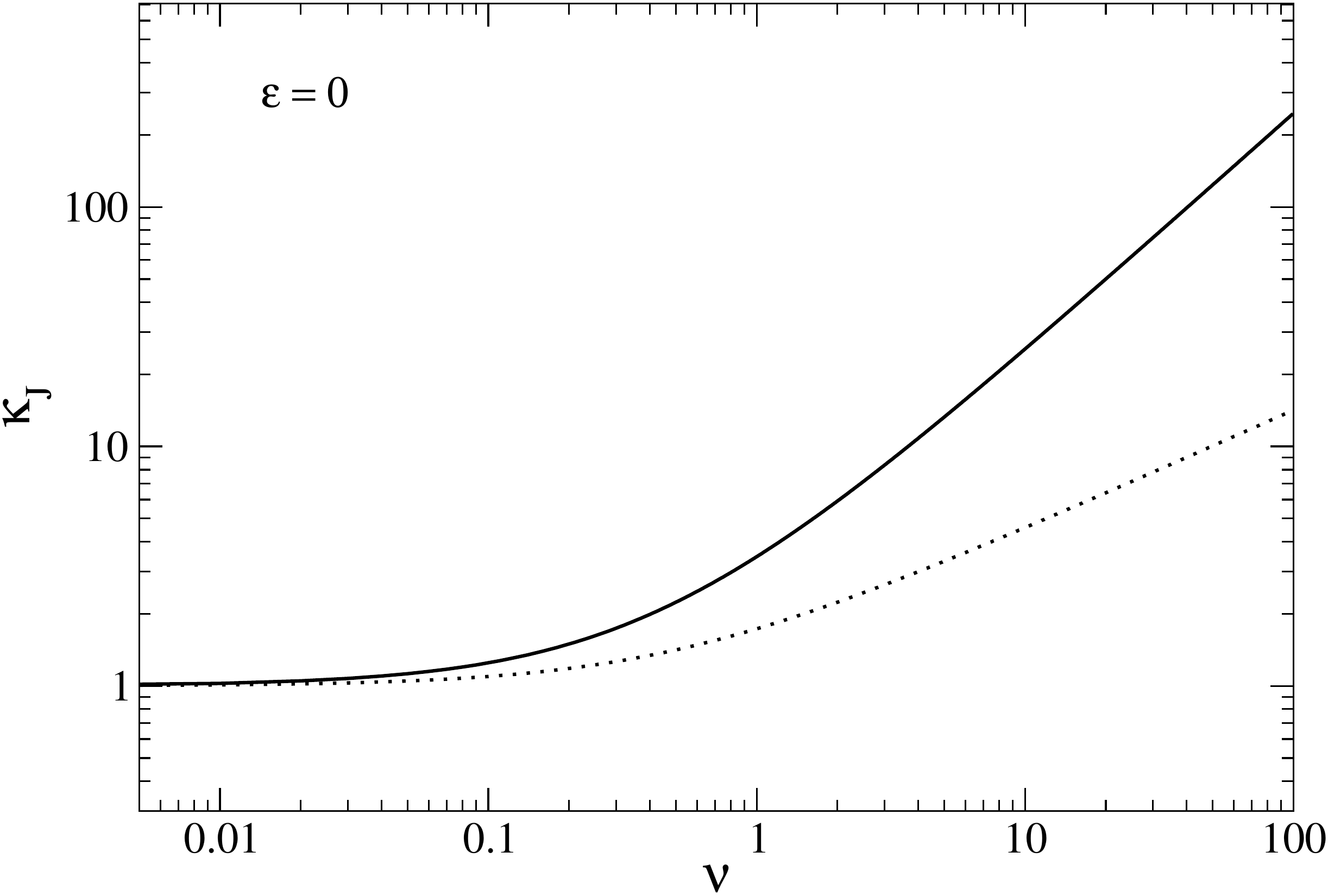}} 
\caption{Jeans wavenumber $\kappa_J$ of the relativistic model in the TF limit
($\epsilon=0$) as a function of the
relativistic parameter  $\nu$ (solid line: exact relativistic  model; dotted
line: simplified relativistic  model).}
\label{kjTF}
\end{figure}

\begin{figure}[h]
\scalebox{0.33}{\includegraphics{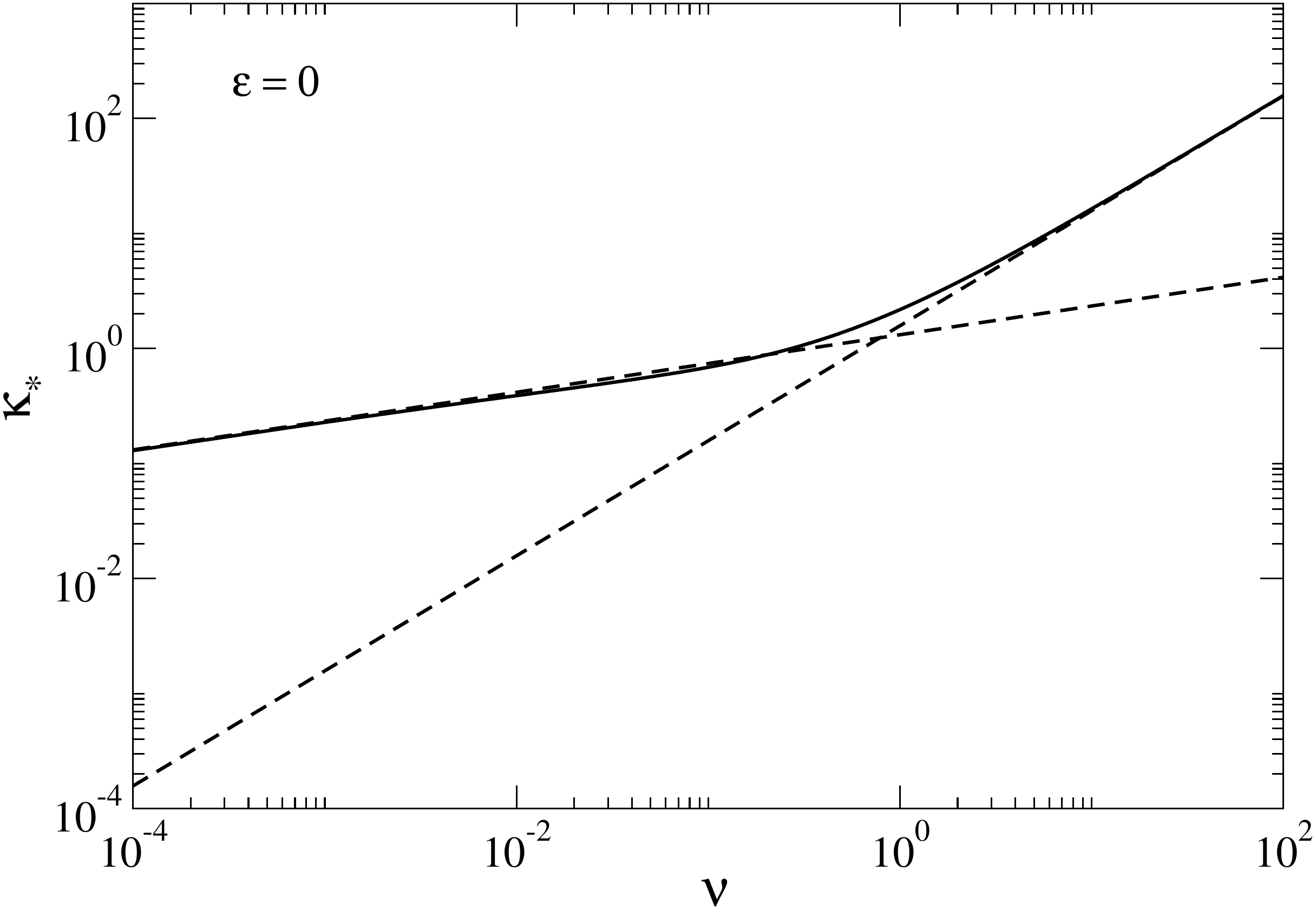}} 
\caption{Most unstable wavenumber $\kappa_*$ of the relativistic model  in
the
TF limit ($\epsilon=0$)
as a function of the
relativistic parameter  $\nu$ (solid line: exact relativistic model;
dashed lines: asymptotic behaviors of the exact relativistic model;
for the simplified relativistic model $\kappa_*=0$).}
\label{nukappaexact}
\end{figure}

\begin{figure}[h]
\scalebox{0.33}{\includegraphics{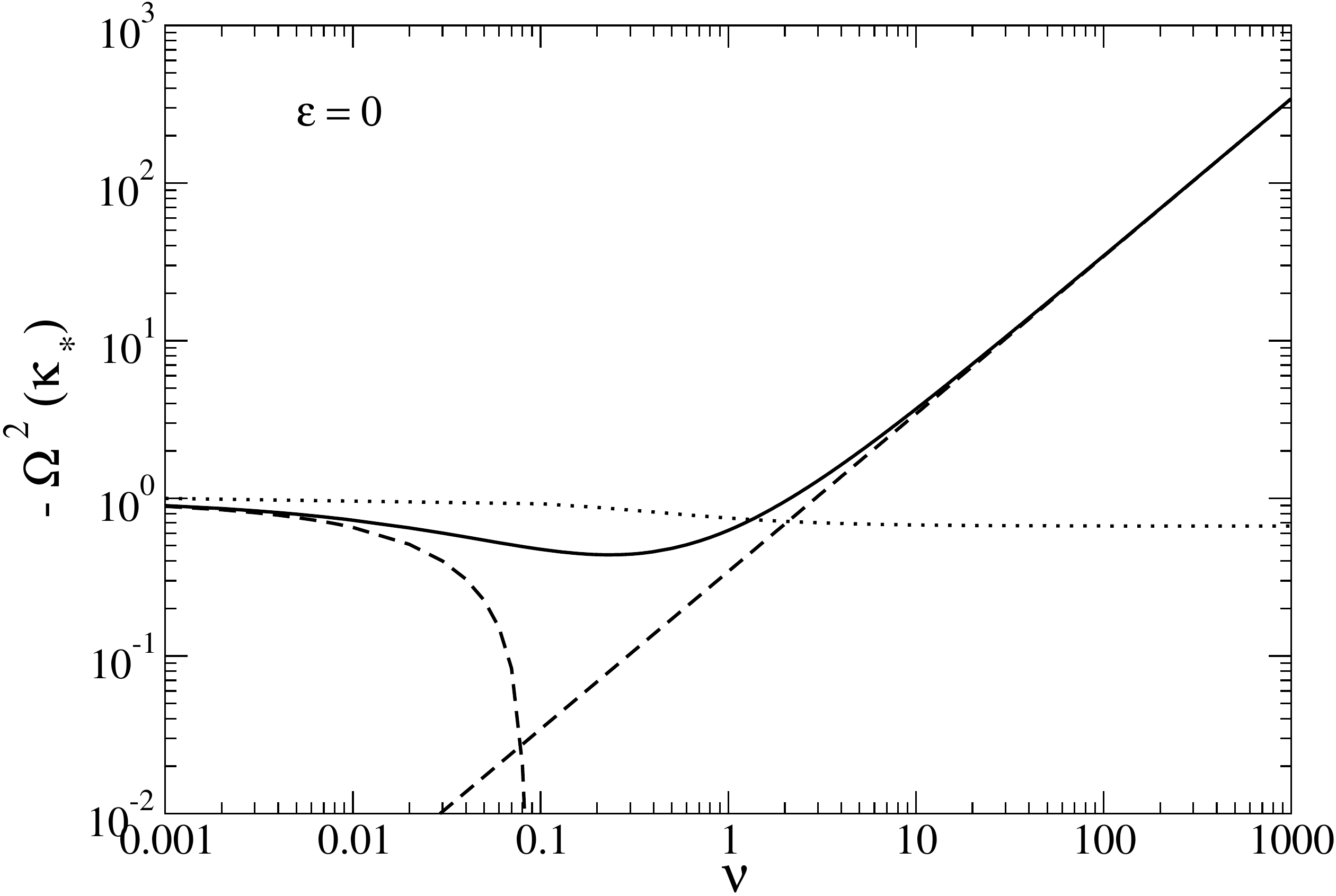}} 
\caption{Maximum growth rate $-\Omega_-^2(\kappa_*)$ of the relativistic model 
in the TF limit ($\epsilon=0$) as a
function of the
relativistic parameter $\nu$ (solid line: exact relativistic model;
dashed lines: asymptotic behaviors of the exact relativistic model; dotted line:
simplified relativistic model).}
\label{nuOmegaexact}
\end{figure}

{\it Remark:} In the case $\hbar=c_s=0$, the
dispersion relation of Eq. (\ref{em1}) reduces to
\begin{eqnarray}
\omega^2=-\frac{c^2k^2}{3+\frac{c^2k^2}{4\pi G\rho}}.
\label{em25}
\end{eqnarray}
For $k\rightarrow 0$:
\begin{eqnarray}
\omega^2\sim -\frac{1}{3}k^2c^2.
\label{em26}
\end{eqnarray}
For $k\rightarrow +\infty$: 
\begin{eqnarray}
\omega^2\rightarrow -4\pi G\rho.
\label{em27}
\end{eqnarray}
The system is unstable at all scales. The maximum growth
rate $\sigma_{\rm max}=(4\pi G\rho)^{1/2}$ is obtained for $k\rightarrow
+\infty$ (infinitely small scales). There is a stabilization at large scales
($k=0$) due to relativistic effects.

\section{The nongravitational limit}
\label{sec_nog}

In this section, we consider the nongravitational limit ($G=0$). This limit may
be relevant in the case of a SF with an attractive self-interaction (e.g. the
axion) that can experience an instability even in the absence of self-gravity.
The nonrelativistic limit $c\rightarrow +\infty$ has been discussed in detail
in Sec. V of \cite{prd1}. In this section, we take relativistic effects into
account. We just give preliminary results that are sufficient for the numerical
applications made in Secs. \ref{sec_anrad} and \ref{sec_app}. A more detailed
treatment of this
problem will be given elsewhere. We note that, in the nongravitational limit,
the simplified relativistic model and the exact relativistic model are
equivalent.

\subsection{The general case}

In the nongravitational limit, the relativistic dispersion relation is given by
\begin{equation}
\frac{\hbar^2}{4m^2c^4}\omega^4-\left (\frac{\hbar^2k^2}{2
m^2c^2}+1+\frac{3c_s^2}{c^2}\right )\omega^2
+ \frac{\hbar^2k^4}{4m^2}+ k^2 c_s^2=0.
\label{nog1}
\end{equation}
When $c_s^2\ge 0$, the system is always stable. When $c_s^2< 0$, there is a
critical wavenumber\footnote{Since self-gravity is neglected, this critical
wavenumber should not be called the Jeans wavenumber. However, we will use
this terminology to unify the notations. This makes
sense if we view Eq. (\ref{nog3}) as the approximation of the exact Jeans
wavenumber in the nongravitational limit $G\rightarrow 0$.}
\begin{eqnarray}
k_J^2=\frac{4m^2|c_s^2|}{\hbar^2}=\frac{16\pi|a_s|\rho}{m}.
\label{nog3}
\end{eqnarray}
We note that this critial wavenumber is independent of $c$. Therefore, it has
the same expression as in the nonrelativistic limit considered in \cite{prd1}.
In the
nonrelativistic limit $c\rightarrow +\infty$, the dispersion relation of Eq.
(\ref{nog1}) becomes
\begin{equation}
\omega^2=\frac{\hbar^2k^4}{4m^2}+c_s^2 k^2 .
\label{nog2}
\end{equation}
In that case, the most unstable wavenumber is
$k_*=(8\pi|a_s|\rho/m)^{1/2}$ and
the maximum growth rate is $\sigma_{\rm max}=4\pi|a_s|\hbar\rho/m^2$
\cite{prd1}. They can
be rewritten as $k_*=k_J/\sqrt{2}$ and $\sigma_{\rm
max}=\alpha\omega_0$ where we have introduced the notations of
Appendix 
\ref{sec_dv}.

\subsection{The noninteracting limit}

In the noninteracting limit ($c_s=0$), the dispersion relation of Eq.
(\ref{nog1})
reduces to
\begin{equation}
\frac{\hbar^2}{4m^2c^4}\omega^4-\left (\frac{\hbar^2k^2}{2
m^2c^2}+1\right )\omega^2
+ \frac{\hbar^2k^4}{4m^2}=0.
\label{nog4}
\end{equation}
The system is always stable. In the nonrelativistic limit $c\rightarrow
+\infty$, the dispersion relation of Eq.
(\ref{nog4}) becomes
\begin{equation}
\omega^2=\frac{\hbar^2k^4}{4m^2}.
\label{nog5}
\end{equation}

\subsection{The TF limit}

In the TF limit, the  dispersion relation of Eq. (\ref{nog1}) reduces to
\begin{equation}
\omega^2=\frac{k^2 c_s^2}{1+\frac{3c_s^2}{c^2}} 
\label{nog6}
\end{equation}
When $c_s^2>0$ and when $c_s^2<-c^2/3$, the system is always stable. 
When $-c^2/3<c_s^2<0$, the system is always unstable. In the  nonrelativistic
limit, the dispersion relation of Eq.
(\ref{nog6}) becomes
\begin{equation}
\omega^2=c_s^2 k^2 .
\label{nog7}
\end{equation}

\section{Astrophysical and cosmological applications in the ultrarelativistic
regime (radiation era)}
\label{sec_anrad}

In this section and in the following one, we use our theoretical results to make
astrophysical and cosmological predictions. We first determine
simplified expressions of the Jeans length and Jeans mass of the SF in different
limits. Then, we apply these results to different types of bosons. In the
present section, we show that large-scale structures cannot form in the
ultrarelativistic regime (early Universe and radiation era) because the Jeans
length is of the order of the Hubble length, except in the case where the
self-interaction between bosons is attractive. In the following section, we show
that large-scale structures can form in the nonrelativistic regime (matter era).

\subsection{The impossibility to form large-scale structures in the radiation
era}
\label{sec_naive}

It is well-known that structure formation cannot take place in the radiation
era. The quick proof is usually based on the following (rough) argument. Using
the standard Jeans relation of Eq. (\ref{intro2}), identifying $\rho$ with the
energy density $\epsilon/c^2$, and computing the speed of sound
$c_s^2=dP/d\rho=P'(\epsilon)c^2$ with the equation of state of radiation
$P=\epsilon/3$ implying $c_s=c/\sqrt{3}$, we obtain  
\begin{eqnarray}
(k_J^2)_{\rm naive}=\frac{12\pi G \epsilon}{c^4}.
\label{naive1}
\end{eqnarray}
As a result, the Jeans length $\lambda_J\sim (c^4/G\epsilon)^{1/2}$ is of the
order of the Hubble length  $\lambda_H$ (see Appendix \ref{sec_hs}). Since
the Hubble length represents the horizon, the Jeans instability cannot take
place. Actually, the correct expression of the Jeans length based on general
relativity is (see Appendix \ref{sec_paddy}): 
\begin{eqnarray}
(k_J^2)_{\rm exact}=\frac{16\pi G \epsilon}{c^4}.
\label{naive2}
\end{eqnarray}
The exact coefficient is different from the one obtained in the naive
approach but the conclusion is the same: $\lambda_J\sim \lambda_H$. As a
result, large-scale structures cannot form in the ultrarelativistic regime,
corresponding to the radiation era. This result has been derived for a fluid.
We now derive the corresponding result for a complex SF.

\subsection{The noninteracting limit}
\label{sec_urni}

In the noninteracting limit ($a_s=0$), the Jeans wavenumber is given by Eq.
(\ref{em16}). In the ultrarelativistic  limit ($c\rightarrow 0$), it reduces
to\footnote{We note that the
simplified relativistic model of \cite{khlopov} is not valid in the
ultrarelativistic limit because it leads to a Jeans
wavenumber given by Eq. (\ref{sm13}), the same as in the
nonrelativistic limit, which is different from Eq.
(\ref{urni1}).}
\begin{eqnarray}
k_J^2=\frac{4\pi G\rho}{c^2}.
\label{urni1}
\end{eqnarray}
This expression is similar to the classical Jeans wavenumber of Eq.
(\ref{intro2}) with the
substitution
$c_s\rightarrow c$.  The Jeans length is
\begin{eqnarray}
\lambda_J=2\pi\left (\frac{c^2}{4\pi G\rho}\right )^{1/2}.
\label{urni2}
\end{eqnarray}
We note that this expression does not depend on the
Planck constant $\hbar$. It is also independent of the
particle mass $m$ which is negligible in the ultrarelativistic regime. We can
express the Jeans wavenumber and the Jeans wavelength
in terms of the energy density $\epsilon$, using
Eq. (\ref{jia9b}) which reduces, in the noninteracting limit ($a_s=0$), to
$\epsilon=\rho c^2$ [see Eq. (\ref{dim1})]. We get
\begin{equation}
k_J^2=\frac{4\pi G\epsilon}{c^4},\qquad \lambda_J=2\pi\left (\frac{c^4}{4\pi
G\epsilon}\right )^{1/2}.
\label{urni2qab}
\end{equation}
The Jeans mass is defined by
\begin{eqnarray}
M_J=\frac{4}{3}\pi\frac{\epsilon}{c^2}\left
(\frac{\lambda_J}{2}\right )^3.
\label{mj}
\end{eqnarray}
Using
Eqs. (\ref{dim1}) and (\ref{urni2qab}), we get
\begin{equation}
M_J=\frac{1}{6}\pi^{5/2}\frac{c^4}{G^{3/2}\epsilon^{1/2}}=\frac{1}{6}\pi^{
5/2}\frac{c^3}{G^{3/2}\rho^{1/2}}.
\label{urni2qa}
\end{equation}
As shown in \cite{shapiro,suarezchavanis3}, the ultrarelativistic regime of
the SF corresponds to the stiff matter era (early Universe) where
$\epsilon\propto a^{-6}$ and to the standard radiation era (due
to photons, neutrinos...) where $\epsilon\propto
a^{-4}$. In the first case, the
Jeans length and
the Jeans mass increase as $a^3$. In the second case, they increase as $a^2$.
Eliminating
the energy density between
Eqs. (\ref{urni2qab}) and (\ref{urni2qa}), we get
\begin{eqnarray}
M_J=\frac{\pi^2}{6}\frac{\lambda_Jc^2}{G}.
\label{urni4}
\end{eqnarray}
This relation is similar to  the mass-radius
relation
\begin{eqnarray}
M=\frac{2q}{(1+q)^2+4q}\frac{Rc^2}{G}
\label{urni5}
\end{eqnarray}
of a general relativistic fluid star described by a linear equation of state
$P=q\epsilon$ confined within a box \cite{aaq}. The radiation case (photon
stars) corresponds to
$q=1/3$ and the stiff matter case (stiff stars) corresponds to $q=1$. We
note that Eq. (\ref{urni4}) displays the relativistic scaling $M\sim Rc^2/G$.

Introducing the scales
$\lambda_C$, $M_C$, $\rho_C$ (see
Appendix \ref{sec_cs}) and the relativistic parameter
$\chi=(\rho/2\rho_C)^{1/2}$ (see
Appendix \ref{sec_dv1})
adapted to the noninteracting limit (see
Appendix
\ref{sec_tsni}), we get
\begin{eqnarray}
\frac{\lambda_J}{\lambda_C}=\frac{2\pi}{\chi},\qquad 
\frac{M_J}{M_C}=\frac{\pi^3}{3\chi},\qquad
\frac{M_J}{M_C}=\frac{\pi^2}{6}\frac{\lambda_J}{\lambda_C}.
\label{urni7}
\end{eqnarray}
Since $\chi\gg 1$ (i.e. $\rho\gg\rho_C$) in the ultrarelativistic limit,
we find that
$\lambda_J\ll\lambda_C$ and $M_J \ll M_C$.

The preceding results are obtained by taking the
noninteracting limit $a_s\rightarrow 0$ before the ultrarelativistic limit
$c\rightarrow 0$. We now consider the case where  the ultrarelativistic limit
$c\rightarrow 0$ is taken before the noninteracting limit $a_s\rightarrow 0$.
We have to consider two cases.

For a repulsive self-interaction ($a_s>0$), introducing the scales
$\lambda_R$, $M_R$, $\rho_R$ (see Appendix \ref{sec_is}) and the
self-interaction parameter $\nu=2\rho/3\rho_R$ (see Appendix \ref{sec_dv2})
adapted to the ultrarelativistic limit (see
Appendix \ref{sec_tsura}), we get
\begin{equation}
\frac{\lambda_J}{\lambda_R}=\frac{2\pi}{\nu^{1/2}},\qquad
\frac{M_J}{M_R}
=\frac{\pi^3}{3\nu^{1/2}},\qquad
\frac{M_J}{M_R}=\frac{\pi^2}{6}\frac{\lambda_J}{\lambda_R}.
\end{equation}
Since $\nu\ll 1$ (i.e. $\rho\ll\rho_R$)  in the noninteracting limit, we find
that
$\lambda_J\gg\lambda_R$ and $M_J \gg M_R$.

For an attractive self-interaction ($a_s<0$), the
ultrarelativistic limit imposes (see \cite{suarezchavanis3} for details) that
the pseudo rest-mass density and the energy
density are given by $\rho_i$ and $\epsilon_i$ defined by Eqs. (\ref{badlu}) and
(\ref{dim4}). This
corresponds to $\nu\sim 1$ (see Appendix \ref{sec_ngsi}). In that case,
the
Jeans scales (\ref{urni1})-(\ref{urni2qa}) are given by
\begin{eqnarray}
k_J^2=\frac{Gm^3}{3|a_s|\hbar^2},\qquad
\lambda_J=2\pi\left (\frac{3|a_s|\hbar^2}{Gm^3}\right )^{1/2},
\label{str1}
\end{eqnarray}
\begin{equation}
M_J=\frac{\pi^3}{\sqrt{3}}\left
(\frac{|a_s|\hbar^2c^4}{G^3m^3}\right
)^{1/2}.
\label{str2}
\end{equation}
Introducing the scales $\lambda_C$, $M_C$, $\rho_C$ (see Appendix \ref{sec_cs})
and the
self-interaction parameter $\sigma=3|a_s|/2r_S$ (see Appendix
\ref{sec_effS})
adapted to the ultrarelativistic limit (see
Appendix \ref{sec_tsurb}), we get
\begin{equation}
\frac{\lambda_J}{\lambda_C}=4\pi\sigma^{1/2},\qquad
\frac{M_J}{M_C}
=\frac{2\pi^3}{3}\sigma^{1/2},\qquad \frac{M_J}{M_C}=\frac{\pi^2}{6}\frac{
\lambda_J } { \lambda_C}.
\end{equation}
Since $\sigma\ll 1$ (i.e. $|a_s|\ll r_S$) in the noninteracting limit, we find
that
$\lambda_J\ll\lambda_C$ and $M_J \ll M_C$.

Introducing the Hubble scales $\lambda_H$ and $M_H$ (see Appendix \ref{sec_hs}),
we obtain 
\begin{eqnarray}
\frac{\lambda_J}{\lambda_H}=\frac{2\sqrt{2}\pi}{\sqrt{3}},
\qquad 
\frac{M_J}{M_H}=\frac{2\sqrt{2}\pi^3}{3\sqrt{3}}.
\label{urni6}
\end{eqnarray}
We note that $\lambda_J\sim \lambda_H$ and $M_J\sim
M_H$. Since the Jeans length is of the
order of the Hubble length (horizon), there is no Jeans
instability. Therefore, large-scale
structures cannot form in the
ultrarelativistic limit (stiff matter and radiation eras).

\begin{figure}[h]
\scalebox{0.33}{\includegraphics{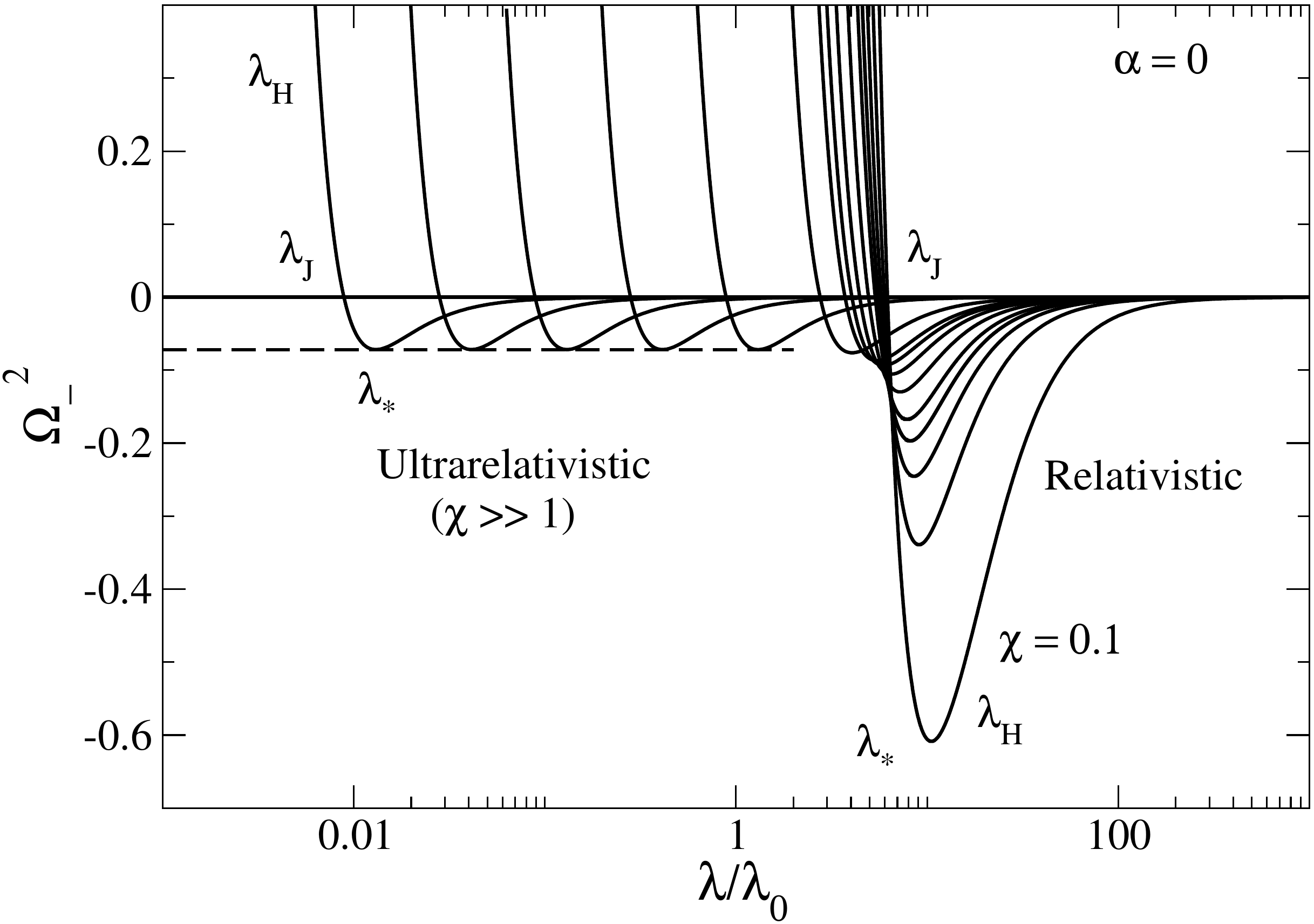}} 
\caption{Growth rate $\Sigma=(-\Omega_{-}^2)^{1/2}$ of the
perturbations in the noninteracting limit ($a_s=0$) as a function of the
wavelength
$\lambda/\lambda_0=2\pi/\kappa$ for different values of the relativistic
parameter between $\chi=0.1$ and $\chi=10^6$. In the ultrarelativistic limit
($\chi\gg 1)$, the maximum growth rate, localized at $\lambda_*\sim 1.47
\lambda_J$, tends to a relatively small
constant value $\Sigma_{\rm max}=0.268$  and the Jeans length is larger than the
Hubble length ($\lambda_J\simeq 5.13\lambda_H$). In the
relativistic regime ($\chi\sim 1$), the Jeans length $\lambda_J$ is slightly
smaller than the Hubble length $\lambda_H$.
}
\label{sigmachi}
\end{figure}

{\it Remark:} In Fig. \ref{sigmachi}, we have plotted the growth
rate of the perturbations as a function of the wavelength  for different values
of the relativistic parameter $\chi$ using the normalization of Appendix
\ref{sec_dv1}. In the ultrarelativistic limit $\chi\rightarrow +\infty$, the
Jeans length $\lambda_J$ and the most unstable wavelength $\lambda_*\simeq 1.47
\lambda_J$ tend to zero while the maximum growth rate tends to a nonzero, but
relatively small, constant value $\sigma_{\rm max}\rightarrow 0.268\sqrt{4\pi
G\rho}$ (see  Appendix \ref{sec_mgrne}). The instability is
relatively localized about $\lambda_*$. However, this instability may not be
physical since the Jeans length is larger than the Hubble length
($\lambda_J\simeq 5.13\lambda_H$).

\subsection{The TF limit}
\label{sec_urtf}

In this section, we consider a SF with a repulsive self-interaction ($a_s>0$).
In the TF limit ($\hbar\rightarrow 0$), the Jeans wavenumber is given by Eq.
(\ref{em22}). In the ultrarelativistic limit ($c\rightarrow
0$),\footnote{This corresponds to
$c_s/c\rightarrow +\infty$ in Eq. (\ref{em22}). We can have $c_s>c$ because 
the pseudo speed of sound $c_s^2=p'(\rho)$ [see Eq. (\ref{jia9})] differs from
the true speed of sound $c_s^2=P'(\epsilon)c^2$. For the equation of state
(\ref{jia9d}), which reduces to $P=\epsilon/3$ (radiation) in the
ultrarelativistic limit
[see Eq. (\ref{dim2})], the true speed of sound is $c_s=c/\sqrt{3}$ and it
satisfies $c_s<c$.} it reduces
to\footnote{We note that the simplified relativistic model of \cite{khlopov}
is not
valid in the ultrarelativistic limit because it leads to a Jeans wavenumber
$k_J^2=8\pi G\rho/c^2$ [see Eq. (\ref{sm17})], independent of $a_s$, which is
different from  Eq. (\ref{urtf1}).}
\begin{eqnarray}
k_J^2=\frac{24\pi G \rho c_s^2}{c^4}=\frac{96\pi^2 a_s \hbar^2 G\rho^2}{m^3
c^4}.
\label{urtf1}
\end{eqnarray}
This expression is
similar to the classical Jeans wavenumber of Eq. (\ref{intro2}) with the
substitution
$c_s\rightarrow c^2/c_s$. The Jeans length is
\begin{eqnarray}
\lambda_J=2\pi\left (\frac{m^3c^4}{96\pi^2 a_s \hbar^2 G\rho^2}\right )^{1/2}.
\label{urtf2}
\end{eqnarray}
We can express the Jeans wavenumber and the Jeans wavelength in terms of the
energy density $\epsilon$, using
Eq. (\ref{jia9b}) which reduces, in the ultrarelativistic limit,
to $\epsilon=({6\pi a_s\hbar^2}/{m^3})\rho^2$ [see Eq. (\ref{dim2})].
We get
\begin{eqnarray}
k_J^2=\frac{16\pi G \epsilon}{c^4},\qquad \lambda_J=2\pi\left (\frac{c^4}{16\pi
G \epsilon}\right )^{1/2}.
\label{urtf4}
\end{eqnarray}
Remarkably, we obtain the same result as the one  obtained for a radiative 
fluid described by the equation of state $P=\epsilon/3$ [see Eq.
(\ref{naive2})]. This equivalence is not trivial since a SF is not an ordinary
fluid and the relation $P=\epsilon/3$ only holds
for the background, not for the perturbations (see Sec. \ref{sec_jia}).
Using Eq. (\ref{urtf4}), the Jeans mass defined by Eq. (\ref{mj}) is
given by
\begin{eqnarray}
M_J=\frac{\pi^{5/2}}{48}\frac{c^4}{G^{3/2}\epsilon^{1/2}}.
\label{urtf5}
\end{eqnarray}
It can also be written as
\begin{eqnarray}
M_J=\frac{\pi^{5/2}}{48}\left (\frac{m^3c^8}{6\pi
a_s\hbar^2G^3}\right )^{1/2}\frac{1}{\rho}.
\label{urtf5b}
\end{eqnarray}
Eliminating the energy density between Eqs. (\ref{urtf4}) and (\ref{urtf5}), we
get 
\begin{eqnarray}
M_J=\frac{\pi^2}{24}\frac{\lambda_Jc^2}{G}.
\label{urtf6}
\end{eqnarray}
As indicated previously, this relation is similar to the mass-radius
relation of a
radiation (photon) star with a linear equation of state $P=\epsilon/3$ confined
within a box \cite{aaq}. In
the present case, this agreement can be explained by the fact that the
equation of state of the SF is $P=\epsilon/3$ for the background. Comparing
Eq. (\ref{urni5})  with $q=1/3$ and Eq. (\ref{urtf6}), we obtain
\begin{eqnarray}
\frac{M_J}{\lambda_J/2}=3.84 \frac{M}{R},
\label{urtf6b}
\end{eqnarray}
where $\lambda_J/2$ is the Jeans radius.

Introducing the scales $\lambda_R$, $M_R$, $\rho_R$ (see Appendix \ref{sec_is})
and the relativistic parameter
$\nu=2\rho/3\rho_R$ (see
Appendix \ref{sec_dv2}) adapted to the TF limit (see Appendix
\ref{sec_tstf}), we get 
\begin{equation}
\frac{\lambda_J}{\lambda_R}=\sqrt{\frac{2}{3}}\pi\frac{1}{\nu},
\quad 
\frac{M_J}{M_R}=\frac{1}{24}\sqrt{\frac{2}{3}}\pi^3\frac{1}{\nu},\quad
\frac{M_J}{M_R}=\frac{\pi^2}{24}\frac{\lambda_J}{\lambda_R}.
\label{urni8}
\end{equation}
Since $\nu\gg 1$ (i.e. $\rho\gg\rho_R$) in the ultrarelativistic limit,
we find that
$\lambda_J\ll\lambda_R$ and $M_J \ll M_R$.

The preceding results are obtained by taking the
TF limit $\hbar\rightarrow 0$ before the ultrarelativistic limit
$c\rightarrow 0$. We now consider the case where  the ultrarelativistic limit
$c\rightarrow 0$ is taken before the TF limit $\hbar\rightarrow 0$. Introducing
the scales $\lambda_R$ and $M_R$ (see Appendix \ref{sec_is}) and the
self-interaction parameter $\nu=2\rho/3\rho_R$ (see
Appendix \ref{sec_dv2}) adapted to the ultrarelativistic limit (see
Appendix \ref{sec_tsura}), we get Eq. (\ref{urni8}) again. Since
$\nu\gg 1$ (i.e. $\rho\gg\rho_R$)  in the TF limit, we find that
$\lambda_J\ll\lambda_R$ and $M_J \ll M_R$.

Introducing the Hubble scales $\lambda_H$ and $M_H$ (see Appendix \ref{sec_hs}),
we obtain
\begin{eqnarray}
\frac{\lambda_J}{\lambda_H}=\sqrt{\frac{2}{3}}\pi,\qquad 
\frac{M_J}{M_H}=\frac{1}{8} \left (\frac{2}{3}\right
)^{3/2}\pi^3.
\label{urni7b}
\end{eqnarray}
We note that $\lambda_J\sim \lambda_H$ and $M_J\sim
M_H$. Since the Jeans length is of the
order of the Hubble length (horizon), there is no Jeans
instability. Therefore, large-scale structures cannot form in the
ultrarelativistic
limit (radiation era).

\begin{figure}[h]
\scalebox{0.33}{\includegraphics{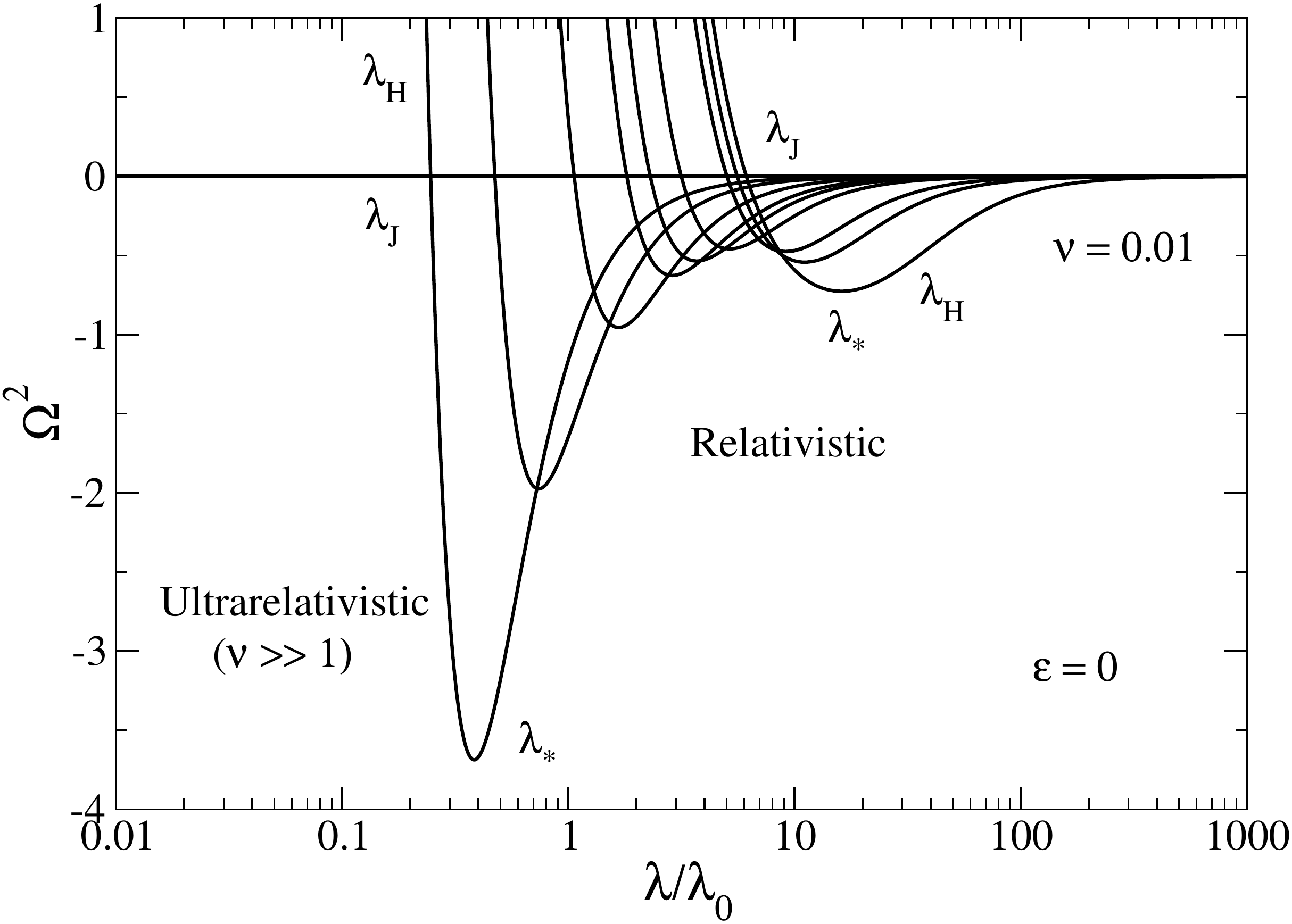}} 
\caption{Growth rate $\Sigma=(-\Omega^2)^{1/2}$ of the
perturbations in the TF limit ($\hbar\rightarrow 0$) as a function of the
wavelength
$\lambda/\lambda_0=2\pi/\kappa$ for different values of the relativistic
parameter between $\nu=0.01$ and $\nu=5$. In the ultrarelativistic
limit
$\nu\rightarrow +\infty$, the maximum growth rate,  localized at $\lambda_*\sim
1.55
\lambda_J$, tends to infinity and the
Jeans length is larger than the
Hubble length ($\lambda_J\simeq 2.565\lambda_H$). In the
relativistic regime ($\nu\sim 1$), the Jeans length $\lambda_J$ is slightly
smaller than the Hubble length $\lambda_H$. We clearly see from this figure
that the maximum growth rate $\sigma_{\rm max}$ reaches a minimum value
$(\sigma_{\rm
max})_{\rm min}=0.662(4\pi G\rho)^{1/2}$ at $\nu=0.23$ as discussed in Sec.
\ref{sec_asc}.
}
\label{sigmanu}
\end{figure}

{\it Remark:} In Fig. \ref{sigmanu}, we have plotted the growth
rate of the perturbations as a function of the wavelength  for different values
of the relativistic parameter $\nu$ using the normalization of Appendix
\ref{sec_dv2}. In the ultrarelativistic limit $\nu\rightarrow +\infty$, the
Jeans length $\lambda_J$ and the most unstable wavelength $\lambda_*\simeq 1.55
\lambda_J$ tend to zero while the maximum growth rate tends to infinity
(see  Appendix  \ref{sec_mgrtfe}). The instability is
relatively localized about $\lambda_*$. However, this instability may not be
physical since the Jeans length is larger than the Hubble length
($\lambda_J\simeq 2.565\lambda_H$).

\subsection{The nongravitational limit}
\label{sec_urng}

In this section, we consider a SF with an attractive self-interaction ($a_s<0$).
In the
nongravitational limit ($G\rightarrow 0$), the Jeans wavenumber is
given by Eq. (\ref{nog3}). For an attractive self-interaction ($a_s<0$), the
ultrarelativistic limit ($c\rightarrow 0$) imposes
(see \cite{suarezchavanis3}  for details) that the pseudo
rest-mass
density and the energy
density are given by Eqs. (\ref{badlu}) and (\ref{dim4}). In that case, the
Jeans wavenumber writes
\begin{eqnarray}
k_J^2=\frac{4m^2c^2}{3\hbar^2}.
\label{urng1}
\end{eqnarray}
The corresponding Jeans length is
\begin{eqnarray}
\lambda_J=\sqrt{3}\pi\frac{\hbar}{mc}.
\label{urng2}
\end{eqnarray}
Using Eqs. (\ref{dim4}) and (\ref{urng2}), the Jeans mass defined by Eq.
(\ref{mj}) is
given by
\begin{eqnarray}
M_J=\frac{\pi^3}{16\sqrt{3}}\frac{\hbar}{|a_s|c}.
\label{urng3}
\end{eqnarray}
Introducing the scales $\lambda_i$, $M_i$, $\rho_i$ (see Appendix
\ref{sec_ngsi}) and the relativistic parameter $\nu=\rho/3\rho_i$ 
(see Appendix \ref{sec_dv2}) adapted to the
nongravitational limit (see Appendix
\ref{sec_tsng}), we obtain
\begin{eqnarray}
\frac{\lambda_J}{\lambda_i}=\sqrt{3}\pi,\qquad 
\frac{M_J}{M_i}=\frac{\pi^3}{16\sqrt{3}}.
\label{fog}
\end{eqnarray}
Since  $\nu\sim 1$ (i.e. $\rho\sim\rho_i$) in the ultrarelativistic
limit, we find that $\lambda_J\sim\lambda_i$ and $M_J\sim M_i$.

The preceding results are obtained by taking the
nongravitational limit $G\rightarrow 0$ before the ultrarelativistic limit
$c\rightarrow 0$. We now consider the case where  the ultrarelativistic limit
$c\rightarrow 0$ is taken before the nongravitational limit $G\rightarrow 0$.

Introducing the scales $\lambda_C$, $M_C$, $\rho_C$ (see Appendix \ref{sec_cs})
and the gravitational parameter
$\sigma=3|a_s|/2r_S$ (see
Appendix \ref{sec_effS}) adapted to the ultrarelativistic limit (see Appendix
\ref{sec_tsurb}), we get 
\begin{eqnarray}
\frac{\lambda_J}{\lambda_C}=\sqrt{3}\pi,\qquad 
\frac{M_J}{M_C}=\frac{\sqrt{3}\pi^3}{64}\frac{1}{\sigma}.
\end{eqnarray}
Since $\sigma\gg 1$ (i.e. $|a_s|\gg r_S$) in the nongravitational limit, we find
that
$\lambda_J\sim\lambda_C$ and $M_J \ll M_C$.

Using Eq. (\ref{dim4}), the Hubble scales $\lambda_H$ and $M_H$ (see Appendix
\ref{sec_hs}) are given by
\begin{equation}
\lambda_H=3\left (\frac{|a_s|\hbar^2}{Gm^3}\right )^{1/2},\qquad
M_H=\frac{3}{2}\left (\frac{|a_s|\hbar^2c^4}{G^3m^3}\right )^{1/2}.
\label{urng4}
\end{equation}
They are of the order of $\lambda_R$ and $M_R$ (see Appendix \ref{sec_is}).
Comparing Eqs. (\ref{urng2}),  (\ref{urng3}) and  (\ref{urng4}), we get
\begin{eqnarray}
\frac{\lambda_J}{\lambda_H}=\frac{\pi}{2\sqrt{\sigma}},\qquad 
\frac{M_J}{M_H}=\frac{\pi^3}{64\sigma^{3/2}}.
\label{urng5}
\end{eqnarray}
Since $\sigma\gg 1$ in the nongravitational limit, we find that
$\lambda_J\ll \lambda_H$ and
$M_J\ll M_H$. Since the Jeans length is smaller than the
Hubble length (horizon), there can be  Jeans instability.
Therefore, structures can form in the
ultrarelativistic limit when $a_s<0$. This is due to the attractive
self-interaction of the bosons, not to self-gravity.

Let us make a numerical application. We consider a boson with a mass
$m=10^{-4}\,
{\rm eV}/c^2$ and a negative scattering length $a_s=-5.8\times 10^{-53}\, {\rm
m}$ (to be specific, we take the same values as for QCD axions \cite{kc}
but we stress that our SF is complex while QCD axions correspond to a real SF).
We consider
the ultrarelativistic limit corresponding to a density
$\rho_i=2.09\times 10^{19}\, {\rm g/m^3}$ [see Eq. (\ref{badlu})]. We find that
$\sigma=3.29\times
10^{14}\gg 1$, implying that we are deep in the nongravitational limit (see
Appendix \ref{sec_tsurb}). We then
obtain $\lambda_J=1.07\times 10^{-2}\, {\rm m}$ and
$M_J=3.41\times 10^{-21}\,
M_{\odot}$. These Jeans scales are much smaller than the Hubble
scales $\lambda_H=1.24\times 10^{5}\, {\rm
m}$ and $M_H=41.9\, M_{\odot}$, implying that the Jeans instability can 
take place. The typical mass and size of the resulting objects can be
compared to the mass and size  $M_{\rm
axiton}\sim 10^{-12}\, M_{\odot}$ and $R_{\rm
axiton}\sim 10^{9}\, {\rm m}$ of
axitons \cite{hr,kt,kt2} that also result from the nongravitational collapse of
a SF with an attractive self-interaction (QCD axion) taking
place in the very early
Universe. We note that axitons correspond to a real SF with an attractive
self-interaction while the objects that we have found correspond to a complex
SF with an attractive self-interaction. They could be called
complaxitons.

{\it Remark:} If we consider ultralight bosons with mass
$m=2.19\times 10^{-22}\, {\rm eV}/c^2$ and negative scattering length
$a_s=-1.11\times 10^{-62}\, {\rm fm}$ (see Sec. \ref{sec_forg}) in the
ultrarelativistic limit where $\rho_i=1.15\times 10^{-9}\, {\rm g/m^3}$, we
obtain $\sigma=2.87\times 10^{7}\gg 1$, $\lambda_J=0.159\, {\rm pc}$ and
$M_J=1.78\times 10^{4}\, M_{\odot}$. These Jeans scales are much smaller than
the Hubble scales $\lambda_H=542\, {\rm pc}$ and $M_H=5.66\times 10^{15}\,
M_{\odot}$ implying that the Jeans instability can 
take place. This suggests that large-scale structures,
corresponding to proto-galaxies (germs), can form in the ultrarelativistic
regime of the SF when $a_s<0$. The resulting galaxies would be much older than
what is usually believed, possibly in agreement with certain recent cosmological
observations
where large-scale structures are observed at high redshifts \cite{nature}.

\section{Astrophysical and cosmological applications in the nonrelativistic
regime (matter era)}
\label{sec_app}

\subsection{Preliminary remarks}

Since large-scale structures cannot form in the ultrarelativistic limit
(except for complaxitons when $a_s<0$), we now
consider the
nonrelativistic limit corresponding to the matter era. In the matter
era $\epsilon=\rho c^2$ [see Eq. (\ref{dim3})], and the DM density (here due to
the SF) behaves as a function of the
scale factor $a$ as\footnote{It is shown in
\cite{shapiro,suarezchavanis1,suarezchavanis3} that the pressure of the SF is
negligible at large scales in the matterlike era so that the homogeneous
SF evolves like pressureless CDM. Note, however, that the pressure of the SF is
important at small scales to stabilize the DM halos and solve the cusp problem.}
\begin{eqnarray}
\rho=\frac{\Omega_{{\rm dm},0}\epsilon_0}{c^2a^3},
\label{appm1}
\end{eqnarray}
where $\epsilon_0=7.64\times 10^{-7}\, {\rm g}\, {\rm m}^{-1}\,{\rm s}^{-2}$ is
the present energy density of the Universe and $\Omega_{{\rm dm},0}=0.2645$ is
the present fraction of DM. Numerically,
\begin{eqnarray}
\frac{\rho}{\rm g/m^3}=2.25\times 10^{-24}\, a^{-3}.
\label{appm2}
\end{eqnarray}
For future reference, we note that the  pulsation defined by
Eq. (\ref{dv1}) evolves with the density as
\begin{eqnarray}
\frac{\omega_0}{{\rm s}^{-1}}=9.16\times 10^{-7}\, \left
(\frac{\rho}{\rm
g/m^3}\right )^{1/2}.
\label{j15}
\end{eqnarray}
The radiation-matter equality occurs at $a_{\rm
eq}=2.95\times 10^{-4}$. This marks the begining of the matter era. At that
moment, the DM density is $\rho_{\rm
eq}=8.77\times 10^{-14}\, {\rm
g}/{\rm m}^3=1.295\times 10^3 M_{\odot}/{\rm pc}^3$ and the pulsation is
$(\omega_0)_{\rm eq}=2.71\times 10^{-13}\, {\rm
s}^{-1}$. In comparision, the present
density of DM
is $\rho_0=2.25\times
10^{-24}\, {\rm g/m^3}$ and the present  pulsation is $(\omega_0)_0=1.37\times
10^{-18}\, {\rm s}^{-1}$.

In the following, we compute the Jeans scales $\lambda_J$ and $M_J$ 
for any value of the density $\rho$ (or scale factor $a$) but we make
numerical applications only at the begining of the matter era, i.e. at $a_{\rm
eq}=2.95\times 10^{-4}$, where the Jeans instability is expected to take place.
For comparison, we also make numerical applications at the present epoch
$a=1$. However, at the present epoch,
nonlinear effects have become important (the DM halos are already formed) so
that the linear Jeans instability analysis is not valid anymore except,
possibly, at very
large scales. In our analysis, we usually compute the physical Jeans
length $\lambda_J$  but, in some cases, we also compute the comoving
Jeans length 
\begin{eqnarray}
\lambda_J^c=\frac{\lambda_J}{a}.
\label{como1}
\end{eqnarray}
The comoving Jeans length plays an important role in the interpretation of the
matter power spectrum \cite{fuzzy,marshrevue}.

\subsection{Possibility to form large-scale structures in the  matter era}
\label{sec_iso}

It is simple to show, for a classical self-gravitating fluid, that structure
formation can take place in the matter era. Using the standard Jeans relation of
Eq. (\ref{intro2}), where $\rho$ is the rest-mass
density, and computing the speed of sound $c_s^2=dP/d\rho$ with the isothermal
equation of state $P=\rho k_B T/m$ implying $c_s=(k_B T/m)^{1/2}$,  we
obtain 
\begin{eqnarray}
k_J^2=\frac{4\pi G\rho}{c_s^2}=\frac{4\pi G\rho m}{k_B T}.
\label{pme2}
\end{eqnarray}
Since $c_s\ll c$ in the matter era, the Jeans length $\lambda_J$ is
much smaller than the Hubble length $\lambda_H$ (see Appendix \ref{sec_hs}).
Therefore, the Jeans
instability can take place in the matter era. We now derive
the corresponding result for a complex SF.

\subsection{The noninteracting limit}
\label{sec_appq}

\subsubsection{The Jeans scales}
\label{sec_appqj}

In this section, we consider a noninteracting SF ($a_s=0$). In the
nonrelativistic limit
$c\rightarrow +\infty$, according to Eq. (\ref{nr7}),
the quantum Jeans length $\lambda_J=2\pi/k_J$ is given by
\begin{eqnarray}
\lambda_J=2\pi\left (\frac{\hbar^2}{16\pi G\rho m^2}\right )^{1/4}.
\label{app1}
\end{eqnarray}
In the
nonrelativistic limit, using Eq.
(\ref{dim3}), the Jeans mass defined by Eq. (\ref{mj}) reduces to
\begin{eqnarray}
M_J=\frac{4}{3}\pi\rho\left (\frac{\lambda_J}{2}\right )^3.
\label{app1a}
\end{eqnarray}
The Jeans mass associated with the Jeans length from Eq. (\ref{app1}) is 
\begin{eqnarray}
M_J=\frac{1}{6}\pi\left
(\frac{\pi^3\hbar^2\rho^{1/3}}{Gm^2}\right
)^{3/4}.
\label{app2}
\end{eqnarray}
Introducing relevant scales, we
get
\begin{eqnarray}
\frac{\lambda_J}{\rm pc}=1.16\times 10^{-12}\, \left (\frac{{\rm
eV}/c^2}{m}\right )^{1/2}\left (\frac{\rm g/m^3}{\rho}\right )^{1/4},
\label{app3}
\end{eqnarray}
\begin{eqnarray}
\frac{M_J}{M_{\odot}}= 1.20\times 10^{-20}\, \left (\frac{{\rm
eV}/c^2}{m}\right )^{3/2}\left (\frac{\rho}{\rm g/m^3}\right )^{1/4}.
\label{app4}
\end{eqnarray}
In the matter era, using Eq. (\ref{appm2}), we find that the Jeans length
increases as
$a^{3/4}$ and the
Jeans mass
decreases as $a^{-3/4}$. The Jeans length
and the Jeans mass
represent the
minimum diameter and the minimum mass of a fluctuation
that can collapse at a given
epoch.\footnote{For a classical fluid with an
isothermal equation of state (see Sec. \ref{sec_iso}), we obtain $\lambda_J=2\pi
(k_B T/4\pi G\rho m)^{1/2}$ and $M_J=(\pi/6)(\pi k_B T/Gm\rho^{1/3})^{3/2}$.
Since $\rho\sim a^{-3}$ and $T\sim a^{-1}$ (temperature of radiation) we find
that $\lambda_J\sim a$ while $M_J\sim 1$ remains constant.}  Eliminating the
density between Eqs.
(\ref{app1}) and (\ref{app2}), we obtain
\begin{eqnarray}
M_{J}\lambda_J=\frac{\pi^4}{6} \frac{\hbar^2}{Gm^2}
.
\label{app5}
\end{eqnarray}
As noted in \cite{prd1},  this relation  is similar to the mass-radius
relation of Newtonian BECDM halos
made of noninteracting bosons:\footnote{This relation can be understood
qualitatively by identifying the halo radius $R$ with the de Broglie wavelength
 $\lambda_{dB}=\hbar/mv$ of a boson with a  velocity  $v\sim
(GM/R)^{1/2}$ equal to the virial
velocity  of the halo.} 
\begin{eqnarray}
MR=9.95\frac{\hbar^2}{Gm^2},
\label{app6}
\end{eqnarray}
where $R$ represents the radius containing $99\%$ of the
mass \cite{rb,membrado,prd2}. Comparing Eqs.
(\ref{app5}) and (\ref{app6}), we find 
\begin{eqnarray}
M_{J}\frac{\lambda_J}{2}=0.820 MR.
\label{app7}
\end{eqnarray}
This similarity is not obvious. Indeed, the Jeans length (\ref{app1}) and
the Jeans
mass (\ref{app2}) are obtained by studying the linear dynamical instability of
an infinite
homogeneous self-gravitating medium while the mass-radius relation (\ref{app6})
is obtained by solving the nonlinear equation of hydrostatic
equilibrium for a single DM halo. Therefore, Eq. (\ref{app5}) applies in the
linear regime of structure formation (when the DM halos start to form),
while Eq. (\ref{app6}) applies in the very
nonlinear regime (when the DM halos are formed). The mass-radius relationships
(\ref{app5}) and (\ref{app6})
are
therefore valid in two extremely different regimes (begining and end of
structure formation). It is therefore intriguing
that they have the same scaling and that they differ only by a numerical factor
$1.64$ of order unity. This coincidence may just be a consequence of
dimensional
analysis.

Introducing the scales $\lambda_C$, $M_C$, $\rho_C$ (see Appendix \ref{sec_cs})
and the relativistic parameter
$\chi=(\rho/2\rho_C)^{1/2}$ (see
Appendix \ref{sec_dv1}) adapted to the noninteracting limit (see Appendix
\ref{sec_tsni}), we get
\begin{equation}
\frac{\lambda_J}{\lambda_C}=\frac{\sqrt{2}\pi}{\chi^{1/2}},
\quad 
\frac{M_J}{M_C}=\frac{\sqrt{2}\pi^3}{12}\chi^{1/2},\quad
\frac{M_J}{M_C}=\frac{\pi^4}{6}\frac{\lambda_C}{\lambda_J}.
\label{ven2}
\end{equation}
Since $\chi\ll 1$ (i.e. $\rho\ll\rho_C$) in the nonrelativistic limit,
we find that $\lambda_J\gg\lambda_C$ and $M_J \ll M_C$. We note that the 
relativistic parameter $\chi$ can be expressed in terms of the Hubble constant
as
$\chi=(3/2)^{1/2}H\hbar/mc^2$.

The preceding results are obtained by taking the
noninteracting limit $a_s\rightarrow 0$ before the nonrelativistic limit
$c\rightarrow +\infty$. We now consider the case where  the nonrelativistic
limit $c\rightarrow +\infty$ is taken before the noninteracting limit
$a_s\rightarrow 0$.

Introducing the scales $\lambda_a$, $M_a$, $\rho_a$ (see Appendix
\ref{sec_cis}) and the self-interaction parameter
$\alpha=(\rho/\rho_a)^{1/2}$ (see Appendix \ref{sec_dv1})  adapted to the
nonrelativistic limit (see Appendix \ref{sec_tsnra}), we
get
\begin{equation}
\frac{\lambda_J}{\lambda_a}=\frac{\sqrt{2}\pi}{\alpha^{1/2}},
\quad 
\frac{M_J}{M_a}=\frac{\sqrt{2}\pi^3}{12}\alpha^{1/2},\quad
\frac{M_J}{M_a}=\frac{\pi^4}{6}\frac{\lambda_a}{\lambda_J}.
\label{alb1}
\end{equation}
Since $\alpha\ll 1$ (i.e. $\rho\ll\rho_a$) in the noninteracting limit,
we find that $\lambda_J\gg\lambda_a$ and $M_J \ll
M_a$.

Introducing the Hubble scales $\lambda_H$ and $M_H$ (see Appendix
\ref{sec_hs}), we obtain 
\begin{eqnarray}
\frac{\lambda_J}{\lambda_H}=\frac{2\pi}{\sqrt{3}}\sqrt{\chi},
\qquad 
\frac{M_J}{M_H}=\frac{\pi^3}{3\sqrt{3}}\chi^{3/2}.
\label{ven1}
\end{eqnarray}
Since $\chi\ll 1$ in the nonrelativistic limit, we find that
$\lambda_J\ll\lambda_H$ and $M_J \ll M_H$. Therefore, large-scale structures can
form in the nonrelativistic regime by Jeans instability since the Jeans length
is
much smaller than the horizon.

We now apply these results to ultralight bosons\footnote{Ultralight bosons are
sometimes called ultralight axions
(ULAs) to distinguishe
them from QCD axions.}  and QCD axions.

\subsubsection{Ultralight axions}
\label{sec_appqb}

We first consider a noninteracting ULA able to form giant
BECs
with the mass and size of DM halos. To determine its mass $m$, we assume
that the most compact DM halos observed in the Universe, namely dwarf
spheroidals (dSphs) like Fornax ($R\sim 1\, {\rm kpc}$, $M\sim 10^8\,
M_{\odot}$, $\overline{\rho}\sim 10^{-18} \, {\rm
g/m^3}$), are pure solitons corresponding to the ground state of the GPP
equations.\footnote{As explained in more detail in Appendix D
of \cite{suarezchavanis3} and in \cite{epjplusarejouter}, large halos have a
core-halo structure with a
solitonic core and a Navarro-Frenk-White (NFW) atmosphere. The core corresponds
to the ground state of the GPP equations and the NFW atmosphere may be the
result of  violent relaxation \cite{lbvr}, gravitational cooling \cite{seidel94}
or
hierarchical clustering. The precise  structure of the atmosphere may be
influenced by incomplete violent relaxation, tidal effects,
stochastic forcing etc.} The
mass-radius relation (\ref{app6}) then gives a
boson mass $m=2.92\times 10^{-22}\, {\rm eV}/c^2$ (see Appendix D of
\cite{suarezchavanis3}). We note that, inversely, the knowledge of the DM
particle mass $m$ does not determine $M$ and $R$ individually, but only their
product $MR$. The individual determination of $M$ and $R$ depends on the epoch
(time, scale factor, or redshift) and can be obtained from the Jeans study. Let
us apply this study at the epoch of
radiation-matter equality.  For a boson mass 
$m=2.92\times 10^{-22}\, {\rm
eV}/c^2$, we find that the dimensionless relativistic parameter  defined by Eq.
(\ref{dv3})  is $\chi=6.12\times 10^{-7}$. The smallness of this value shows
that we are in the nonrelativistic limit (the transition between the
ultrarelativistic limit and the nonrelativistic limit takes place at
$\rho_C=0.118 \, {\rm
g}/{\rm m}^3 \gg\rho_{\rm eq}=8.77\times 10^{-14}\, {\rm
g}/{\rm m}^3$). To evaluate the  Jeans length and the Jeans mass  at the
epoch of radiation-matter equality we use Eqs.  (\ref{app1}) and (\ref{app2})
 and obtain $\lambda_J=125\, {\rm pc}$ and $M_J=1.31\times
10^{9}\, M_{\odot}$. In
comparison,
$\lambda_H=4.40\times 10^4\, {\rm pc}$, $M_H=4.59\times 10^{17}\,
M_{\odot}$, $\lambda_C=2.20\times 10^{-2}\, {\rm pc}$ and $M_C=4.59\times
10^{11}\, M_{\odot}$.  The
relativistic corrections are negligible since $\chi\ll 1$. We note
that the Jeans length $\lambda_J$ at the begining of the structure formation
process (radiation-matter equality epoch) is one order of magnitude smaller than
the current radius $R$ of dwarf DM halos like Fornax and the Jeans
mass $M_J$ is one order of magnitude larger than their
current mass $M$. This
suggests
that the system loses mass during the nonlinear process of halo
formation and increases in size. This explains why the current density of the
dwarf DM halos $\overline{\rho}\sim 10^{-18} \, {\rm
g/m^3}$ is five orders of magnitude smaller than the background density at the
epoch of
radiation-matter equality $\rho_{\rm eq}=8.77\times 10^{-14}\, {\rm
g}/{\rm m}^3$. We also note that the comoving Jeans length
$\lambda_J^c$ defined by Eq. (\ref{como1}) is, at the epoch of
radiation-matter equality, equal to  $\lambda_J^c=0.424\, {\rm Mpc}$.

\begin{figure}[h]
\scalebox{0.33}{\includegraphics{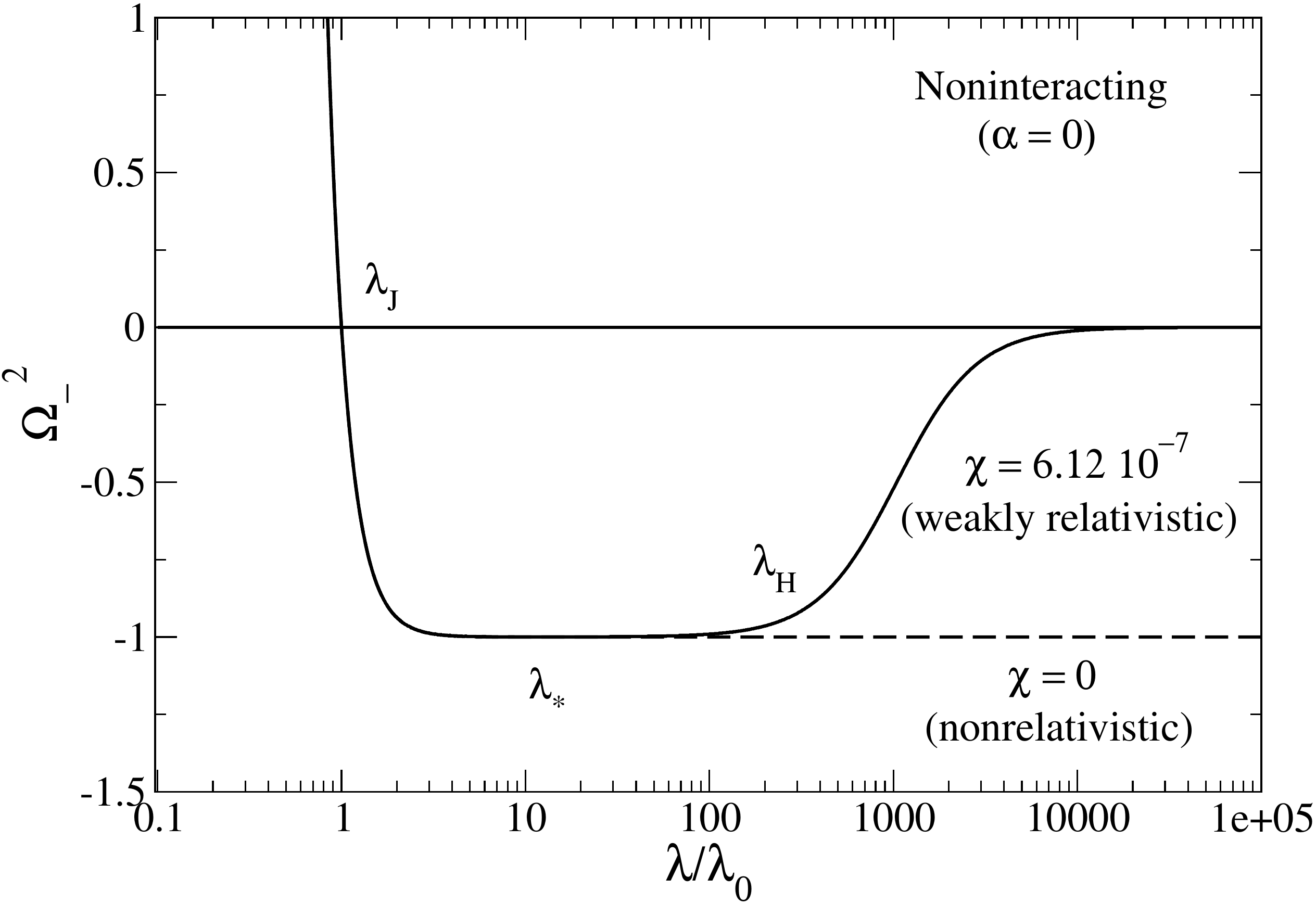}} 
\caption{Growth rate $\Sigma=(-\Omega_{-}^2)^{1/2}$ of the perturbations in the
noninteracting limit ($a_s=0$) as a function of the wavelength
$\lambda/\lambda_0=2\pi/\kappa$ for the relativistic parameter
$\chi=6.12\times 10^{-7}$. In the weakly relativistic regime
($\chi\ll 1$), the Jeans length
$\lambda_J$ is much smaller than the Hubble length $\lambda_H$. The growth
rate shows a plateau with typical value
$\Sigma_{\rm max}$ in the range $10\lambda_J\le \lambda\le \lambda_H$.
Therefore, structures can form through Jeans instability in this
range of scales with almost the same growth rate. The Newtonian case is shown as
a dashed line for comparison. In that case, the plateau extends to infinity.}
\label{optimumCHI}
\end{figure}

To evaluate the maximum growth rate at the epoch of
radiation-matter equality, we can use the nonrelativistic
result of Eq. (\ref{nr5b}). We obtain $\sigma_{\rm max}=2.71\times
10^{-13}\, {\rm s}^{-1}$ corresponding to a characteristic time $\sigma_{\rm
max}^{-1}=3.69\times 10^{12}\, {\rm s}=1.17\times 10^5\, {\rm yrs}$. The
relativistic corrections are negligible since $\chi\ll 1$. By
contrast, in order to determine the most unstable wavelength
$\lambda_*$, we need to take into account relativistic corrections even though
$\chi\ll 1$. Indeed, in the Newtonian approximation ($\chi=0$), the most
unstable wavelength is infinite ($k_*=0$ or $\lambda_*\rightarrow +\infty$).
However, when relativistic corrections are taken into account, we find that the
maximum growth rate has a finite value.\footnote{We note, in contrast, that the
simplified model of
Sec. \ref{sec_sm} gives $\lambda_*\rightarrow +\infty$ like the
nonrelativistic model of Sec. \ref{sec_nr}.} When
$\chi\rightarrow 0$, we get (see Appendix \ref{sec_mgrne}):
\begin{eqnarray}
\frac{\lambda_*}{\lambda_J}\sim \frac{1}{A\chi^{1/6}},
\label{app8}
\end{eqnarray}
where $A=0.953184...$ For $\chi=6.12\times
10^{-7}$, we
obtain
$\lambda_*=11.4\lambda_J=1.43\, {\rm kpc}$. Therefore, the
maximum growth rate is reached at a length $\lambda_*$ equal to about ten times
the Jeans length $\lambda_J$.\footnote{Since
$\chi=6.12\times 10^{-7}$ is very small, it may seem surprising
that $\lambda_*$ has a relatively small value ($\sim 10\lambda_J$) while
$\lambda_*\rightarrow +\infty$ as $\chi\rightarrow 0$. The reason is that
$\lambda_*$ diverges as $\chi^{-1/6}$, where $1/6$ is a small exponent. This is
why the value of $\lambda_*$ is relatively small for $\chi=6.12\times 10^{-7}$,
and why
relativistic effects are important
to determine this quantity, while they can be safely neglected to compute other
quantities such as $\lambda_J$, $M_J$, $\sigma_{\rm max}$ etc.} This is an
interesting result because this length
is precisely of the order of the size of dwarf DM halos like Fornax.  In Fig.
\ref{optimumCHI}, we plot the growth rate
$\sigma=(-\omega_{-}^2)^{1/2}$ of the perturbation as a function of the
wavelength $\lambda=2\pi/k$ for a relativistic
parameter $\chi=6.12\times
10^{-7}$ using the normalization of Appendix
\ref{sec_dv1}. For such a small value of $\chi$, this figure displays a
plateau starting at about $10\lambda_J$ and ending at
about $\lambda_H$. Below the Jeans length $\lambda_J=125\, {\rm pc}$ there is no
instability
and above the Hubble length $\lambda_H=4.40\times 10^4\, {\rm pc}$ the growth
rate $\sigma(\lambda)$ substantially  decreases because of general
relativity (see the Remark at the end of Sec. \ref{sec_empa}). Therefore, the
Jeans instability can take place, with almost the same growth rate $\sim
\sigma_{\rm max}=2.71\times
10^{-13}\, {\rm s}^{-1}$, in the range
$10\lambda_J\le\lambda\le \lambda_H$. By comparison, in the nonrelativistic
model
($\chi=0$), the maximum growth rate corresponds to $\lambda_*=\infty$ and the
plateau extends to infinity. As a result, there is no upper limit on the size of
the clusters that can undergo Jeans' instability. Therefore, our relativistic
treatment solves one of the problems of the classical Jeans
theory discussed in the Introduction.

{\it Remark:} If we compute the Jeans length and the Jeans mass at
the present epoch, we find $\chi=3.10\times 10^{-12}$, $\lambda_J=55.5\, {\rm
kpc}$, $M_J=2.95\times 10^{6}\, M_{\odot}$ and
$\lambda_*=86.9\lambda_J=4.82\times
10^{3}\, {\rm
kpc}$. In comparison $\lambda_H=8.69\times 10^{6}\, {\rm
kpc}$,
$M_H=9.06\times 10^{22}\, M_{\odot}$, $\lambda_C=2.20\times 10^{-5}\, {\rm
kpc}$, and $M_C=4.59\times 10^{11}\,
M_{\odot}$. For ULAs with a
mass $m=2.92\times 10^{-22}\, {\rm eV}/c^2$, the Jeans
length is of the order of the galactic size.
Therefore, ULAs  can form DM halos of relevant
size.

\subsubsection{QCD axions}
\label{sec_appqqcd}

We now consider QCD axions with mass $m=10^{-4}\, {\rm eV}/c^2$
and scattering length $a_s=-5.8\times 10^{-53}\, {\rm m}$. In the
nonrelativistic regime, the axions can be described
by a complex SF governed by the GPP equations. We apply the
Jeans study at the 
epoch of radiation-matter equality. The dimensionless self-interaction 
parameter defined by Eq. (\ref{dv4}) is $\alpha=7.82\times
10^{-10}$. The smallness of this value shows that we are in the noninteracting
limit (the transition between the
nongravitational limit and the noninteracting limit takes place at
$\rho_a=1.44\times 10^{5} \, {\rm
g}/{\rm m}^3 \gg\rho_{\rm eq}=8.77\times 10^{-14}\, {\rm
g}/{\rm m}^3$). Therefore, we can neglect the self-interaction of the
axions and take $a_s=0$.\footnote{This approximation is
valid in the linear regime of structure formation when considering the growth
of perturbations in a homogeneous Universe (Jeans problem). In the nonlinear
regime of structure formation (equilibrium states $=$ DM halos) the attractive
self-interaction of the
QCD axions  must be taken into account and
leads to mini axion stars  with a maximum mass $M_{\rm
max}=6.5\times
10^{-14}\, M_{\odot}$ and a minimum radius $R_*=3.3\times 10^{-4}\,
R_{\odot}=230\,
{\rm km}$ \cite{phi6}.} On
the other hand, we find that the dimensionless
relativistic parameter has a small value $\chi=1.79\times 10^{-24}\ll 1$ so that
we are in the nonrelativistic regime (the
transition between the
ultrarelativistic limit and the nonrelativistic limit takes place at
$\rho_C=1.38\times 10^{34} \, {\rm
g}/{\rm m}^3 \gg\rho_{\rm eq}=8.77\times 10^{-14}\, {\rm
g}/{\rm m}^3$). We
then obtain $\lambda_J=2.13\times
10^{-7}\, {\rm pc}=9.45\, R_{\odot}=6.57\times 10^9 {\rm m}$,
$M_J=6.53\times 10^{-18}\,
M_{\odot}=1.30\times 10^{13} {\rm kg}$, $\sigma_{\rm
max}=2.71\times 10^{-13}{\rm s}^{-1}$, and $\lambda_*=9.52\times 10^3\,
\lambda_J=2.03\times 10^{-3}\, {\rm
pc}$.  In comparison,
$\lambda_H=4.40\times 10^4\, {\rm pc}$, $M_H=4.59\times 10^{17}\,
M_{\odot}$, $\lambda_C=6.41\times 10^{-20}\, {\rm pc}$, $M_C=1.33\times
10^{-6}\, M_{\odot}$, $\lambda_a=1.34\times 10^{-12}\, {\rm pc}$,
and $M_a=6.38\times 10^{-14}M_{\odot}$. We note
that the Jeans length and the Jeans mass
are much smaller than the typical size and mass of DM halos. As a result, QCD
axions behave essentially as CDM and cannot solve the CDM
crisis. They may form mini axion stars \cite{bectcoll} that could be the
constituents
of DM halos (in the form of mini-MACHOs), but they cannot form DM halos. As a
result, QCD axions (or mini axion stars) can be regarded as possible CDM
particle candidates.
We also note that the comoving Jeans length at the
epoch of
radiation-matter equality is $\lambda_J^c=7.22\times 10^{-4}\, {\rm pc}$.

{\it Remark:} If we compute the Jeans length and the Jeans mass at
the present epoch, we find $\alpha=3.96\times 10^{-15}$, $\chi=9.05\times
10^{-30}$, $\lambda_J=9.46\times
10^{-5}\, {\rm pc}=19.5\, {\rm AU}$, $M_J=1.46\times 10^{-20}\,
M_{\odot}$, and $\lambda_*=7.27\times 10^{4}\lambda_J=6.88 \, {\rm pc}$. In
comparison
$\lambda_H=8.69\times
10^{9}\, {\rm pc}$, $M_H=9.06\times 10^{22}\, M_{\odot}$, $\lambda_C=6.41\times
10^{-20}\, {\rm pc}$, $M_C=1.33\times 10^{-6}\,
M_{\odot}$, $\lambda_a=1.34\times 10^{-12}\, {\rm pc}$,
and $M_a=6.38\times 10^{-14}M_{\odot}$. For
QCD axions the Jeans length is of the order of the Solar System size and the
Jeans mass is of the order of the asteroid mass. Therefore,
on all scales relevant in cosmology, the QCD axion fluid can be treated as a
pressureless fluid equivalent to CDM.

\subsection{The TF limit}
\label{sec_appc}

\subsubsection{The Jeans scales}

In this section, we consider a SF with a repulsive self-interaction ($a_s>0$) in
the TF
limit $\hbar\rightarrow 0$ and in the nonrelativistic limit
$c\rightarrow +\infty$. The Jeans wavenumber is given by Eq. (\ref{nr10}) where
the velocity of sound is given by Eq. (\ref{jia9}). The Jeans
length
$\lambda_J=2\pi/k_J$ is 
\begin{eqnarray}
\lambda_J=2\pi\left (\frac{a_s\hbar^2}{G m^3}\right )^{1/2}.
\label{app9}
\end{eqnarray}
We note that the Jeans length is independent of the density.
The associated Jeans mass from Eq. (\ref{app1a}) is 
\begin{eqnarray}
M_J=\frac{1}{6}\pi\rho \left (\frac{4\pi^2 a_s\hbar^2}{G
m^3}\right )^{3/2}.
\label{app10}
\end{eqnarray}
Introducing relevant scales, we get
\begin{eqnarray}
\frac{\lambda_J}{\rm pc}=34.9\, \left (\frac{a_s}{\rm
 fm}\right )^{1/2}\left (\frac{{\rm
eV}/c^2}{m}\right )^{3/2},
\label{app11}
\end{eqnarray}
\begin{eqnarray}
\frac{M_J}{M_{\odot}}=3.30\times 10^{20}\, \left
(\frac{a_s}{\rm
 fm}\right )^{3/2}\left (\frac{{\rm
eV}/c^2}{m}\right )^{9/2} \frac{\rho}{\rm g/m^3}.
\label{app12}
\end{eqnarray}
In the matter era, using Eq. (\ref{appm2}), we find that the Jeans length
is constant while the Jeans mass decreases as $a^{-3}$.
As noted in \cite{prd1}, the relation of
Eq. (\ref{app9}) is
similar to the relation\footnote{This is the radius of a polytrope of index
$n=1$ corresponding to the equation of state (\ref{jia8}).} 
\begin{eqnarray}
R=\pi\left (\frac{a_s\hbar^2}{G m^3}\right )^{1/2}
\label{app13}
\end{eqnarray}
determining the radius of a self-interacting DM halo in the TF
approximation \cite{leekoh,goodman,arbey,bohmer,prd1}. Comparing Eqs.
(\ref{app9}) and (\ref{app13}), we find that
\begin{eqnarray}
\frac{\lambda_J}{2}=R.
\label{app14}
\end{eqnarray}
As before, this coincidence is essentially a consequence of dimensional
analysis.

Introducing the scales $\lambda_R$, $M_R$, $\rho_R$ (see Appendix \ref{sec_is})
and the relativistic parameter $\nu=2\rho/3\rho_R$  (see Appendix
\ref{sec_dv2}) adapted to the TF limit (see  Appendix \ref{sec_tstf}),
we get
\begin{eqnarray}
\frac{\lambda_J}{\lambda_R}=2\pi,\qquad 
\frac{M_J}{M_R}=\frac{\pi^3}{3}\nu.
\label{soir2}
\end{eqnarray}
Since $\nu\ll 1$ (i.e. $\rho\ll \rho_R$) in the nonrelativistic limit, we find
that $\lambda_J\sim\lambda_R$ and $M_J \ll M_R$.

The preceding results are obtained by taking the
TF limit $\hbar\rightarrow 0$ before the nonrelativistic limit
$c\rightarrow +\infty$. We now consider the case where  the nonrelativistic
limit $c\rightarrow +\infty$ is taken before the TF limit $\hbar\rightarrow 0$.

Introducing the scales $\lambda_a$ and $M_a$ (see Appendix \ref{sec_cis})
and the self-interaction parameter $\alpha=(\rho/\rho_a)^{1/2}$ (see
Appendix \ref{sec_dv1}) adapted to the nonrelativistic limit  (see Appendix
\ref{sec_tsnra}), we get 
\begin{eqnarray}
\frac{\lambda_J}{\lambda_a}=2\pi,\qquad 
\frac{M_J}{M_a}=\frac{\pi^3}{3}\alpha^2.
\label{sunq}
\end{eqnarray}
Since $\alpha\gg 1$ (i.e. $\rho\gg \rho_a$) in the TF limit, we find
that
$\lambda_J\sim\lambda_a$ and $M_J \gg M_a$.

Introducing the Hubble scales $\lambda_H$ and $M_H$ (see Appendix
\ref{sec_hs}), we obtain 
\begin{eqnarray}
\frac{\lambda_J}{\lambda_H}=\frac{2\sqrt{2}\pi}{\sqrt{3}}\nu^{
1/2},\qquad 
\frac{M_J}{M_H}=\frac{1}{8}\left (\frac{8}{3}\right )^{3/2}\pi^3 \nu^{3/2}.
\label{soir1}
\end{eqnarray}
Since $\nu\ll 1$ in the nonrelativistic limit, we
find that
$\lambda_J\ll\lambda_H$ and $M_J \ll M_H$. Therefore, large-scale structures can
form in
the nonrelativistic regime by Jeans instability since the Jeans length is
much smaller than the horizon.

We now apply these results to self-interacting bosons.

\subsubsection{Self-interacting bosons}
\label{sec_sib}

We consider a self-interacting  boson able to form giant BECs
with the mass and size of DM halos.  To
determine the ratio $a_s/m^3$, we 
use the same argument as in Sec. \ref{sec_appqb}.  We assume that
dSphs like Fornax ($R\sim 1\, {\rm kpc}$, $M\sim 10^8\,
M_{\odot}$, $\overline{\rho}\sim 10^{-18} \, {\rm
g/m^3}$) are purely solitonic, corresponding to the ground state of the
GPP equations. In the TF limit, the relation from Eq. (\ref{app13})  gives
$a_s/m^3=3.28\times
10^{3}\, {\rm fm}/({\rm eV/c^2})^3$ (see Appendix D of \cite{suarezchavanis3}).
Other arguments developed in \cite{shapiro} and in Appendix D of
\cite{suarezchavanis3} can be used to put constraints on the individual values
of $a_s$ and $m$. We shall consider two models (see Appendix D of
\cite{suarezchavanis3}) corresponding to
$(m,a_s)=(1.10\times 10^{-3}\, {\rm eV/c^2}, 4.41\times 10^{-6}\, {\rm fm})$ and
$(m,a_s)=(3\times 10^{-21}\, {\rm eV/c^2}, 1.11\times 10^{-58}\, {\rm fm})$.
We note that the knowledge of the DM particle mass $m$ and scattering length
$a_s$ only determines the radius $R$ of the halo (through the ratio $a_s/m^3$),
not its mass $M$. The individual determination of $M$ and
$R$ depends on the epoch (time, scale factor, or redshift) and can be obtained
from the Jeans study. Let us apply this study at the epoch of radiation-matter
equality. For the two models considered
above, the values of the dimensionless self-interaction 
parameter defined by Eq. (\ref{dv4}) are  $\alpha=4.91\times
10^{20}$ and $\alpha=1.66\times
10^{3}$ respectively.  The greatness of these values shows that we are in the
TF limit (the transition between the TF limit and the noninteracting limit takes
place at
$\rho_a=3.64\times 10^{-55} \, {\rm
g}/{\rm m}^3 \ll\rho_{\rm eq}=8.77\times 10^{-14}\, {\rm
g}/{\rm m}^3$ and $\rho_a=3.18\times 10^{-20} \, {\rm
g}/{\rm m}^3 \ll\rho_{\rm eq}=8.77\times 10^{-14}\, {\rm
g}/{\rm m}^3$ respectively). On the other hand, for a
boson with ratio $a_s/m^3=3.28\times 10^{3}\, {\rm fm}/({\rm eV/c^2})^3$, we
find
that the
dimensionless relativistic parameter defined by Eq. (\ref{dv10}) is
$\nu=7.88\times
10^{-5}$. The smallness of this value shows that we are in the  nonrelativistic
limit (the transition between the
ultrarelativistic limit and the nonrelativistic limit takes place at
$\rho_R=7.41\times 10^{-10} \, {\rm
g}/{\rm m}^3 \gg\rho_{\rm eq}=8.77\times 10^{-14}\, {\rm
g}/{\rm m}^3$). To
evaluate
the  Jeans length and the Jeans mass  at the
epoch of
radiation-matter equality we use Eqs.  (\ref{app9}) and (\ref{app10}) and obtain
$\lambda_J=2.00\, {\rm kpc}$ and $M_J=5.44\times
10^{12}\, M_{\odot}$. In comparison,
$\lambda_H=4.40\times 10^4\, {\rm pc}$, $M_H=4.59\times 10^{17}\,
M_{\odot}$, $\lambda_a=318\, {\rm pc}$,
$M_a=2.21\times 10^{-30}\, M_{\odot}$ (model I), $M_a=2.66\times
10^{5}\, M_{\odot}$ (model II), $\lambda_R=318\, {\rm pc}$, and
$M_R=6.68\times
10^{15}\, M_{\odot}$. The
relativistic corrections are negligible since $\nu\ll 1$.  We note that the
Jeans length $\lambda_J$ at the begining of the structure formation process 
(radiation-matter equality epoch) is of the same order as the current radius
$R$ of dwarf DM halos like Fornax (as a consequence of Eq. (\ref{app14})). By
contrast, the Jeans
mass $M_J$ is four orders of magnitude
larger than their current mass $M$. This suggests that the system loses
mass
during the nonlinear process of halo
formation but keeps the same size.  This explains why the current density of
dwarf DM halos is much smaller than the background density at the epoch of
radiation-matter equality. We also note that the comoving Jeans
length at the epoch of
radiation-matter equality is $\lambda_J^c=6.78\, {\rm Mpc}$.

\begin{figure}[h]
\scalebox{0.33}{\includegraphics{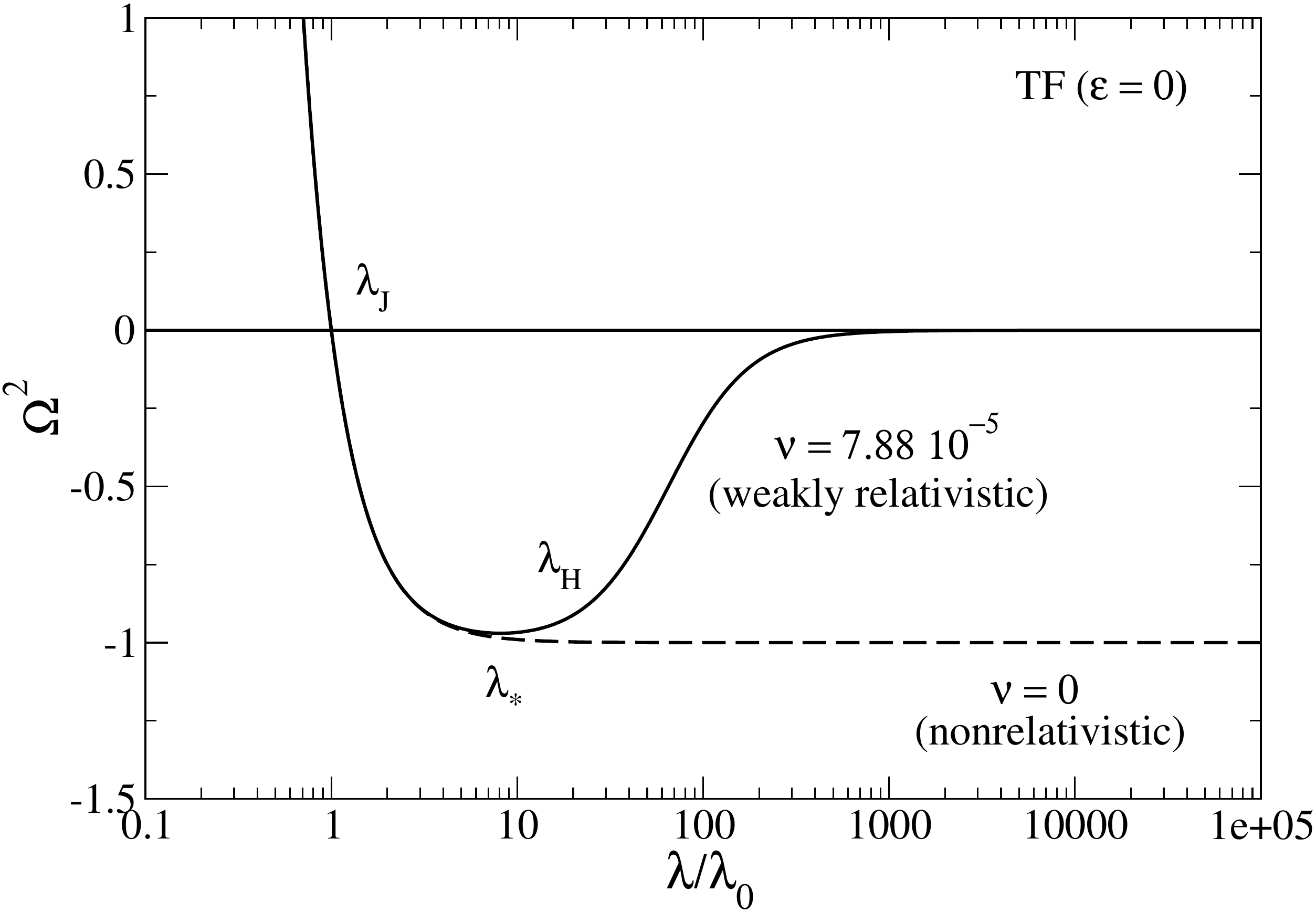}} 
\caption{Growth rate $\Sigma=\sqrt{-\Omega^2}$ of the perturbations in the TF
limit ($\hbar\rightarrow 0$) as a function of the
wavelength
$\lambda/\lambda_0=2\pi/\kappa$ for the relativistic parameter
$\nu=7.88\times 10^{-5}$ (solid line). The Newtonian case is shown for
comparison (dashed line). The interpretation is the same as in Fig.
\ref{optimumCHI} but the plateau is less developed in the present
case because $\lambda_J$ is
closer to $\lambda_H$.}
\label{optimumNU}
\end{figure}

To evaluate the maximum growth rate at the epoch of
radiation-matter equality, we can use the nonrelativistic
result of Eq. (\ref{nr5b}). We obtain $\sigma_{\rm max}=2.71\times 10^{-13}\,
{\rm s}^{-1}$ corresponding to a
characteristic time $\sigma_{\rm max}^{-1}=3.69\times 10^{12}\,
{\rm s}=1.17\times 10^5\, {\rm yrs}$. The
relativistic corrections are negligible since $\nu\ll
1$. However, in order to determine the most unstable
wavelength
$\lambda_*$, we need to take into account relativistic corrections even though
$\nu\ll 1$ (for the same reason as that given in Sec. \ref{sec_appqb}). When
$\nu\rightarrow 0$, we get (see Appendix \ref{sec_mgrtfe}):
\begin{eqnarray}
\frac{\lambda_*}{\lambda_J}\sim\frac{1}{(3\nu)^{1/4}}.
\label{app15}
\end{eqnarray}
For $\nu=7.88\times 10^{-5}$, we obtain
$\lambda_*=8.06\, \lambda_J=16.1\, {\rm kpc}$. Therefore, the maximum growth
rate is reached at a length $\lambda_*$ equal to about ten times the Jeans
length $\lambda_J$. In Fig. \ref{optimumNU}, we have plotted the
growth rate
$\sigma$ of the perturbation as a
function of the wavelength $\lambda=2\pi/k$ for a relativistic
parameter $\nu=7.88\times 10^{-5}$ using the normalization of Appendix
\ref{sec_dv2}. The
conclusions are essentially the same as those reached
in Sec. \ref{sec_appqb}.

{\it Remark:} If we compute the Jeans length and the Jeans mass at
the present epoch, we find $\nu=2.02\times 10^{-15}$, $\lambda_J=2.00\, {\rm
kpc}$, $M_J=1.39\times 10^{2}\, M_{\odot}$
and
$\lambda_*=3.58\times 10^3 \lambda_J=7.17\times
10^{3}\, {\rm
kpc}$. In comparison $\lambda_H=8.69\times 10^{6}\, {\rm
kpc}$,
$M_H=9.06\times 10^{22}\, M_{\odot}$, $\lambda_a=318\, {\rm pc}$,
$M_a=2.21\times 10^{-30}\, M_{\odot}$ (model I), $M_a=2.66\times
10^{5}\, M_{\odot}$ (model II), $\lambda_R=318\,
{\rm pc}$, and $M_R=6.68\times 10^{15}\,
M_{\odot}$. For
self-interacting bosons, the Jeans length is of the
order of the
galactic size. However, we find
that the Jeans mass is considerably smaller than the mass of DM halos. This
suggests
that the linear Jeans analysis may not be applicable at the present epoch which
corresponds to a very nonlinear regime. For the two models considered
above, the values of the dimensionless self-interaction 
parameter defined by Eq. (\ref{dv4}) are  $\alpha=2.49\times
10^{15}$ and $\alpha=8.42\times
10^{-3}$ respectively.  This shows that, at the present epoch, the TF
limit is clearly valid for model I while it is not valid for
model II. This is consistent with the remark made in Appendix D
of
\cite{suarezchavanis3} according to which the second model, when applied to a DM
halo (equilibrium state) obtained in the nonlinear regime of structure
formation, is
close to the transition between the noninteracting limit and the TF limit.
Therefore, a nonperturbative calculation  similar to the one
performed in \cite{prd2}, taking into account
both the self-interaction and the quantum potential, is necessary in that case.

\subsection{The nongravitational limit}
\label{sec_ng}

\subsubsection{The Jeans scales}

In this section, we consider a SF with an attractive self-interaction ($a_s<0$)
in the
nongravitational limit ($G=0$) and in the nonrelativistic limit
$c\rightarrow +\infty$.\footnote{The case of a nonrelativistic self-gravitating
SF with an attractive
self-interaction is
treated in \cite{prd1}.}  According to Eq.
(\ref{nog3}), the Jeans length
$\lambda_J=2\pi/k_J$ is given by
\begin{eqnarray}
\lambda_J=2\pi\left (\frac{m}{16\pi|a_s|\rho}\right )^{1/2}.
\label{ng1}
\end{eqnarray}
The associated Jeans mass  from Eq. (\ref{app1a}) is 
\begin{eqnarray}
M_J=\frac{\pi}{6}\frac{1}{\rho^{1/2}} \left (\frac{\pi m}{4|a_s|}\right )^{3/2}.
\label{ng2}
\end{eqnarray}
Introducing relevant scales, we get
\begin{equation}
\frac{\lambda_J}{\rm pc}=3.83\times 10^{-26}\, \left (\frac{\rm
 fm}{|a_s|}\right )^{1/2}\left (\frac{m}{{\rm
eV}/c^2}\right )^{1/2}\left (\frac{\rm g/m^3}{\rho}\right )^{1/2},
\label{ng4}
\end{equation}
\begin{equation}
\frac{M_J}{M_{\odot}}=4.36\times 10^{-61}\, \left
(\frac{\rm
 fm}{|a_s|}\right )^{3/2}\left (\frac{m}{{\rm
eV}/c^2}\right )^{3/2} \left (\frac{\rm g/m^3}{\rho}\right )^{1/2}.
\label{ng5}
\end{equation}
In the matter era,  using Eq. (\ref{appm2}), we find that the 
Jeans length and the Jeans mass both
increase 
as $a^{3/2}$.  
Eliminating the density between Eqs. (\ref{ng1}) and (\ref{ng2}), we
obtain
\begin{eqnarray}
M_J=\frac{\pi^2}{24}\frac{m}{|a_s|} \lambda_J.
\label{ng3}
\end{eqnarray}
As noted in \cite{prd1}, this relation is similar to the
mass-radius relation of a nongravitational BEC with an attractive
self-interaction  \cite{prd2}:
\begin{eqnarray}
M=0.275 \frac{m}{|a_s|}R,
\label{ng3b}
\end{eqnarray}
where $R$ represents the radius containing $99\%$ of the mass. We
have
\begin{eqnarray}
\frac{M}{R}=0.334\frac{M_J}{\lambda_J/2}.
\label{ng3bb}
\end{eqnarray}
We recall, however, that the equilibrium states of a
nongravitational BEC with an attractive self-interaction are unstable (see
\cite{prd1} for detail). Therefore, only the relations (\ref{ng1})-(\ref{ng3})
obtained from
the linear  Jeans analysis make sense. They determine the onset of collapse of
a homogeneous distribution of BECs  due to their attractive self-interaction. 
Their evolution in the nonlinear regime (collapse) requires a specific
analysis \cite{bectcoll,eby,
tkachevprl,helfer,dkr,phi6}.

Introducing the scales $\lambda_i$, $M_i$, $\rho_i$ (see Appendix
\ref{sec_ngsi}) and the relativistic parameter $\nu=\rho/3\rho_i$  (see Appendix
\ref{sec_dv2}) adapted to the
nongravitational limit (see Appendix
\ref{sec_tsng}), we obtain
\begin{equation}
\frac{\lambda_J}{\lambda_i}=\pi \frac{1}{\nu^{1/2}},\quad 
\frac{M_J}{M_i}=\frac{\pi^3}{24}\frac{1}{\nu^{1/2}},\quad
\frac{M_J}{M_i}=\frac{\pi^2}{24}\frac{\lambda_J}{\lambda_i}.
\label{ywo}
\end{equation}
Since $\nu\ll 1$ (i.e. $\rho\ll \rho_i$) in the nonrelativistic limit, we
find that $\lambda_J\gg\lambda_i$ and $M_J\gg M_i$.

The preceding results are obtained by taking the
nongravitational limit $G\rightarrow 0$ before the nonrelativistic limit
$c\rightarrow +\infty$. We now consider the case where  the nonrelativistic
limit $c\rightarrow +\infty$ is taken before the nongravitational limit
$G\rightarrow 0$.

Introducing the scales $\lambda_a$, $M_a$, $\rho_a$ (see Appendix
\ref{sec_cis}) and the self-interaction parameter
$\alpha=(\rho/\rho_a)^{1/2}$
(see Appendix \ref{sec_dv1})  adapted to the nonrelativistic limit (see Appendix
\ref{sec_tsnrb}), we get
\begin{eqnarray}
\frac{\lambda_J}{\lambda_a}=\frac{\pi}{\alpha},\qquad 
\frac{M_J}{M_a}=\frac{\pi^3}{24\alpha},\qquad
\frac{M_J}{M_a}=\frac{\pi^2}{24}\frac{\lambda_J}{\lambda_a}.
\label{wqo}
\end{eqnarray}
Since $\alpha\gg 1$ (i.e. $\rho\gg\rho_a$) in the nongravitational limit, we
find that
$\lambda_J\ll\lambda_a$ and $M_J \ll M_a$.

Introducing the Hubble scales $\lambda_H$ and $M_H$ (see Appendix
\ref{sec_hs}), we obtain
\begin{eqnarray}
\frac{\lambda_J}{\lambda_H}=\frac{\pi}{\sqrt{2\sigma}},\qquad
\frac{M_J}{M_H}=\frac{\pi^3}{16\sqrt{2}}\frac{1}{\sigma^{3/2}}
\end{eqnarray}
Since $\sigma\gg 1$ in the nongravitational limit, we find that
$\lambda_J\ll \lambda_H$ and $M_J\ll M_H$. Therefore, large-scale structures can
form in the nonrelativistic regime since the Jeans length is
much smaller than the horizon. We stress that this instability is due to
the attractive self-interaction of the bosons ($a_s<0$), not to
self-gravity.

We now apply these results to ultralight bosons with attractive
self-interaction (ULAs).

\subsubsection{Ultralight axions}
\label{sec_forg}

\begin{figure}[h]
\scalebox{0.33}{\includegraphics{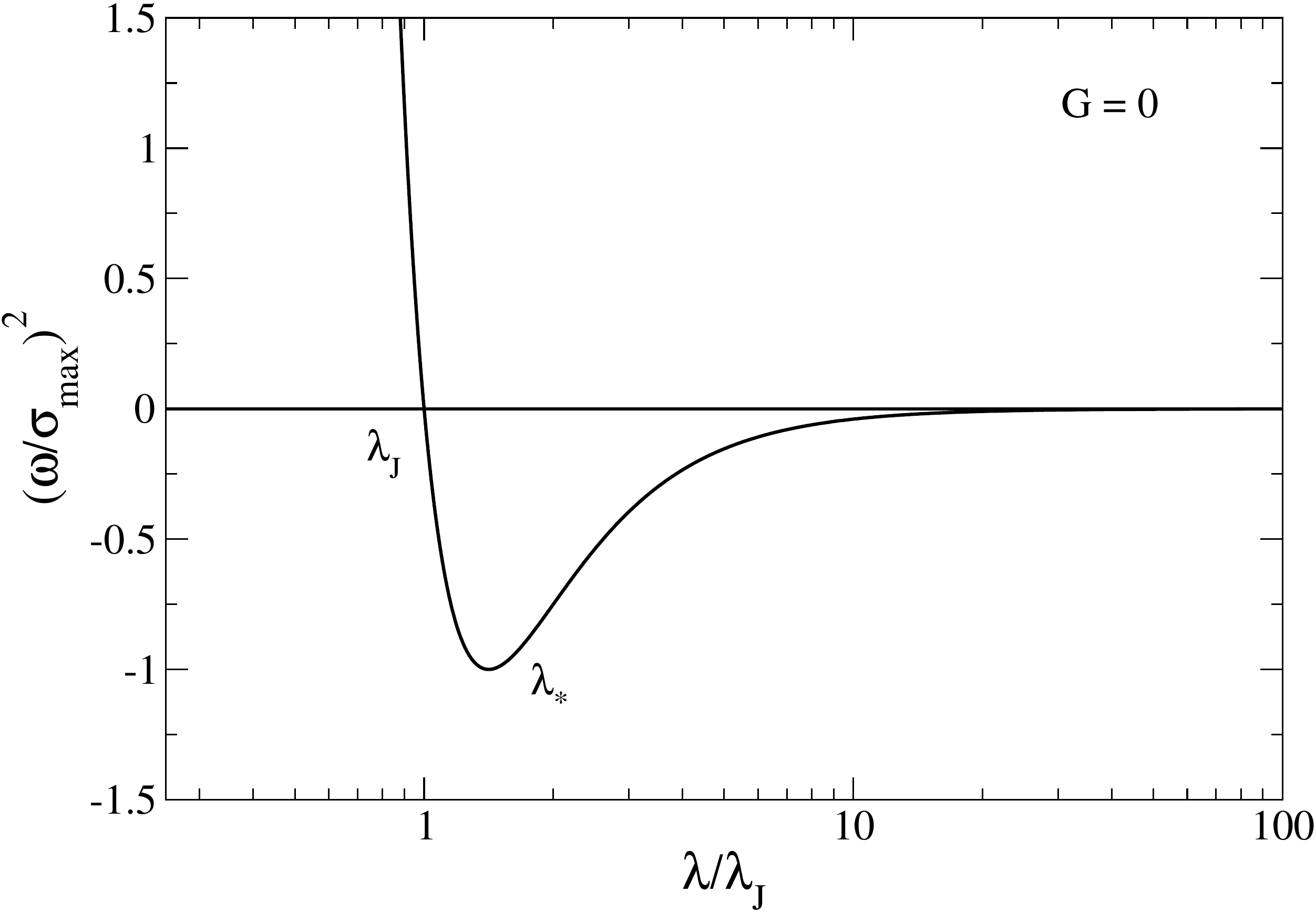}} 
\caption{Growth rate $\sigma=(-\omega^2)^{1/2}$ of the
perturbations for nonrelativistic self-attracting bosons ($a_s<0$) in the
nongravitational limit ($G=0$) as a function of the wavelength $\lambda$ (see
Eq. (\ref{nog2})). The growth rate is normalized by the
maximum growth rate $\sigma_{\rm max}$ and the wavelength by the Jeans length
$\lambda_J$. The instability is relatively peaked about the most unstable
wavelength $\lambda_*=\sqrt{2}\lambda_J$. Therefore, structures  can form  with
a typical size $\sim\lambda_*$.}
\label{Gzero}
\end{figure}

We consider ULAs with mass $m=2.19\times
10^{-22}\, {\rm eV}/c^2$ and negative scattering length $a_s=-1.11\times
10^{-62}\, {\rm fm}$ corresponding to a ratio $|a_s|/m^3=1.06\times
10^{3}\, {\rm fm}/({\rm eV/c^2})^3$. As shown in Appendix D of
\cite{suarezchavanis3} (see
also \cite{bectcoll}), these ULAs can form giant BECs whose maximum
mass and
corresponding radius are of the order of the mass and radius of dSphs like
Fornax ($R\sim 1\, {\rm kpc}$, $M\sim 10^8\,
M_{\odot}$, $\overline{\rho}\sim 10^{-18} \, {\rm
g/m^3}$). At the epoch of radiation-matter equality, the value of the
dimensionless self-interaction parameter defined by Eq. (\ref{dv4}) is
$\alpha=31.2$ so that we are in the nongravitational limit (the
transition between the nongravitational limit and the
noninteracting limit takes place at $\rho_a=9.02\times 10^{-17} \, {\rm
g}/{\rm m}^3 \ll\rho_{\rm eq}=8.77\times 10^{-14}\, {\rm
g}/{\rm m}^3$).\footnote{This approximation is
valid in the linear regime of structure formation when considering the growth
of perturbations in a homogeneous Universe (Jeans problem). In the nonlinear
regime of structure formation (equilibrium states $=$ DM halos) self-gravity
must be taken into
account and leads to axionic clusters, being possibly the cores of large DM
halos, with a maximum mass $M_{\rm
max}=10^8\, M_{\odot}$ and a minimum radius $R_*=1\, {\rm kpc}$ \cite{phi6}.}
On the other hand, we find
that the
dimensionless relativistic parameter defined by Eq. (\ref{dv10}) is
$\nu=2.54\times 10^{-5}$. The smallness of this value shows that we are in the 
nonrelativistic
limit (the transition between the
ultrarelativistic limit and the nonrelativistic limit takes place at
$\rho_i=1.15\times 10^{-9} \, {\rm
g}/{\rm m}^3 \gg\rho_{\rm eq}=8.77\times 10^{-14}\, {\rm
g}/{\rm m}^3$). Using Eqs. (\ref{ng1}) and (\ref{ng2}), we obtain
$\lambda_J=18.2\, {\rm pc}$ and $M_J=4.08\times 10^{6}\,
M_{\odot}$.  The most unstable wavelength and the  maximum growth rate  (see
Sec. \ref{sec_nog}) are $\lambda_*=25.7\, {\rm
pc}$ and $\sigma_{\rm
max}=8.46\times 10^{-12}\, {\rm s}^{-1}$.  In
comparison $\lambda_i=2.92\times 10^{-2}\, {\rm pc}$, $M_i=1.59\times 10^{4}\,
M_{\odot}$, $\lambda_a=181\, {\rm pc}$, $M_a=9.86\times 10^{7}\,
M_{\odot}$, $\lambda_H=4.40\times 10^4\, {\rm pc}$, and $M_H=4.59\times
10^{17}\,
M_{\odot}$. These results are illustrated in Fig. \ref{Gzero}. We also note that
the comoving Jeans
length at the epoch of
radiation-matter equality is $\lambda_J^c=61.7\, {\rm kpc}$.
We note that the
Jeans length $\lambda_J$ and the Jeans mass $M_J$ at the begining of the
structure formation process 
(radiation-matter equality epoch) are two orders of magnitude smaller than the
current radius
$R$ and mass $M$ of dwarf DM halos like Fornax. This suggests
that the system gains
mass and increases in size
during the nonlinear process of halo
formation. This could be achieved through hierarchical
clustering.

{\it Remark:} At the present epoch, we find that $\alpha=1.58\times 10^{-4}$
(corresponding to $\rho_0=2.25\times
10^{-24}\, {\rm g/m^3}\ll\rho_a=9.02\times 10^{-17} \, {\rm
g}/{\rm m}^3$) and $\nu=6.53\times 10^{-16}$ (corresponding to
$\rho_i=1.15\times 10^{-9} \, {\rm
g}/{\rm m}^3 \gg\rho_{0}=2.25\times 10^{-24}\, {\rm
g}/{\rm m}^3$), so that we are in the noninteracting and nonrelativistic limit
of Sec. \ref{sec_appq}.

\section{Jeans type instability of fermionic dark matter}
\label{sec_fermi}

In the preceding sections, we have considered the Jeans instability of a
completely condensed self-gravitating relativistic boson gas at $T=0$. In this
section, to make a comparison, we consider the same problem for a completely
degenerate self-gravitating relativistic fermion gas at $T=0$. In the case of
BECs, gravitational collapse is prevented by the Heisenberg uncertainty
principle or by the self-repulsion of the bosons (when $a_s>0$).
In the case of fermions, gravitational collapse is prevented by the Pauli
exclusion principle (like in the case of white dwarfs and neutron stars).

\subsection{The ultrarelativistic limit}
\label{sec_fermiur}

In the ultrarelativistic limit, a gas of fermions with spin $s=1/2$ has an
equation of state \cite{chandrabook}:
\begin{eqnarray}
P=\frac{1}{8}\left (\frac{3}{\pi}\right )^{1/3}\frac{hc}{m^{4/3}}\rho^{4/3},
\label{fermi11}
\end{eqnarray} 
where $\rho=nm$ is the rest-mass density. This corresponds to a polytrope
of index $n=3$.  Expressed
in terms of the energy density \cite{chandrabook}:
\begin{eqnarray}
\epsilon=\frac{3}{8}\left (\frac{3}{\pi}\right
)^{1/3}\frac{hc}{m^{4/3}}\rho^{4/3},
\label{fermi11b}
\end{eqnarray} 
we obtain
\begin{eqnarray}
P=\frac{1}{3}\epsilon.
\label{fermi12}
\end{eqnarray} 
Since the equation of state of ultrarelativistic fermions is the
same as radiation our results will be identical to those obtained in Sec.
\ref{sec_urtf} for ultrarelativistic bosons.

Using the results of Appendix \ref{sec_paddy}, we find that the Jeans
wavenumber and the Jeans length are given by
\begin{eqnarray}
k_J^2=\frac{16\pi G \epsilon}{c^4},\qquad \lambda_J=2\pi\left (\frac{c^4}{16\pi
G \epsilon}\right )^{1/2}.
\label{fermi13}
\end{eqnarray}
Expressing the energy density in terms of the rest-mass density
from Eq. (\ref{fermi11b}), we obtain 
\begin{eqnarray}
\lambda_J=\frac{1}{\sqrt{3}}\left (\frac{\pi}{3}\right )^{1/6}\left
(\frac{c^3}{G\hbar}\right )^{1/2}\left (\frac{m}{\rho}\right )^{2/3}.
\label{fermi13r}
\end{eqnarray}
Using Eq. (\ref{fermi13}), the Jeans mass (\ref{mj}) is
given by 
\begin{eqnarray}
M_J=\frac{\pi^{5/2}}{48}\frac{c^4}{G^{3/2}\epsilon^{1/2}}.
\label{fermi14}
\end{eqnarray}
It can also be written as
\begin{eqnarray}
M_J=\frac{\pi^2}{24\sqrt{3}}\left (\frac{\pi}{3}\right
)^{1/6}\left (\frac{c^7}{\hbar G^{3}}\right )^{1/2}\left (\frac{m}{\rho}\right
)^{2/3}.
\label{fermi14r}
\end{eqnarray}
Eliminating the energy density between Eqs. (\ref{fermi13}) and (\ref{fermi14}),
we
get
\begin{eqnarray}
M_J=\frac{\pi^2}{24}\frac{c^2\lambda_J}{G}.
\label{fermi15}
\end{eqnarray}
This relation is similar to the mass-radius
relation of a general relativistic star with a linear equation of
state $P=\epsilon/3$  confined within a box \cite{aaq}. In
the present case, this agreement can be explained by the fact that the
equation of state of an ultrarelativistic Fermi gas is $P=\epsilon/3$ (like
radiation). Comparing
Eq. (\ref{urni5})  with $q=1/3$ and Eq. (\ref{fermi15}), we obtain
\begin{eqnarray}
\frac{M_J}{\lambda_J/2}=3.84\frac{M}{R}.
\label{fermi16}
\end{eqnarray}

Introducing the scales $\lambda_F$, $M_F$, $\rho_F$ (see Appendix
\ref{sec_chas}) and the relativistic parameter
$\mu=(\rho/\rho_F)^{2/3}$ appropriate to fermions (see
Appendix \ref{sec_fermita}), we obtain
\begin{equation}
\frac{\lambda_J}{\lambda_F}=\frac{1}{\sqrt{3}}\left
(\frac{\pi}{3}\right
)^{1/6}\frac{1}{\mu},\qquad 
\frac{M_J}{M_F}=\frac{\pi^2}{24\sqrt{3}}\left (\frac{\pi}{3}\right
)^{1/6}\frac{1}{\mu},
\label{fermi18}
\end{equation}
\begin{eqnarray}
\frac{M_J}{M_F}=\frac{\pi^2}{24}\frac{\lambda_J}{\lambda_F}.
\label{fermi18b}
\end{eqnarray}
Since $\mu\gg 1$ (i.e. $\rho\gg\rho_F$) in the ultrarelativistic limit, we find
that 
$\lambda_J\ll\lambda_F$ and $M_J \ll M_F$.

Introducing the Hubble scales $\lambda_H$ and $M_H$ (see Appendix \ref{sec_hs}),
we obtain
\begin{eqnarray}
\frac{\lambda_J}{\lambda_H}=\frac{2\pi}{\sqrt{6}},\qquad 
\frac{M_J}{M_H}= \frac{1}{8}\left (\frac{2}{3}\right
)^{3/2}\pi^3.
\label{fermi17}
\end{eqnarray}
We note that $\lambda_J\sim \lambda_H$ and $M_J\sim
M_H$. Since the Jeans length is of the
order of the Hubble length (horizon), there is no Jeans instability.
Therefore, large-scale structures cannot form in the
ultrarelativistic
limit.

{\it Remark:} The previous results are based on a general relativistic
treatment. If we use a Newtonian treatment, we find from Eqs. (\ref{intro2})
and (\ref{fermi11}) that the  Jeans length is given by
\begin{eqnarray}
\lambda_J=\sqrt{\frac{\pi}{6}}\left (\frac{3}{\pi}\right )^{1/6}\left
(\frac{hc}{G}\right )^{1/2}\frac{1}{m^{2/3}\rho^{1/3}}.
\label{mac1}
\end{eqnarray}
The corresponding Jeans mass, defined by Eq. (\ref{app1a}), is
\begin{eqnarray}
M_J=\frac{\pi^2}{6\sqrt{72}}\left (\frac{hc}{G}\right )^{3/2}\frac{1}{m^2}.
\label{mac2}
\end{eqnarray}
This relation  is similar to the Chandrasekhar mass of Newtonian
ultrarelativistic self-gravitating fermions\footnote{This is the mass of a
polytrope of index $n=3$ corresponding to the equation of state
(\ref{fermi11}) \cite{chandrabook}.}
\begin{eqnarray}
M=0.376\left (\frac{hc}{G}\right )^{3/2}\frac{1}{m^2}.
\label{mac3}
\end{eqnarray}
Comparing Eqs.
(\ref{mac2}) and (\ref{mac3}), we find that
\begin{eqnarray}
M_{J}=0.516 M.
\label{mac4}
\end{eqnarray}
Introducing the scales $\lambda_F$, $M_F$, $\rho_F$ (see Appendix
\ref{sec_chas}) and the relativistic parameter $\mu=(\rho/\rho_F)^{2/3}$ 
appropriate to fermions (see Appendix \ref{sec_fermitb}), we get
\begin{equation}
\frac{\lambda_J}{\lambda_F}=\frac{\pi^{5/6}}{3^{1/3}\sqrt{\mu}},\qquad 
\frac{M_J}{M_F}=\frac{\pi^{7/2}}{18}.
\label{mac5}
\end{equation}
Since $\mu\gg 1$ (i.e. $\rho\gg\rho_F$) in the ultrarelativistic limit, we find
that 
$\lambda_J\ll\lambda_F$ and $M_J \sim M_F$. We shall not discuss the Hubble
length here because the Newtonian treatment is not valid in the
ultrarelativistic limit.

\subsection{The nonrelativistic limit}
\label{sec_ferminr}

In the nonrelativistic limit, a gas of fermions with spin $s=1/2$ has an
equation of state \cite{chandrabook}:
\begin{eqnarray}
P=\frac{1}{20}\left (\frac{3}{\pi}\right )^{2/3}\frac{h^2}{m^{8/3}}\rho^{5/3},
\label{fermi1}
\end{eqnarray} 
where $\rho=nm$ is the rest-mass density. This corresponds to a polytrope
of index $n=3/2$. Using Eq. (\ref{intro2}), the Jeans
length is given by  
\begin{eqnarray}
\lambda_J=\frac{1}{2}\left (\frac{\pi}{3}\right
)^{1/6}\frac{h}{G^{1/2}m^{4/3}\rho^{1/6}}.
\label{fermi2}
\end{eqnarray} 
The associated Jeans mass, defined by Eq. (\ref{app1a}), is
\begin{eqnarray}
M_J=\frac{1}{16}\left (\frac{\pi}{3}\right
)^{3/2}\frac{h^3\rho^{1/2}}{G^{3/2}m^{4}}.
\label{fermi3}
\end{eqnarray} 
Introducing relevant scales, we get
\begin{eqnarray}
\frac{\lambda_J}{\rm pc}=1.94\times 10^{3}\, \left (\frac{{\rm
eV}/c^2}{m}\right )^{4/3}\left (\frac{\rm g/m^3}{\rho}\right )^{1/6},
\label{fermi4}
\end{eqnarray}
\begin{eqnarray}
\frac{M_J}{M_{\odot}}= 5.63\times 10^{25}\, \left (\frac{{\rm
eV}/c^2}{m}\right )^{4}\left (\frac{\rho}{\rm g/m^3}\right )^{1/2}.
\label{fermi5}
\end{eqnarray}
In the matter era, using Eq. (\ref{appm2}), we find that the Jeans length
increases as
$a^{1/2}$ and the
Jeans mass
decreases as $a^{-3/2}$.  Eliminating the density between Eqs.
(\ref{fermi2}) and (\ref{fermi3}), we obtain
\begin{eqnarray}
M_{J}\lambda_J^3=\frac{\pi^2}{1152}\frac{h^6}{G^3m^8}.
\label{fermi6}
\end{eqnarray}
This relation  is similar to the mass-radius
relation of nonrelativistic self-gravitating fermions\footnote{This is the
mass-radius
relation of a polytrope of index $n=3/2$ corresponding to the equation of state
(\ref{fermi1}) \cite{chandrabook}.}
\begin{eqnarray}
MR^3=1.49\times 10^{-3}\frac{h^6}{G^3m^8}.
\label{fermi7}
\end{eqnarray}
Comparing Eqs.
(\ref{fermi6}) and (\ref{fermi7}), we find that 
\begin{eqnarray}
M_{J}\left (\frac{\lambda_J}{2}\right )^3=0.719 MR^3.
\label{fermi8}
\end{eqnarray}

Introducing the scales $\lambda_F$, $M_F$, $\rho_F$ (see Appendix
\ref{sec_chas}) and the relativistic parameter $\mu=(\rho/\rho_F)^{2/3}$ 
appropriate to fermions (see Appendix \ref{sec_fermitb}), we get
\begin{equation}
\frac{\lambda_J}{\lambda_F}=\pi\left (\frac{\pi}{3}\right
)^{1/6}\frac{1}{\mu^{1/4}},\qquad 
\frac{M_J}{M_F}=\frac{\pi^3}{2}\left (\frac{\pi}{3}\right
)^{3/2}\mu^{3/4},
\label{fermi10}
\end{equation}
\begin{equation}
\frac{M_J}{M_F}=\frac{\pi^8}{18} \left (\frac{\lambda_F}{\lambda_J}\right )^3.
\label{fermi10sam}
\end{equation}
Since $\mu\ll 1$ (i.e. $\rho\ll\rho_F$) in the nonrelativistic limit, we find
that 
$\lambda_J\gg\lambda_F$ and $M_J \ll M_F$.  We note that the 
relativistic parameter $\mu$ can be expressed in terms of the Hubble constant as
$\mu=(3/8\pi)^{2/3}(H^2\hbar^3/Gm^4c^3)^{2/3}$.

Introducing the Hubble scales $\lambda_H$ and $M_H$ (see
Appendix \ref{sec_hs}), we obtain 
\begin{equation}
\frac{\lambda_J}{\lambda_H}=2\sqrt{2}\pi\left
(\frac{\pi}{3}\right
)^{2/3} \mu^{1/2},\qquad 
\frac{M_J}{M_H}=\frac{2\sqrt{2}}{9}\pi^5\mu^{3/2}.
\label{fermi9}
\end{equation}
Since $\mu\ll 1$ in the nonrelativistic limit,
we find that
$\lambda_J\ll\lambda_H$ and $M_J \ll M_H$. Therefore, large-scale structures can
form in
the nonrelativistic regime by Jeans instability since the Jeans length is
much smaller than the horizon.

Let us consider a fermionic particle, like a  sterile neutrino,
able to form giant fermion balls
with the mass and size of DM halos. To determine its mass $m$, we assume
that the most compact DM halos observed in the Universe, namely dwarf
spheroidals (dSphs) like Fornax ($R\sim 1\, {\rm kpc}$, $M\sim 10^8\,
M_{\odot}$, $\overline{\rho}\sim 10^{-18} \, {\rm
g/m^3}$), are completely degenerate, corresponding to the ground state of
the self-gravitating Fermi gas.\footnote{As explained in more
detail in Refs.
\cite{kingclassique,kingfermionic,epjplusarejouter} and in Appendix D of Ref.
\cite{suarezchavanis3}, large fermionic
halos have a core-halo structure with a fermion ball surrounded by a NFW
atmosphere. The fermion ball corresponds to the ground state of the
self-gravitating Fermi gas at $T=0$ and the NFW atmosphere may be the result of
a violent
relaxation \cite{lbvr,kingclassique,kingfermionic,epjplusarejouter} or
hierarchical
clustering. The precise  structure of the atmosphere may be influenced
by incomplete violent relaxation,
tidal effects, stochastic forcing etc.} 
The mass-radius relation (\ref{fermi7}) then gives a
fermion mass $m=170\, {\rm eV}/c^2$ (see Appendix D of
\cite{suarezchavanis3}). We note that, inversely, the knowledge of the DM
particle mass $m$ does not determine $M$ and $R$ individually, but only their
product $MR^3$. The individual determination of $M$ and $R$ depends on the epoch
(time, scale factor, or redshift) and can be obtained from the Jeans study. Let
us  apply this study at the epoch of
radiation-matter equality. For a fermion mass 
$m=170\, {\rm
eV}/c^2$, we find that the dimensionless relativistic parameter  defined by Eq.
(\ref{fermi20b})  is $\mu=5.89\times 10^{-7}$. The smallness of this value shows
that we are in the nonrelativistic limit (the transition between the
ultrarelativistic limit and the nonrelativistic limit takes place at
$\rho_F=1.94\times 10^{-4} \, {\rm
g}/{\rm m}^3 \gg\rho_{\rm eq}=8.77\times 10^{-14}\, {\rm
g}/{\rm m}^3$). 
To evaluate the  Jeans length and the Jeans mass  at the
epoch of radiation-matter equality we use Eqs.  (\ref{fermi2}) and
(\ref{fermi3})
 and obtain $\lambda_J=309\, {\rm pc}$ and $M_J=2.00\times
10^{10}\, M_{\odot}$.
In comparison,
$\lambda_H=4.40\times 10^4\, {\rm pc}$, $M_H=4.59\times 10^{17}\,
M_{\odot}$, $\lambda_F=2.70\, {\rm pc}$ and
$M_F=5.64\times
10^{13}\, M_{\odot}$. The
relativistic corrections are negligible since $\mu\ll
1$.\footnote{In this section, we have used Newtonian gravity. If we want to
describe the stabilization of the system at the Hubble scale (similarly to the
case of bosons), we must extend our results to the context of general
relativity.} We
also note that the comoving Jeans length is, at the epoch of
radiation-matter equality, equal to $\lambda_J^c=1.05\, {\rm Mpc}$. 
The
dicussion is essentially
the same as for bosons (see Secs. \ref{sec_appqb} and
\ref{sec_sib}). Indeed, at a qualitative level, fermions and bosons
behave relatively similarly, as noted in \cite{kingclassique,kingfermionic}. The
main differences are quantitative: (i) the mass of the fermionic DM particle is
much larger than its bosonic counterpart;
(ii) the solitonic/BEC core in bosonic DM halos is replaced by a fermion ball in
fermionic DM halos.

{\it Remark:} If we compute the Jeans length and the Jeans mass at
the present epoch, we find $\mu=5.13\times 10^{-14}$, $\lambda_J=18.0\, {\rm
kpc}$, $M_J=1.01\times 10^{5}\, M_{\odot}$. In
comparison $\lambda_H=8.69\times 10^{6}\, {\rm
kpc}$,
$M_H=9.06\times 10^{22}\, M_{\odot}$, $\lambda_F=2.70\, {\rm pc}$ and
$M_F=5.64\times
10^{13}\, M_{\odot}$.

\section{Conclusion}

The main idea when considering the formation of the
large-scale structures of the
Universe is that of gravitational instability. Usually, it is assumed that there
exist small primordial perturbations which gradually increase in amplitude to
form the structures that are being observed at present at the scale of galaxies
and galaxy
clusters. An overdense region is expected to attract material from its
surroundings and become even denser. The denser they become the more they
accrete, ending in an instability which can finally cause the collapse of a
fluctuation to a gravitationally bound object. The knowledge of the Jeans
length $\lambda_J$ gives an estimate of the minimum size of the objects
that can form by gravitational instability.

In this paper, we have studied in detail the Jeans instability of a complex
self-interacting SF in general relativity. This study is rather academic since
the expansion of the Universe is neglected but it remains fundamentally
interesting and important. In particular, it allows us to isolate
characteristic length, mass and density scales that play a
crucial role in any treatment of perturbations and structure formation in
cosmology. This
study was initiated in the seminal work of Khlopov {\it et al.} \cite{khlopov}
but our study goes beyond certain approximations made by these authors and
completes their work. 

Our approach is rather original with respect to other works on SFs. Indeed, 
instead of solving the field equations as usually done, we have used a
hydrodynamic
representation of the KGE equations introduced in our
previous work \cite{suarezchavanis1}. We stress that for
a complex SF this hydrodynamic representation is exact in the sense that the
hydrodynamic equations are equivalent to the KGE equations.\footnote{It is
important to remark that there is no approximation in our study of the
quantum relativistic Jeans problem. In Ref. \cite{suarezchavanis1} we have
written the hydrodynamic equations in a weak field approximation but since we
consider the linear instability of the SF, this approximation is fully
justified at the linear order and is not a limitation of our study. The
resulting linearized
equations are therefore exact. By contrast, our hydrodynamic approach cannot be
extended to the case of a real SF without making approximations
(see Sec. II of \cite{phi6} for details) so that other
approaches such as
those developed by other authors (see footnote 8) are more relevant in that
case. Our hydrodynamic approach is, however, valid for a real SF
in the nonrelativistic regime.} This hydrodynamic approach allows us to treat
the problem of structure
formation on the same footing as the original Jeans
 study for a classical
self-gravitating collisional gas \cite{jeans1902}. Our approach is, however,
much more general
since it includes quantum mechanics ($\hbar$) and relativity ($c$) in addition
to self-gravity ($G$) and pressure due to the self-interaction ($a_s$).

In this paper, we have studied the general dispersion relation $\omega(k)$ of a
relativistic and quantum fluid obtained in \cite{suarezchavanis1}
and
deduced from it the Jeans length $\lambda_J$, the most unstable wavelength
$\lambda_*$, and the maximum growth rate $\sigma_{\rm max}$. 
We have determined how the Jeans length and the Jeans
mass vary as a function of the density of the Universe (the density of the
Universe, as well as the scale factor or the redshift, can be considered
as a measure of time in the cosmic history of the SF). This
allowed
us to analyze the cosmological evolution of the Jeans scales.
We have stressed the analogy, previously noticed
in \cite{prd1}, between the Jeans mass-radius relation
$M_J(\lambda_J)$ in the linear regime of structure formation  and the
mass-radius relation $M(R)$ of boson stars and dark matter halos in the
nonlinear regime of structure formation (this analogy
will be further developed in a future work \cite{prep}). We have
considered different limits (ultrarelativistic, nonrelativistic,
noninteracting, TF, nongravitational) and
we have given precise conditions of validity of these different limits in
terms of dimensionless parameters depending on the characteristics of the SF
(mass and scattering length) and on the density of the Universe. We have given
the numerical values of
the Jeans mass and Jeans length at the epoch of radiation-matter equality
(and at the current epoch for comparison) for different types of bosonic
particles (QCD axions, ultralight axions, self-interacting bosons...). Our study
therefore refines and completes previous works on the subject.

We have shown that structure formation is impossible in the early Universe
(stiff matter era and radiationlike era) because the Jeans length is always
larger than the Hubble length (horizon). This corresponds to
the ultrarelativistic limit of our formalism. There is, however, an exception
when the SF has an attractive self-interaction. In that case, the instability
is caused by the self-interaction, not by the self-gravity which is usually
negligible. If the boson has a mass
$m=10^{-4}\, {\rm eV}/c^2$ and a negative scattering length $a_s=-5.8\times
10^{-53}\, {\rm m}$ (similarly to the QCD axion), one can form
small objects with a size $\lambda_J= 1.07\, {\rm cm}$ and a mass
$M_J=3.41\times 10^{-21}M_{\odot}$. These
objects, corresponding to a complex SF
with an attractive self-interaction, are the
counterparts of the axitons \cite{hr,kt,kt2} with size $R_{\rm
axiton}\sim 10^{9}\, {\rm m}$  and mass $M_{\rm
axiton}\sim 10^{-12}\, M_{\odot}$ corresponding
to a real SF with an attractive self-interaction (QCD axion).
On the other hand,  if the boson
has an ultrasmall  mass
$m=2.19\times 10^{-22}\, {\rm eV}/c^2$ and a negative scattering length
$a_s=-1.11\times 10^{-62}\, {\rm fm}$, one can form big objects 
(protogalaxies) with a size $\lambda_J=0.159\, {\rm pc}$ and a mass
$M_J=1.78\times 10^{4}\, M_{\odot}$. These objects could be the
germs in the process of galaxy formation. Since these germs appear in the
ultrarelativistic regime, the resulting galaxies would be much older than what
is usually believed, possibly in agreement with certain recent cosmological
observations where large-scale structures are observed at high redshifts
\cite{nature}.

We have shown that structure formation can take
place in the matter era due to gravitational attraction. This corresponds to the
nonrelativistic limit of our
formalism. We have obtained the following results.\footnote{We remain
voluntarily qualitative here because
the exact values of the Jeans scales depend on the epoch considered and
they determine only the minimum size of the fluctuations that can experience
gravitational
instability in the linear regime, not the scales of
the structures that are formed in the nonlinear regime.
Therefore, we just give orders of magnitude to show how the Jeans
scales depend on the type of bosons. More precise numerical applications are
made in the main text.} For QCD axions with a mass $m=10^{-4}\,
{\rm eV}/c^2$ and a scattering length $a_s=-5.8\times 10^{-53}\, {\rm m}$ we
have shown that the self-interaction can be neglected in the matter era. The
Jeans length is of the order of the Solar
System size and the Jeans mass is of the order of the asteroid mass. Therefore,
at the galactic level, QCD axions
behave essentially as CDM and cannot solve the CDM crisis. They
can form mini-MACHOSs (mini axion stars) that could be the constituents of DM
halos, but they cannot form DM halos themselves. By contrast, for
noninteracting ULAs with $m=2.92\times
10^{-22}\, {\rm eV/c^2}$, for self-repulsive bosons with a
ratio $a_s/m^3=3.28\times
10^{3}\, {\rm fm}/({\rm eV/c^2})^3$ that are in the TF regime, and for ULAs with
a mass $m=2.19\times
10^{-22}\, {\rm eV/c^2}$ and a negative scattering length $a_s\sim
-1.11\times 10^{-62}\, {\rm fm}$ (corresponding to a ratio $|a_s|/m^3=1.06\times
10^{3}\, {\rm fm}/({\rm eV/c^2})^3$) that are in the nongravitational regime we
find that
the Jeans length and the Jeans mass are
of the order of the galactic size. Therefore, they are good candidates to
form DM halos. The study of the Jeans
instability is not sufficient to
discriminate these different particles but other considerations
\cite{shapiro,suarezchavanis3} suggest that
axions must be self-interacting.\footnote{Concerning the
linear regime of structure formation (Jeans problem), we have
shown that, in the matter era, the self-interaction of the QCD axion is
negligible while the self-interaction of ULAs is usually important and must be
taken into account. In the nonlinear regime of structure formation (DM halos)
the self-interaction of the QCD axion and of ULAs is always important even if
it looks extremely small at first sight (see Appendix L of \cite{phi6}). The
importance of the self-interaction of the bosons has not always been fully
appreciated since many works \cite{ch2,ch3,hui,marshrevue} neglect it from
the start.}

We have shown that, in the matter era, general relativistic
effects become important at very large scales, of the order of the Hubble
length $\lambda_H$, and tend to stabilize the system. Therefore,
structure formation can take place for $\lambda_{J}\le \lambda\le \lambda_H$.
Below the Jeans length there is no instability because gravity is too weak and
above the Hubble length the
growth rate decreases substantially because of general relativity. In between,
the growth rate is almost
constant, leading to a plateau $[10\lambda_{J},\lambda_H]$ where $\sigma\simeq
\sigma_{\rm max}$. This
implies that the optimal wavelength $\lambda_*$ is finite. By contrast, in the
nonrelativistic model, the plateau  $[10\lambda_{J},+\infty[$ extends to
infinity and the  optimal wavelength $\lambda_*\rightarrow +\infty$. Therefore,
relativistic
effects can solve some problems of the classical Jeans study, providing
a maximum scale (of the order of the horizon) above which there is no
gravitational  instability anymore. Although these
results have already been obtained in other contexts, it is interesting to
recover them in a rigorous manner for a complex SF in a static
background. In particular, our study extends the previous study of Khlopov {\it
et al.} \cite{khlopov} where these effects were neglected.

The SF model may have profound cosmological implications. DM is
usually described by hydrodynamical equations without the quantum potential. In
the context of CDM models with vanishing temperature and vanishing pressure, the
usual
Jeans analysis predicts that all scales are unstable. Consequently,  the Jeans
scale $\lambda_J$ is zero. This is the intrinsic reason why the CDM model
generates
cuspy dark matter halo profiles and an abundance of low mass halos. However,
these cusps and satellites are not observed
\cite{burkert,firmani,kauffmann}. These problems (small-scale crisis of
the CDM model)
can be solved if the DM in the Universe is made up of a SF. In that
case, the wave properties of the dark matter can stabilize gravitational
collapse, providing halo cores and suppressing small-scale structures. Indeed,
if DM is a SF, there exists a nonzero Jeans length. Stability below
the Jeans scale is guaranteed by the Heisenberg uncertainty principle or by
the pressure arising from the repulsive interaction of the bosons. This
non-thermal quantum pressure stabilizes the system against gravitational
collapse. For wavelengths smaller than the Jeans length, the evolution cannot
bring the small
perturbations in the early Universe to the nonlinear regime, and the
inhomogeneities are erased (they remain oscillating modes). These modes are
expected to induce a cutoff in the mass power spectrum for the distribution of
galaxies in the Universe. For wavelengths larger than the Jeans length, the SF
follows the evolution of the standard CDM scenario. Therefore, the SF and the
CDM model differ at small scales while they are indistinguishable
at large scales. Cosmological simulations are required to determine the
viability of the SF/BECDM model.

The study in this work was motivated by the proposal that DM could be made out
of SFs. Still, the nature of DM remains unknown. There exist other theories
according to which DM could be made of massive neutrinos (see,
e.g., \cite{vega,ruffini,kingclassique,kingfermionic} and references therein).
In these theories, gravitational collapse
is prevented by the Pauli exclusion principle for fermions. We
have made a comparison between the Jeans problem for fermions and bosons.
Qualitatively, the two types of particles behave similarly and are able to
account for the observations. The main differences between fermionic and bosonic
DM are quantitative. In particular, the mass of the fermionic DM particle
($m=170\, {\rm eV}/c^2$) is much larger than the mass of the bosonic
DM particle ($m=2.92\times 10^{-22}\, {\rm eV}/c^2<m<1.10\times 10^{-3}\, {\rm
eV/c^2}$) because the quantum pressure is due to
the Pauli exclusion principle instead of the Heisenberg uncertainty
principle or the repulsive scattering of the bosons (see Appendix D of
\cite{suarezchavanis3}). On
the other hand, the gravitational collapse of self-gravitating fermions leads to
DM halos with a core-halo structure made of a ``fermion ball'' surrounded by an
isothermal (or NFW) atmosphere \cite{vega,ruffini,kingclassique,kingfermionic}
while the
gravitational collapse of self-gravitating BECs leads to DM halos with
a core-halo structure made of a solitonic/BEC core surrounded by
an isothermal
(or NFW) atmosphere \cite{ch2,ch3,epjplusarejouter}. Since it does not seem
possible at
the present stage to make a clear
distinction (from observations) between a fermionic core and a bosonic core, it
is not easy to
determine  whether DM is made of bosons or fermions.

\acknowledgments{A. S. acknowledges CONACyT for the postdoctoral grant
received (No. 231276).}

\appendix

\section{Dimensionless variables}
\label{sec_dv}

In order to simplify the calculations and make the figures, it is convenient to
introduce dimensionless variables. Different choices are possible depending on
the parameters that we fix to construct the reference scales. We present below
the two normalizations used in this paper, but other normalizations
are possible.

\subsection{First normalization: $G$ and $\hbar$ fixed}
\label{sec_dv1}

We introduce the reference pulsation and the reference wavenumber 
\begin{eqnarray}
\omega_0=\sqrt{4\pi G\rho},\qquad k_0=\left (\frac{16\pi G\rho
m^2}{\hbar^2}\right )^{1/4}.
\label{dv1}
\end{eqnarray}
The pulsation $\omega_0$ coincides with the inverse dynamical time and
the 
wavenumber $k_0$ coincides with the quantum Jeans wavenumber (\ref{nr7}) in the
nonrelativistic limit. We
define the dimensionless variables
\begin{eqnarray}
\Omega=\frac{\omega}{\omega_0},\qquad
\kappa=\frac{k}{k_0},
\label{dv2}
\end{eqnarray}
\begin{eqnarray}
\alpha=\left (\frac{m^2c_s^4}{4\pi G\rho\hbar^2}\right
)^{1/2},\qquad \chi=\left (\frac{4\pi G\rho\hbar^2}{m^2c^4}\right )^{1/2}.
\label{dv3}
\end{eqnarray}
We note that $\alpha\chi=c_s^2/c^2$. We also note that
$\chi=\omega_0/\omega_C$, where $\omega_C=c/\lambda_C=mc^2/\hbar$ is the
Compton pulsation ($\lambda_C=\hbar/mc$ is the Compton wavelength). In this
normalization, $G$ and $\hbar$ are
used to construct the reference scales $\omega_0$ and $k_0$, so they are
assumed to be ``fixed''. Then, the parameter
$\alpha\propto c_s^2$ measures the speed of sound (self-interaction) and the
parameter $\chi\propto
1/c^2$ measures the importance of relativistic effects. The noninteracting limit
($c_s\rightarrow 0$)  corresponds to $\alpha\rightarrow 0$ and
the nonrelativistic limit ($c\rightarrow +\infty$) corresponds to
$\chi\rightarrow 0$. 
 For a quartic SF potential of the form of Eq. (\ref{jia1}), using the
expression of the speed of sound given by Eq. (\ref{jia9}), we can write the
parameter $\alpha$ as
\begin{eqnarray}
\alpha=\left (\frac{4\pi\rho\hbar^2a_s^2}{Gm^4}\right
)^{1/2}.
\label{dv4}
\end{eqnarray}

{\it Remark:} Using this normalization amounts to taking $4\pi
G=\rho=2m=\hbar=1$, $\alpha=c_s^2/2$ and $\chi=2/c^2$ in the original equations.

\subsection{Second normalization: $G$ and $c_s$ fixed}
\label{sec_dv2}

We introduce the reference pulsation and the reference wavenumber
\begin{eqnarray}
\omega_0=\sqrt{4\pi G\rho},\qquad k_0=\left (\frac{4\pi G\rho}{c_s^2}\right
)^{1/2}.
\label{dv7}
\end{eqnarray}
The pulsation $\omega_0$ coincides with the inverse dynamical time and
the 
wavenumber $k_0$ coincides with the classical Jeans wavenumber (\ref{nr10}) in
the nonrelativistic limit. We
define the dimensionless variables
\begin{eqnarray}
\Omega=\frac{\omega}{\omega_0},\qquad
\kappa=\frac{k}{k_0},
\label{dv8}
\end{eqnarray}
\begin{eqnarray}
\nu=\frac{c_s^2}{c^2},\qquad \epsilon=\frac{1}{\alpha}=\left
(\frac{4\pi G\rho\hbar^2}{m^2c_s^4}\right )^{1/2}.
\label{dv9}
\end{eqnarray}
In this normalization, $G$ and $c_s$ are
used to construct the reference scales $\omega_0$ and $k_0$, so they are
assumed to be ``fixed''.  Then, the parameter
$\nu\propto 1/c^2$ measures the importance of relativistic effects and the
parameter
$\epsilon\propto \hbar$ measures the importance of quantum effects. The
nonrelativistic limit ($c\rightarrow +\infty$) corresponds to
$\nu\rightarrow 0$ and the TF limit ($\hbar\rightarrow 0$)  corresponds to
$\epsilon\rightarrow 0$.
For a quartic SF potential of the form of Eq. (\ref{jia1}), using the
expression of the speed of sound given by Eq. (\ref{jia9}), we can write the
parameter
$\nu$ as
\begin{eqnarray}
\nu=\frac{4\pi |a_s|\hbar^2\rho}{m^3c^2}.
\label{dv10}
\end{eqnarray}

{\it Remark:} Using this normalisation amounts to taking $4\pi
G=\rho=c_s=m=1$, $\nu=1/c^2$ and $\epsilon=1/\alpha=\hbar$ in the original
equations.

\section{Optimal wavenumber and maximum growth rate in the noninteracting limit}
\label{sec_mgrni}

In this Appendix, we determine the most unstable wavenumber $k_*$ and the 
maximum growth rate $\sigma_{\rm max}$ in the
noninteracting
limit ($a_s=0$) for the simplified and exact relativistic models. We
give their
asymptotic behaviors in the nonrelativistic ($c\rightarrow +\infty$) and
ultrarelativistic  ($c\rightarrow 0$) limits.

\subsection{Simplified relativistic model}
\label{sec_mgrnis}

Using the normalization of Appendix \ref{sec_dv1}, the dispersion relation of
Eq. (\ref{sm9}) takes the
form
\begin{eqnarray}
\frac{1}{4}\chi^2\Omega^4-(1+\chi\kappa^2)\Omega^2+\kappa^4-1=0.
\label{ydsm1}
\end{eqnarray}
Its solutions are given by
\begin{equation}
\Omega^2_{\pm}(\kappa)=\frac{2}{\chi^2}\left[
1+\chi\kappa^2\pm(1+2\chi\kappa^2+\chi^2)^ {
1/2}\right].
\label{ydsm2}
\end{equation}
The functions $\Omega^2_{\pm}(\kappa)$ are plotted in Figs. \ref{dispsimple} and
\ref{dispsimpleMOINS}.

We first consider the branch (+). The minimum pulsation, corresponding to
$\kappa_*=0$, is given by $(\Omega_+)_{\rm min}=(\Omega_+^2(0))^{1/2}$ with
\begin{equation}
\Omega_{+}^2(0)=\frac{2}{\chi^2}\left(1+\sqrt{1+\chi^2}\right ).
\label{ydsm4b}
\end{equation}
This function is plotted in Fig. \ref{Wplus}.
For $\chi\rightarrow 0$ (nonrelativistic limit):
\begin{eqnarray}
\Omega_+^2(0)\sim \frac{4}{\chi^2}.
\label{ydsm8}
\end{eqnarray}
For $\chi\rightarrow +\infty$ (ultrarelativistic limit):
\begin{eqnarray}
\Omega_+^2(0)\sim \frac{2}{\chi}.
\label{ydsm9}
\end{eqnarray}

We now consider the branch (-). The Jeans wavenumber is given by 
\begin{eqnarray}
\kappa_J^2=1.
\label{ydsm7}
\end{eqnarray}
The maximum growth rate, corresponding to
$\kappa_*=0$, is given by $\Sigma_{\rm max}=(-\Omega_-^2(0))^{1/2}$ with
\begin{equation}
\Omega_{-}^2(0)=\frac{2}{\chi^2}\left(1-\sqrt{1+\chi^2}\right ).
\label{ydsm4c}
\end{equation}
This function is plotted in Fig. \ref{maxgrowthrateOm}.
For $\chi\rightarrow 0$ (nonrelativistic limit):
\begin{eqnarray}
\Omega_-^2(0)\simeq -1+\frac{\chi^2}{4}.
\label{ydsm10}
\end{eqnarray}
For $\chi\rightarrow +\infty$ (ultrarelativistic limit):
\begin{eqnarray}
\Omega_-^2(0)\sim-\frac{2}{\chi}.
\label{ydsm11}
\end{eqnarray}

\subsection{Exact relativistic model}
\label{sec_mgrne}

Using the normalization of Appendix \ref{sec_dv1}, the dispersion relation of
Eq. (\ref{em11}) takes the form
\begin{eqnarray}
\frac{1}{4}\chi^2\Omega^4-\left(\frac{1+\gamma}{3\gamma+1}
\chi\kappa^2+1\right)\Omega^2\nonumber\\
+\frac{1}{1+3\gamma}\left\lbrack(1-\gamma)\kappa^4-1\right\rbrack=0
\label{ydem7}
\end{eqnarray}
with
\begin{eqnarray}
\gamma=\frac{\chi}{2\kappa^2}.
\label{ydem8}
\end{eqnarray}
The functions $\Omega^2_{\pm}(\kappa)$ are  plotted in Figs.
\ref{dispexact} and \ref{dispexactMOINS}.

We first consider the branch (+). The minimum pulsation, corresponding to
$\kappa=0$, is given by $(\Omega_+)_{\rm min}=(\Omega_+^2(0))^{1/2}$ with
\begin{eqnarray}
\Omega_+^2(0)= \frac{4}{\chi^2}.
\label{ydsm8b}
\end{eqnarray}
This function is plotted in Fig. \ref{Wplus}.

We now consider the branch (-). The Jeans wavenumber is 
\begin{eqnarray}
\kappa_J^2=\frac{1}{4}\left (\chi+\sqrt{16+\chi^2}\right ).
\label{ydem12}
\end{eqnarray}
This function is plotted in
Fig. \ref{kj}.
For $\chi\rightarrow 0$ (nonrelativistic limit):
\begin{eqnarray}
\kappa_J^2\rightarrow 1.
\label{ydem12nr}
\end{eqnarray}
For $\chi\rightarrow +\infty$ (ultrarelativistic limit):
\begin{eqnarray}
\kappa_J^2\sim \frac{\chi}{2}.
\label{ydem12ur}
\end{eqnarray}
The wavenumber $\kappa_*$ corresponding to
the  maximum growth rate is obtained from the condition
$(\Omega^2)'(\kappa_*)=0$ where $\Omega^2(\kappa)$ is the solution of the
second degree equation
(\ref{ydem7}). Instead of solving Eq. (\ref{ydem7}) for $\Omega^2(\kappa)$
and taking the
derivative of this function with respect to $\kappa$, we proceed as follows. We
multiply Eq. (\ref{ydem7})
by $2\kappa^2(3\gamma+1)$, differentiate the resulting expression with respect
to $\kappa$, and cancel the terms where $(\Omega^2)'(\kappa)=0$ appear. In this
way
we obtain
\begin{equation}
\chi^2\Omega^4-2
\left(4\chi\kappa^2+\chi^2+2\right)\Omega^2+4\left(3\kappa^4-\chi\kappa
^2-1\right)=0.
\label{g1}
\end{equation}
The most unstable wavenumber $\kappa_*$ and the corresponding maximum growth
rate $\Sigma_{\rm max}=(-\Omega^2(\kappa_*))^{1/2}$ are then determined by
Eqs. (\ref{ydem7}) and (\ref{g1}). The functions
$\kappa_*(\chi)$ and $\Sigma_{\rm max}(\chi)$ are
plotted  in Figs.
\ref{maxgrowthrateK} and \ref{maxgrowthrateOm}. Their asymptotic behaviors for
small and large
$\chi$ can be determined analytically.  

For $\chi\rightarrow 0$ (nonrelativistic limit):
\begin{eqnarray}
\kappa_*\sim A \,\chi^{1/6},
\label{g2a}
\end{eqnarray}
\begin{eqnarray}
\Omega^2(\kappa_*) \simeq -1+B\, \chi^{2/3},
\label{g2b}
\end{eqnarray}
where the coefficients $A$ and $B$ are determined by the algebraic equations
\begin{eqnarray}
-B&+&\frac{3}{2A^2}+A^4=0,\nonumber\\
-4 B&+&12 A^4=0.
\label{g4}
\end{eqnarray}
We obtain $A=0.953184...$ and $B=2.47645...$.

For $\chi\rightarrow\infty$ (ultrarelativistic limit):
\begin{eqnarray}
\kappa_*\simeq\chi^{1/2}\left(C+\frac{D}{\chi^2}\right),
\label{g6b}
\end{eqnarray}
\begin{eqnarray}
\Omega^2_-(\kappa_*)\simeq A-\frac{B}{\chi^2},
\label{g6a}
\end{eqnarray}
where the coefficients $A$, $B$, $C$ and $D$ are determined by the algebraic
equations
\begin{eqnarray}
3A^2&-&4AC^2+2A^2C^2-4C^4-8AC^4+8C^6=0,\nonumber\\
-2A&+&A^2-4C^2-8AC^2+12C^4=0,\nonumber\\
-\frac{AB}{2}&+&\frac{BC^2(1+2C^2)}{3+2C^2}\nonumber\\
&+&\frac{2C^2(-3-2C^2-6CD+16C^3D+8C^5D)}{(3+2C^2)^2}\nonumber\\
&-&A\left[1+\frac{2(3CD+12C^3D+4C^5D)}{(3+2C^2)^2}\right]=0,\nonumber\\
-4&-&4A+2B-2AB+8BC^2-8CD-16ACD\nonumber\\
&+&48C^3D=0.
\label{g7}
\end{eqnarray}
We obtain $A=-0.0717968...$, $B=0.464102...$, $C=0.481717...$ and
$D=0.898895...$.

The
function $-\Omega^2(\kappa_{*})[\chi]$ starts from $1$ at
$\chi=0$,
decreases and
tends to $0.0717968...$ as $\chi\rightarrow
+\infty$.

\section{Optimal wavenumber and maximum growth rate in the TF case}
\label{sec_mgrtf}

In this Appendix, we determine the most unstable wavenumber $k_*$
and the maximum growth rate $\sigma_{\rm max}$ in the
TF limit ($\hbar\rightarrow 0$) for the simplified and exact relativistic
models. We give their
asymptotic behaviors in the nonrelativistic ($c\rightarrow +\infty$) and
ultrarelativistic  ($c\rightarrow 0$) limits.

\subsection{Simplified relativistic model}
\label{sec_mgrtfsm}

Using the normalization of Appendix \ref{sec_dv2},  the dispersion relation of
Eq. (\ref{sm14}) takes the form
\begin{eqnarray}
\Omega^2=\frac{\kappa^2-1-2\nu}{1+3\nu}.
\label{ydsm12}
\end{eqnarray}
This function is plotted in Fig. \ref{dispsimpleTF}.
The Jeans wavenumber is given by
\begin{eqnarray}
\kappa_J^2=1+2\nu.
\label{ydsm13}
\end{eqnarray}
This function is plotted in Fig. \ref{kjTF}.
For $\nu\rightarrow 0$ (nonrelativistic limit):
\begin{eqnarray}
\kappa_J^2\rightarrow 1.
\label{ydsm13nr}
\end{eqnarray}
For $\nu\rightarrow +\infty$ (ultrarelativistic limit):
\begin{eqnarray}
\kappa_J^2\sim 2\nu.
\label{ydm13ur}
\end{eqnarray}
The maximum growth rate, corresponding to
$\kappa=0$, is given by $\Sigma_{\rm max}=(-\Omega^2(0))^{1/2}$ with
\begin{eqnarray}
\Omega^2(0)=-\frac{1+2\nu}{1+3\nu}.
\label{ydsm14}
\end{eqnarray}
This function is plotted in Fig. \ref{nuOmegaexact}.
For $\nu\rightarrow 0$ (nonrelativistic limit): 
\begin{eqnarray}
\Omega^2(0)\simeq -1+\nu.
\label{ydsm15}
\end{eqnarray}
For $\nu\rightarrow +\infty$ (ultrarelativistic limit):
\begin{eqnarray}
\Omega^2(0)\simeq -\frac{2}{3}-\frac{1}{9\nu}.
\label{ydsm16}
\end{eqnarray}

\subsection{Exact relativistic model}
\label{sec_mgrtfe}

Using the normalization of Appendix \ref{sec_dv2},  the dispersion relation of
Eq. (\ref{em17})  
takes the form
\begin{eqnarray}
\Omega^2=\frac{1}{\gamma}(1+2\nu)\frac{1}{1+3\gamma}\frac{\nu-(3\nu+1)\gamma}{
1+3\nu}
\label{ydem13}
\end{eqnarray}
with
\begin{eqnarray}
\gamma=\frac{\nu}{\kappa^2}(1+2\nu).
\label{ydem14}
\end{eqnarray}
This function is plotted in Fig.
\ref{dispexactTF}. 
The Jeans wavenumber is
\begin{eqnarray}
\kappa_J^2=(1+2\nu)(1+3\nu).
\label{ydem17}
\end{eqnarray}
This function is plotted in Fig. \ref{kjTF}. For $\nu\rightarrow 0$
(nonrelativistic limit):
\begin{eqnarray}
\kappa_J^2\rightarrow 1.
\label{ydem17nr}
\end{eqnarray}
For $\nu\rightarrow +\infty$ (ultrarelativistic limit):
\begin{eqnarray}
\kappa_J^2\sim 6\nu^2.
\label{ydem17ur}
\end{eqnarray}
The
maximum growth rate, obtained from the condition
$(\Omega^2)'(\kappa_*)=0$, corresponds to an optimal wavenumber $\kappa_*$
given by
\begin{equation}
\kappa_*^2=\frac{3(3\nu+1)(1+2\nu)\nu}{3\nu+\sqrt{3\nu(6\nu+1)}}.
\label{ydem18}
\end{equation}
An equivalent expression is
\begin{equation}
\kappa_*^2=3\nu(1+2\nu)\left (\sqrt{2+\frac{1}{3\nu}}-1\right ).
\label{ydem19}
\end{equation}
The maximum growth rate $\Sigma_{\rm max}=\sqrt{-\Omega^2(\kappa_*)}$ is then
obtained by substituting Eq. (\ref{ydem18}) or Eq. (\ref{ydem19}) into Eq.
(\ref{ydem13}). The functions $\kappa_*(\nu)$ and $\Sigma_{\rm max}(\nu)$ are
plotted in Figs. \ref{nukappaexact} and \ref{nuOmegaexact}.

For $\nu\rightarrow 0$ (nonrelativistic limit):
\begin{equation}
\kappa_*^2\sim \sqrt{3\nu},
\label{ydem20}
\end{equation}
\begin{equation}
\Omega^2(\kappa_{*})\simeq -1+2\sqrt{3\nu}+...
\label{ydem21}
\end{equation}

For $\nu\rightarrow +\infty$ (ultrarelativistic limit): 
\begin{equation}
\kappa_*^2\sim \frac{6\nu^2}{1+\sqrt{2}},
\label{ydem22}
\end{equation}
\begin{equation}
\Omega^2(\kappa_{*})\sim 2(2\sqrt{2}-3)\nu.
\label{ydem23}
\end{equation}

The
function $-\Omega^2(\kappa_{*})[\nu]$ starts from $1$ at
$\nu=0$,
decreases, reaches a minimum value $-\Omega^2(\kappa_{*})\simeq 0.438$ at
$\nu\simeq 0.23$, increases and tends to $+\infty$ as $\nu\rightarrow
+\infty$.

\section{Characteristic scales}
\label{sec_chs}

In this Appendix, we introduce characteristic length, mass and density
scales that play an important
role in the astrophysical and cosmological  applications considered in this
paper (see Secs.
\ref{sec_anrad}-\ref{sec_fermi}).

\subsection{The Hubble scales}
\label{sec_hs}

We introduce the Hubble length
\begin{eqnarray}
\lambda_H=\frac{c}{H},
\label{hs1}
\end{eqnarray}
where $H=\dot a/a$ is the Hubble constant. The Hubble length represents the
distance travelled by a photon with velocity $c$ during the Hubble time
$t_H=1/H$ which gives the
typical age of the Universe. Therefore, the 
Hubble length $\lambda_H=ct_H$ represents the typical size of the visible
Universe (particle horizon).  In a flat Universe, the Hubble
constant is related
to the energy density $\epsilon$ by  the Friedmann equation \cite{weinbergbook}:
\begin{eqnarray}
H^2=\frac{8\pi G}{3c^2}\epsilon.
\label{hs2}
\end{eqnarray}
We introduce the Hubble mass $M_H=(4/3)\pi(\epsilon/c^2)\lambda_H^3$. It
represents the typical mass of the visible Universe. Using Eqs. (\ref{hs1})
and (\ref{hs2}), we get
\begin{eqnarray}
M_H=\frac{c^3}{2GH}.
\label{hs3}
\end{eqnarray}
The Hubble length and the Hubble mass can also be written in terms of the
energy density as
\begin{eqnarray}
\lambda_H=\left (\frac{3c^4}{8\pi G\epsilon}\right )^{1/2},\qquad
M_H=\left (\frac{3c^8}{32\pi G^3\epsilon}\right )^{1/2}.
\label{hs4}
\end{eqnarray}
They are connected to each other by the relation
$\lambda_H=2GM_H/c^2$. This relation is similar to
the Schwarzschild radius of a black hole
\begin{eqnarray}
R_S=\frac{2GM}{c^2}.
\label{schwar}
\end{eqnarray}
Therefore, the Hubble length and the Hubble mass of the Universe display the
same scaling as a Schwarzschild black hole (see Appendix F.5 of
\cite{bectcoll}).

{\it Remark:} the current values of the Hubble scales are
$H_0=2.18\times 10^{-18}\, {\rm s}^{-1}$, $t_H=1/H_0=14.5\, {\rm
Gyrs}$, $\lambda_H=1.37\times 10^{26}\, {\rm m}$, $\epsilon_0/c^2=8.45\times
10^{-24}\, {\rm g/m^3}$, and $M_H=9.26\times 10^{55}\, {\rm g}$.

\subsection{The scales associated with noninteracting relativistic
self-gravitating
 BECs}
\label{sec_cs}

We introduce the length and mass scales associated with noninteracting
relativistic self-gravitating  BECs: 
\begin{eqnarray}
\lambda_C=\frac{\hbar}{mc},\qquad M_C=\frac{\hbar c}{Gm}=\frac{M_P^2}{m},
\label{cs1}
\end{eqnarray}
where $M_P=(\hbar c/G)^{1/2}$ is the Planck mass. These scales appear in the
works of Kaup \cite{kaup} and Ruffini and
Bonazzola  \cite{rb} on boson stars. $M_C$ is of the order of
the maximum mass of a noninteracting relativistic self-gravitating  BEC and
$\lambda_C$, which coincides with  the Compton length, is
of the order of the corresponding (minimum) radius. They are connected to each
other by the relativistic scaling $M_C=c^2\lambda_C/G$. We also introduce the
density scale  
\begin{equation}
\rho_C= \frac{m^2c^4}{8\pi G\hbar^2},
\label{tsni1}
\end{equation}
which is of the order of $M_C/\lambda_C^3$. We note that the relativistic
parameter defined in Appendix \ref{sec_dv1} can be
written as
\begin{eqnarray}
\chi=\left (\frac{\rho}{2\rho_{C}}\right )^{1/2}.
\label{tsni2b}
\end{eqnarray}

The mass-radius relation $M(R)$ (parametrized by the central energy density
$\epsilon_0$) 
of noninteracting self-gravitating
BECs in the context of general relativity is represented in Fig. 3
of \cite{seidel90}. In the
nonrelativistic limit, the mass-radius
relation is given by Eq. (\ref{app6}). The system becomes relativistic when its
radius
$R$ approaches the Schwarzschild radius $R_S$. This determines the
maximum mass of a noninteracting boson star above which there is no equilibrium
state.  The spiral is made of unstable equilibrium states. Combining
Eqs. (\ref{app6}) and
(\ref{schwar}) we obtain the scalings of Eqs. (\ref{cs1}) and (\ref{tsni1}). By
numerically solving the KGE equations, one gets the exact values $M_{\rm
max}=0.633 M_C$ and $R_{\rm min}=6.03\lambda_C$, where $R$ is the radius
containing $95\%$ of the mass \cite{kaup,rb}.

\subsection{The scales associated with self-repulsive relativistic
self-gravitating BECs}
\label{sec_is}

We introduce the length and mass scales associated with
relativistic self-gravitating  BECs with a repulsive
$|\varphi|^4$ self-interaction
($a_s>0$): 
\begin{eqnarray}
\lambda_R=\left (\frac{a_s\hbar^2}{Gm^3}\right )^{1/2},\qquad M_R=\left
(\frac{a_s\hbar^2 c^4}{G^3m^3}\right )^{1/2}.
\label{is1}
\end{eqnarray}
Introducing the dimensionless self-interaction parameter (see Appendix B of
\cite{suarezchavanis3}):
\begin{eqnarray}
\frac{\lambda}{8\pi}=\frac{a_s}{\lambda_C}=\frac{a_s mc}{\hbar},
\end{eqnarray}
they can be rewritten as
\begin{eqnarray}
\lambda_R=\sqrt{\frac{\lambda}{8\pi}}\frac{M_P}{m}\lambda_C,\qquad
M_R=\sqrt{\frac{\lambda}{8\pi}}\frac{M_P^3}{m^2}.
\end{eqnarray}
These scales appear in the works of Colpi {\it et al.} \cite{colpi} 
and Tkachev \cite{tkachev} on boson
stars and in the work of Chavanis and Harko \cite{chavharko} on neutron
stars with a superfluid core (BEC stars). $M_R$ is of the order of the maximum
mass of a
relativistic self-gravitating BEC in the TF limit and $\lambda_R$ is
of the order of the corresponding (minimum) radius. 
They are connected to
each
other by the relativistic scaling $M_R=c^2\lambda_R/G$.
We also introduce the  density scale
\begin{equation}
\rho_R=\frac{m^3c^2}{6\pi a_s\hbar^2},
\label{tstf1}
\end{equation}
which is of the order of $M_R/\lambda_R^3$. We note that the relativistic
parameter defined in Appendix \ref{sec_dv2} can be
written
as
\begin{eqnarray}
\nu=\frac{2\rho}{3\rho_R}.
\label{tstf2b}
\end{eqnarray}

The mass-radius relation $M(R)$ (parametrized by the central energy density
$\epsilon_0$)
of self-gravitating
BECs with a repulsive self-interaction in the context of general relativity is
represented in  Fig. 9 of \cite{chavharko}. In the nonrelativistic limit, the
radius of the system is independent of its mass and given by Eq.
(\ref{app13}). The system becomes relativistic when its radius
$R$ approaches the Schwarzschild radius $R_S$. This determines the maximum
mass of a self-repulsive boson star above which there is no equilibrium state.
The spiral is made of unstable equilibrium states.
Combining Eqs. 
(\ref{app13}) and (\ref{schwar}) we obtain the scalings of Eqs.
(\ref{is1})-(\ref{tstf1}). By numerically solving the KGE equations \cite{colpi}
or the equivalent hydrodynamic equations \cite{chavharko}, one gets the exact
values $M_{\rm max}=0.307 M_R$, $R_{\rm
min}=1.92\lambda_R$ and $(\epsilon_0)_{\rm
max}=1.19\rho_R c^2$.

\subsection{The scales associated with self-interacting nonrelativistic
self-gravitating
BECs}
\label{sec_cis}

We introduce the length and mass scales associated with nonrelativistic
self-gravitating BECs with a $|\psi|^4$ self-interaction:
\begin{eqnarray}
\lambda_a=\left (\frac{|a_s|\hbar^2}{Gm^3}\right
)^{1/2},\qquad M_a=\frac{\hbar}{\sqrt{Gm|a_s|}}.
\label{cis1}
\end{eqnarray}
These scales appear in the works of Chavanis \cite{prd1,prd2,bectcoll} on
self-gravitating BECs with attractive or repulsive self-interactions. For a
repulsive
self-interaction ($a_s>0$), they determine the transition between the
noninteracting limit and
the TF limit. For an attractive
self-interaction ($a_s<0$), $M_a$
is of the order of the maximum mass of a self-gravitating BEC and $\lambda_a$ is
of the order of the corresponding (minimum) radius. We also introduce the
density scale 
\begin{equation}
\rho_{a}=\frac{Gm^4}{4\pi\hbar^2a_s^2},
\label{tsnr1}
\end{equation}
which is of the order of $M_a/\lambda_a^3$. 
We note that the self-interaction parameter defined in Appendix \ref{sec_dv1}
can be written
as 
\begin{eqnarray}
\alpha=\left (\frac{\rho}{\rho_{a}}\right )^{1/2}.
\label{nrl2b}
\end{eqnarray}

The mass-radius relation $M(R)$ (parametrized by the central density $\rho_0$)
of nonrelativistic self-gravitating
BECs with a repulsive self-interaction is represented in Fig. 4 of \cite{prd2}.
In the noninteracting limit,  the mass-radius
relation is given by Eq. (\ref{app6}). In the TF limit, the
radius of the system is independent on its mass and given by Eq.
(\ref{app13}). At the transition, combining Eqs. 
(\ref{app6}) and (\ref{app13}), we obtain the scalings of Eqs.
(\ref{cis1}) and (\ref{tsnr1}).

The mass-radius relation $M(R)$ (parametrized by the central density $\rho_0$)
of nonrelativistic
self-gravitating
BECs with an attractive self-interaction is represented in Fig. 6 of
\cite{prd2}.
In the noninteracting limit,  the mass-radius
relation is given by Eq. (\ref{app6}). In the nongravitational limit, the
mass-radius relation is given by Eq. (\ref{ng3b}) but these equilibrium states
are
unstable. At the transition,
combining Eqs. (\ref{app6}) and (\ref{ng3b}), we obtain the scalings of Eqs.
(\ref{cis1}) and (\ref{tsnr1}).  They determine the maximum
mass of Newtonian self-attracting boson stars above which there is no
equilibrium state. By numerically solving the GPP
equations or the equivalent hydrodynamic equations \cite{prd2}, one gets the
exact
values $M_{\rm max}=1.012 M_a$, $R_{\rm
min}=5.5\lambda_a$ and $(\rho_0)_{\rm
max}=0.5\rho_a$.

\subsection{The scales associated with self-attracting nongravitational
BECs}
\label{sec_ngsi}

We introduce the length and mass scales associated with
nongravitational BECs with an attractive self-interaction ($a_s<0$):
\begin{eqnarray}
\lambda_i=\frac{\hbar}{mc},\qquad
M_i=\frac{\hbar}{|a_s|c}.
\label{ngsi1}
\end{eqnarray}
These scales appear in Sec. \ref{sec_urng} of this
paper. They correspond to the Jeans length and Jeans mass of a
nongravitational BEC with attractive self-interaction in the ultrarelativistic
limit. We note
that
$\lambda_i$ coincides with the Compton wavelength of the particle.
To the best
of our knowledge, the mass scale $M_i$ has not been introduced before.
We also introduce the density scale
\begin{eqnarray}
\rho_i=\frac{m^3c^2}{12\pi |a_s|\hbar^2},
\label{lun1}
\end{eqnarray}
which is of the order of $M_i/\lambda_i^3$.
We note that the relativistic parameter defined in Appendix \ref{sec_dv2} can be
written
as 
\begin{eqnarray}
\nu=\frac{\rho}{3\rho_i}.
\label{lun2b}
\end{eqnarray}

{\it Remark:} these scales are embedded in the KG equation (\ref{jia4}).
Comparing the first and second terms, we obtain the length scale
$\lambda_i=\hbar/mc$. Comparing the second and third terms, and using Eq.
(\ref{jia7}), we obtain the density scale $\rho_i={m^3c^2}/{12\pi
|a_s|\hbar^2}$. From these two characteristic scales, we obtain
the mass scale $M_i\sim \rho_i\lambda_i^3={\hbar}/{|a_s|c}$.

\subsection{The effective Schwarzschild radius of a boson}
\label{sec_effS}

Following Ref. \cite{suarezchavanis3}, we introduce the effective
Schwarzschild radius of a boson
\begin{equation}
r_{S}=\frac{2Gm}{c^2}
\label{lun3}
\end{equation}
and the self-interaction parameter 
\begin{equation}
\sigma=\frac{3|a_s|c^2}{4Gm}=\frac{3|a_s|}{2r_S}.
\label{lun3b}
\end{equation}
We also introduce a density scale
\begin{equation}
\rho_{h}=\frac{|a_s| m c^6}{G^2\hbar^2}.
\label{mys}
\end{equation}

\subsection{The scales associated with relativistic self-gravitating  fermions}
\label{sec_chas}

We introduce the length and mass scales associated with 
relativistic self-gravitating fermions: 
\begin{equation}
\lambda_F=\left (\frac{\hbar^3}{Gc}\right )^{1/2}\frac{1}{m^2},\qquad
M_F=\left (\frac{\hbar c}{G}\right )^{3/2}\frac{1}{m^2}=\frac{M_P^3}{m^2}.
\label{chas1}
\end{equation}
These scales appear in the work of Chandrasekhar \cite{chandramax} on white
dwarf stars and in the work of Oppenheimer and Volkoff \cite{ov} on neutron
stars. $M_F$ is of the
order of
the maximum mass of a 
relativistic self-gravitating fermion star and $\lambda_F$ is
of the order of the corresponding (minimum) radius. They are connected to each
other by the relativistic scaling $M_F=c^2\lambda_F/G$.
We also introduce the density scale
\begin{equation}
\rho_{F}= \frac{m^4c^3}{\hbar^3},
\label{fermi19}
\end{equation}
which is of the order of $M_{F}/\lambda_{F}^3$. We define the relativistic
parameter for fermions by
\begin{equation}
\mu=\left (\frac{\rho\hbar^3}{m^4c^3}\right )^{2/3}=\left
(\frac{\rho}{\rho_{F}}\right )^{2/3}.
\label{fermi20b}
\end{equation}

The mass-radius relation $M(R)$ (parametrized by the central energy density
$\epsilon_0$)
of self-gravitating fermions in the
context of general relativity is represented in Fig. 7
of \cite{ns}. In the nonrelativistic limit, the mass-radius
relation is given by Eq. (\ref{fermi1}). The system becomes relativistic when
its radius
$R$ approaches the Schwarzschild radius $R_S$.  This determines the maximum
mass of a fermion star  above which there is no equilibrium state.  The spiral
is made of unstable equilibrium states. Combining
Eqs. 
(\ref{fermi1}) and (\ref{schwar}) we obtain the scalings of Eqs. (\ref{chas1})
and
(\ref{fermi19}). By numerically solving the equations of hydrostatic
equilibrium in general relativity, one gets the exact
values $M_{\rm max}=0.384 M_F$, $R_{\rm
min}=3.36\lambda_F$ and $(\epsilon_0)_{\rm
max}=2.33\times 10^{-2} \rho_F c^2$ \cite{ov,ns}.

\section{The general relativistic Jeans instability of a fluid}
\label{sec_paddy}

In this Appendix, we  consider the Jeans instability of a self-gravitating fluid
with an equation of state $P(\epsilon)$ in the framework of general
relativity. We assume that this equation of state is valid
for both homogeneous and inhomogeneous distributions.  The
exact Jeans wavenumber obtained from the Einstein equations slightly perturbed
from a spatially homogeneous distribution is \cite{paddy}:
\begin{eqnarray}
k_J^2=\frac{4\pi G\epsilon}{c_s^2c^2}\left (1-6\frac{c_s^2}{c^2}+8w-3w^2\right
),
\label{paddy1}
\end{eqnarray}
where $c_s^2=P'(\epsilon)c^2$ is the speed of sound in the homogeneous fluid and
$w=P/\epsilon$ is the
equation of state parameter.

For a nonrelativistic fluid ($c_s\ll c^2$, $w\ll 1$ and $\epsilon=\rho c^2$),
Eq. (\ref{paddy1}) reduces to the standard Jeans formula of Eq. (\ref{intro2}).
For the equation of state of radiation $P=\epsilon/3$, we get
\begin{eqnarray}
k_J^2=\frac{16\pi G\epsilon}{c^4}.
\label{paddy3}
\end{eqnarray}
For the equation of state of stiff matter $P=\epsilon$, we obtain
\begin{eqnarray}
k_J^2=0.
\label{paddy4}
\end{eqnarray}
Since $\lambda_J\rightarrow +\infty$, there is no Jeans instability in a stiff
fluid.

\section{Transition scales in cosmology}
\label{sec_ts}

The Jeans mass-radius relations $M_J(\lambda_J)$  (parametrized by the density
of the universe  $\rho$) in cosmology are similar to the mass-radius relations
$M(R)$ (parametrized by the central density $\rho_0$) of boson stars and DM
halos (see Appendix \ref{sec_chs}). This similarity (noted in Sec.
V of \cite{prd1}) is not obvious since the first relations  $M_J(\lambda_J)$
are valid in the linear regime of structure formation (Jeans problem) while the
second relations $M(R)$ are valid  in the nonlinear regime of structure
formation (DM halos $=$ equilibrium states). In the main text, we have given
asymptotic
expressions of  $M_J(\lambda_J)$ in different limits of physical interest. In
this Appendix, we regroup these results in order to offer a clear relation
between these different limits.

\subsection{Transition scales in the noninteracting limit}
\label{sec_tsni}

We consider a SF in the noninteracting limit. The following results can be
deduced from Eq. (\ref{em16}) using the scales of Appendix \ref{sec_cs}.

The ultrarelativistic era (Sec. \ref{sec_urni}) corresponds to:

(i) $\chi\gg 1$.

(ii) $\rho\gg \rho_C$ (high densities/early Universe).

(iii)  $\lambda_J\sim \lambda_H\ll\lambda_C$ and $M_J\sim
M_H\ll M_C$.

The nonrelativistic era (Sec. \ref{sec_appq}) corresponds to:

(i) $\chi\ll 1$. 

(ii) $\rho\ll \rho_C$ (low densities/late Universe). 

(iii) $\lambda_C\ll\lambda_J\ll\lambda_H$ and $M_J\ll M_C\ll M_H$.

\begin{figure}[h]
\scalebox{0.33}{\includegraphics{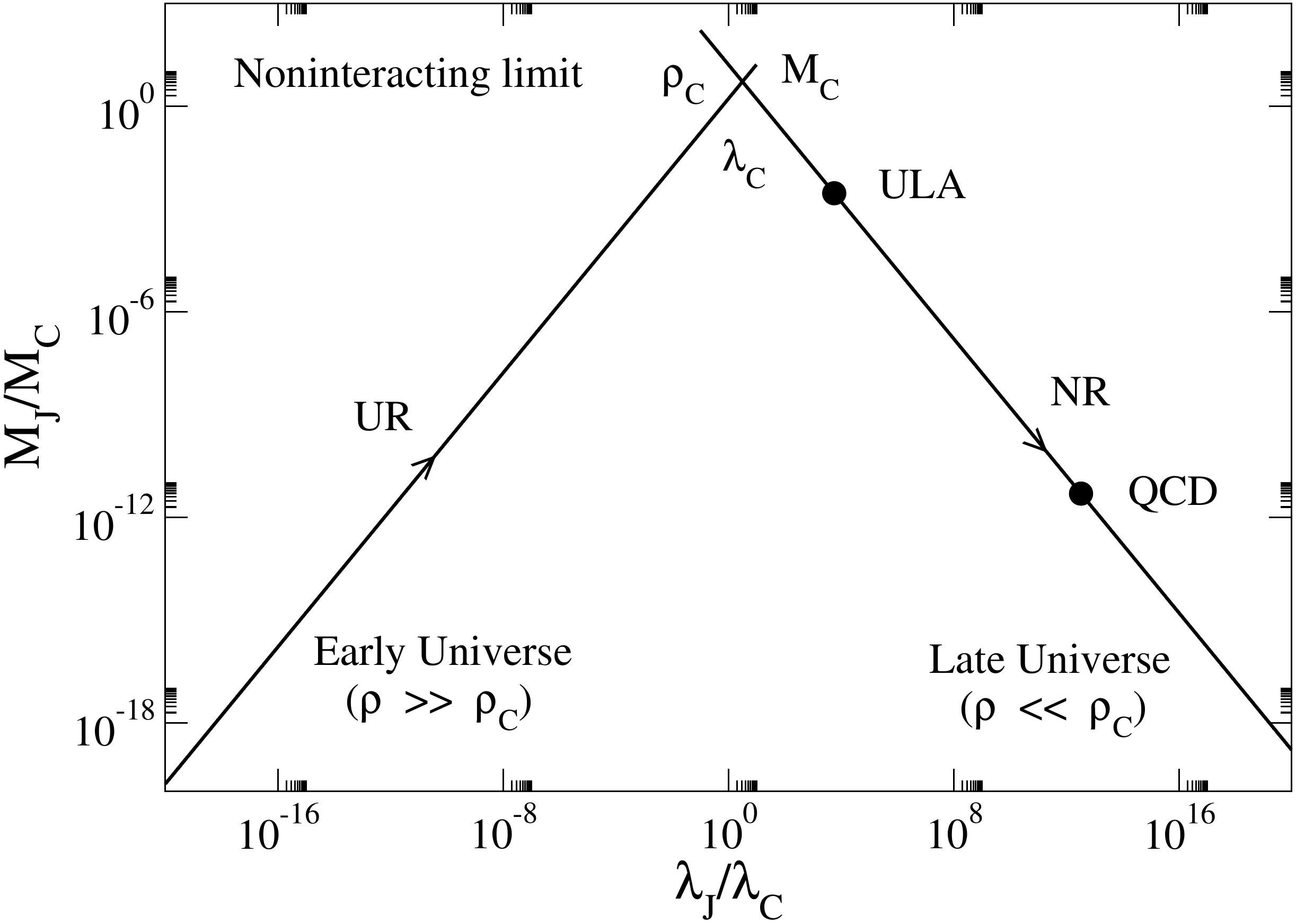}} 
\caption{Jeans mass-radius relation $M_J(\lambda_J)$  parametrized by the
density of the Universe $\rho$ in the noninteracting
limit. The arrows indicate the cosmic evolution of the
Universe from the ultrarelativistic (UR) regime at high densities
[see Eq. (\ref{urni7})] to the
nonrelativistic  (NR) regime at low densities [see Eq. (\ref{ven2})]. This curve
can be compared with
the mass-radius relation $M(R)$ (parametrized by the central energy density
$\epsilon_0$) of noninteracting self-gravitating
BECs in the context of general relativity (see Fig. 3
of \cite{seidel90}). The bullets correspond to 
ULAs and QCD axions at the epoch of radiation-matter equality considered in
Secs. \ref{sec_appqb} and \ref{sec_appqqcd}.}
\label{noninteracting}
\end{figure}

The Jeans mass-radius relation $M_J(\lambda_J)$ presents a maximum $M_C$ at
$\lambda_C$, corresponding to a density $\rho_C$ (see Fig.
\ref{noninteracting}). The transition density
$\rho_C$ where $\lambda_J\sim \lambda_C$ and $M_J\sim M_C$ separates
the ultrarelativistic regime from the nonrelativistic regime. It corresponds to
the density called $\rho_v(0)$ in
\cite{suarezchavanis3} separating the stiff matter era (slow oscillation
regime  $\omega=mc^2/\hbar \ll H$) from the matterlike era (fast oscillation
regime $\omega=mc^2/\hbar \gg H$).  This is also the density below which the
Jeans length is smaller than the Hubble length (horizon), allowing the
formation of the large-scale structures of the universe.

\subsection{Transition scales in the TF limit}
\label{sec_tstf}

We consider a SF with a repulsive self-interaction ($a_s>0$) in the TF limit. 
The following results can be
deduced from Eq. (\ref{em22}) using the scales of Appendix \ref{sec_is}.

The ultrarelativistic era (Sec. \ref{sec_urtf}) corresponds to:

(i) $\nu\gg 1$.

(ii) $\rho\gg \rho_R$ (high densities/early
Universe).

(iii) $\lambda_J\sim\lambda_H\ll \lambda_R$ and $M_J\sim M_H\ll M_R$.

The nonrelativistic era (Sec. \ref{sec_appc}) corresponds to:

(i) $\nu\ll 1$.

(ii) $\rho\ll \rho_R$ (low densities/late Universe). 

(iii) $\lambda_J\sim\lambda_R\ll\lambda_H$ and $M_J\ll M_R\ll M_H$.

\begin{figure}[h]
\scalebox{0.33}{\includegraphics{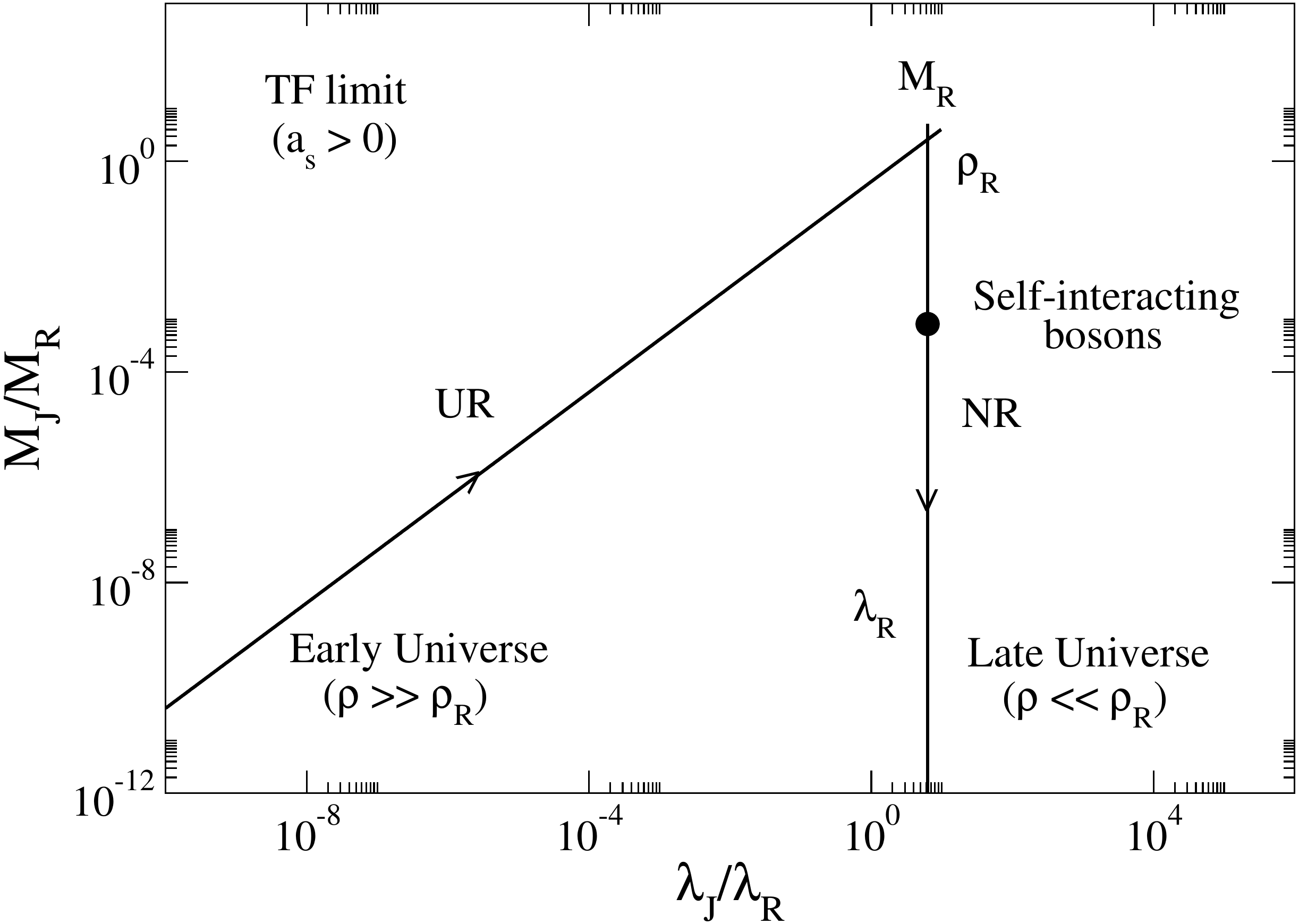}} 
\caption{Jeans mass-radius relation $M_J(\lambda_J)$  parametrized by the
density of the Universe $\rho$ in the TF
limit. The arrows indicate the cosmic evolution of the
Universe from the ultrarelativistic (UR) regime at high densities [see Eq.
(\ref{urni8})] to the
nonrelativistic (NR) regime  at low densities [see Eq. (\ref{soir2})]. This
curve can be compared with
the mass-radius relation $M(R)$ (parametrized by the central energy density
$\epsilon_0$)
of self-gravitating
BECs with a repulsive self-interaction in the context of general relativity
(see Fig. 9 of \cite{chavharko}).  The bullet corresponds to 
self-interacting bosons at the epoch of radiation-matter equality considered in
Sec. \ref{sec_sib}.}
\label{tf}
\end{figure}

The Jeans mass-radius relation $M_J(\lambda_J)$ presents a maximum $M_R$ at
$\lambda_R$, corresponding to a density $\rho_R$ (see Fig. \ref{tf}). The Jeans
length is always
smaller than $\lambda_R$. The transition density
$\rho_R$ where $\lambda_J\sim \lambda_R$ and $M_J\sim M_R$ separates
the ultrarelativistic regime from the nonrelativistic regime.
The transition density
$\rho_R$ corresponds to the density called $\rho_t$ in
\cite{suarezchavanis3} separating the radiationlike era from
the matterlike era. This is also the density below which the
Jeans length is smaller than the Hubble length (horizon), allowing the
formation of the large-scale structures of the universe.

\subsection{Transition scales in the nongravitational limit}
\label{sec_tsng}

We consider a SF with an attractive self-interaction ($a_s<0$) in the
nongravitational limit. The following results can be
deduced from Eqs. (\ref{dim4}) and (\ref{nog3}) using the scales of Appendix
\ref{sec_ngsi}.

The ultrarelativistic era  (Sec. \ref{sec_urng}) corresponds to:

(i) $\nu\sim 1$.

(ii) $\rho\sim \rho_{i}$ (high densities/early
Universe).

(iii) $\lambda_J\sim\lambda_i\ll\lambda_H$ and $M_J\sim
M_i\ll M_H$.

The nonrelativistic era (Sec. \ref{sec_ng}) corresponds to:

(i) $\nu\ll 1$.

(ii) $\rho\ll \rho_{i}$ (low densities/late
Universe).

(iii) $\lambda_i\ll\lambda_J\ll \lambda_H$ and $M_i\ll
M_J\ll M_H$.

\begin{figure}[h]
\scalebox{0.33}{\includegraphics{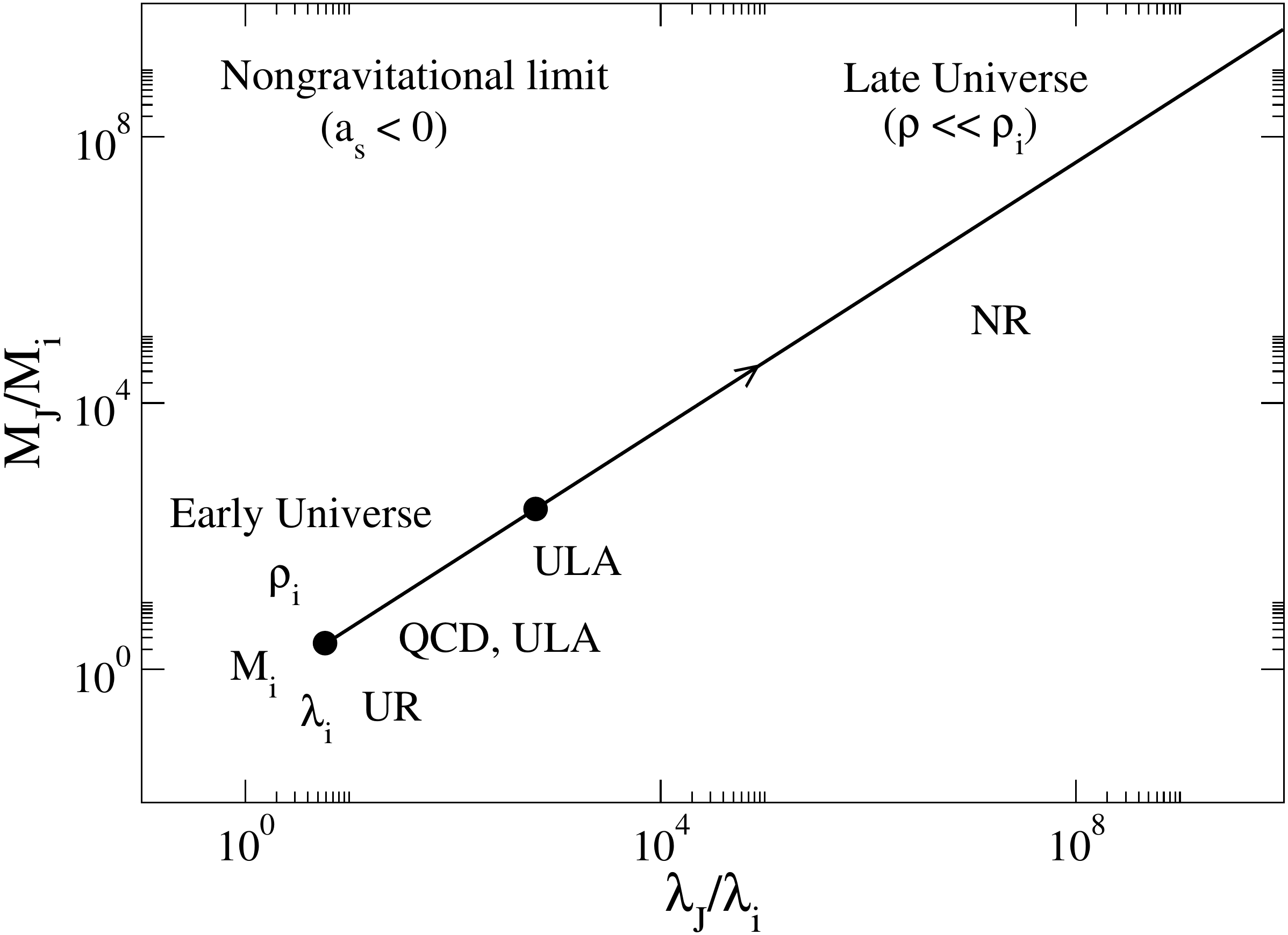}} 
\caption{Jeans mass-radius relation $M_J(\lambda_J)$  parametrized by the
density of the Universe $\rho$ in the nongravitational
limit. The arrows indicate the cosmic evolution of the
Universe from the ultrarelativistic (UR) regime at high densities [see
Eq. (\ref{fog})] to the
nonrelativistic (NR) regime  at low densities [see Eq. (\ref{ywo})].
The
first bullet corresponds to QCD-like axions and ULAs in the very early universe
considered
in Sec. \ref{sec_urng} while  the
second bullet corresponds to ULAs at the epoch
of radiation-matter equality considered in Sec. \ref{sec_forg}.}
\label{nongravitational}
\end{figure}

The Jeans mass-radius relation $M_J(\lambda_J)$ increases monotonically
starting from a minimum value $M_i$ at
$\lambda_i$, with a density $\rho_i$ (see Fig. \ref{nongravitational}). The
density
$\rho_i$ corresponds to the density introduced in
\cite{suarezchavanis3} at which a complex SF with an attractive self-interaction
emerges in the ultrarelativistic era.

\subsection{Transition scales in the nonrelativistic limit}
\label{sec_tsnr}

We consider a SF in the nonrelativistic limit.  The following results can be
deduced from Eq. (\ref{nr5})  using the scales of Appendix
\ref{sec_cis}.

\subsubsection{$a_s>0$}
\label{sec_tsnra}

When $a_s>0$, the TF era  (Sec. \ref{sec_appc}) corresponds to:

(i) $\alpha\gg 1$.

(ii)  $\rho\gg \rho_{a}$ (high densities/early
Universe).

(iii) $\lambda_J\sim\lambda_a\ll \lambda_H$ and $M_a\ll M_J\ll M_H$.

The noninteracting era (Sec. \ref{sec_appq}) corresponds to:

(i) $\alpha\ll 1$.

(ii) $\rho\ll \rho_{a}$ (low densities/late Universe).

(iii) $\lambda_a\ll\lambda_J\ll\lambda_H$ and $M_J\ll M_a,M_H$
($M_H\ll M_a$ if $\rho\gg \rho_h$ and $M_H\gg M_a$ if
$\rho\ll \rho_h$).

\begin{figure}[h]
\scalebox{0.33}{\includegraphics{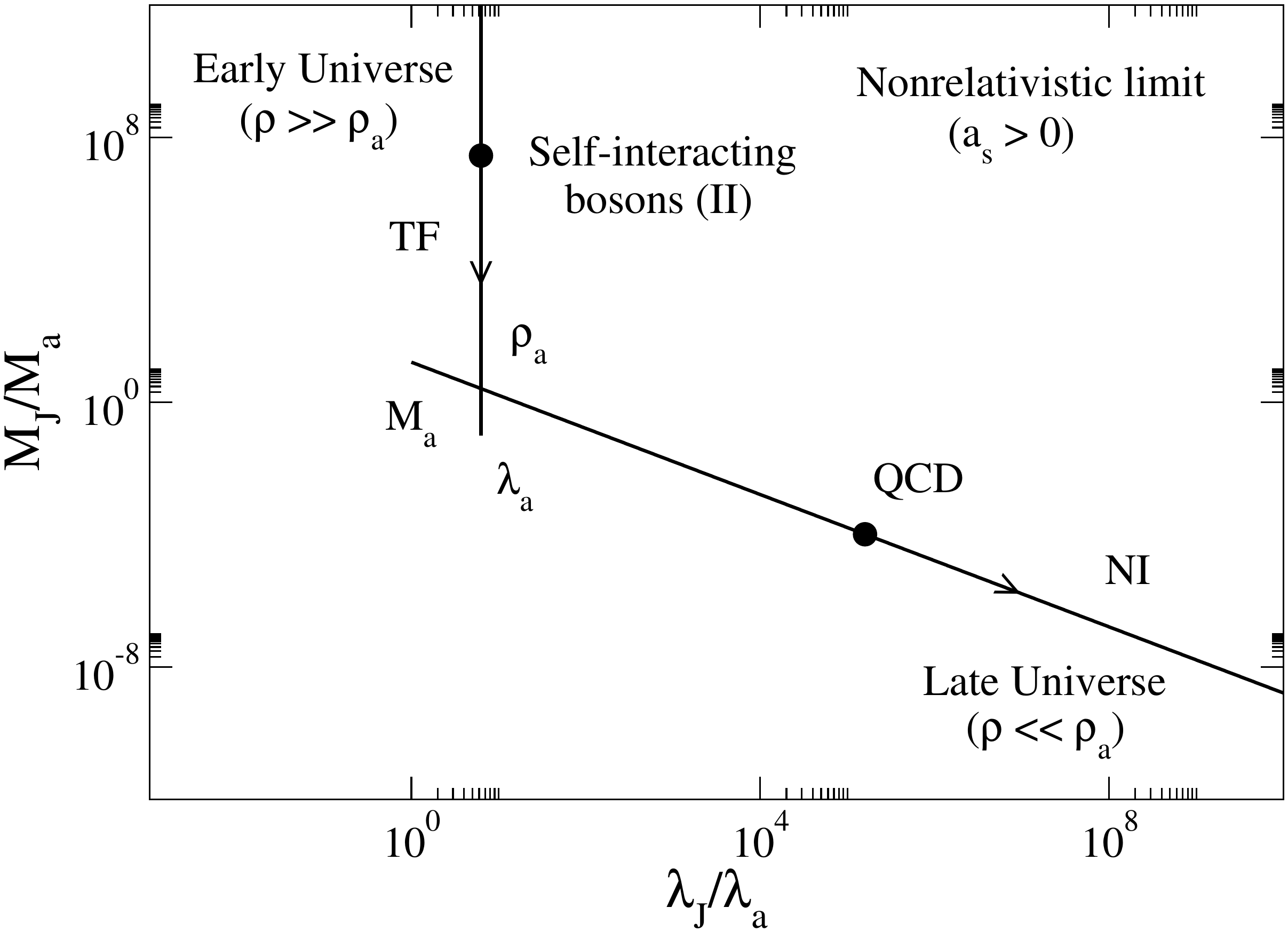}} 
\caption{Jeans mass-radius relation $M_J(\lambda_J)$  parametrized by the
density of the Universe $\rho$ in the nonrelativistic
limit ($a_s>0$). The arrows indicate the cosmic
evolution of the
Universe from the TF regime at high densities [see Eq. (\ref{sunq})] to the
noninteracting (NI) regime at low densities [see Eq. (\ref{alb1})]. This curve
can be compared with
the mass-radius relation $M(R)$ (parametrized by the central density
$\rho_0$) of nonrelativistic self-gravitating
BECs with a repulsive self-interaction (see Fig. 4 of
\cite{prd2}). The bullets correspond to QCD axions and
self-interacting bosons (model II) at the epoch of
radiation-matter equality considered in
Secs.  \ref{sec_appqqcd} and \ref{sec_sib}. The self-interacting bosons (model
I) are outside the frame since $M_J/M_a=2.49\times 10^{42}$.}
\label{nonrelativisticPOS}
\end{figure}

The Jeans mass-radius relation $M_J(\lambda_J)$ decreases
monotonically (see Fig. \ref{nonrelativisticPOS}). The Jeans length is always
larger than $\lambda_a$. The
transition density $\rho_a$ where $\lambda_J\sim\lambda_a$ and $M_J\sim
M_a$ separates the TF regime from the noninteracting regime.

\subsubsection{$a_s<0$}
\label{sec_tsnrb}

When $a_s<0$, the
nongravitational era (Sec. \ref{sec_ng}) corresponds to:

(i) $\alpha\gg 1$.

(ii)   $\rho\gg \rho_a$  (high densities/early
Universe).

(iii) $\lambda_J\ll\lambda_a, \lambda_H$ ($\lambda_H\ll \lambda_a$ if $\rho\gg
\rho_i$ and $\lambda_H\gg \lambda_a$ if $\rho\ll
\rho_i$) and $M_J\ll M_a$ ($M_H\ll M_a$ if $\rho\gg \rho_h$ and $M_H\gg M_a$ if
$\rho\ll \rho_h$).

The noninteracting era (Sec. \ref{sec_appq}) corresponds to:

(i) $\alpha\ll 1$.

(ii) $\rho\ll \rho_{a}$ (low densities/late Universe).

(iii) $\lambda_a\ll\lambda_J\ll\lambda_H$ and $M_J\ll M_a,M_H$ ($M_H\ll M_a$ if
$\rho\gg \rho_h$ and $M_H\gg M_a$ if
$\rho\ll \rho_h$).

\begin{figure}[h]
\scalebox{0.33}{\includegraphics{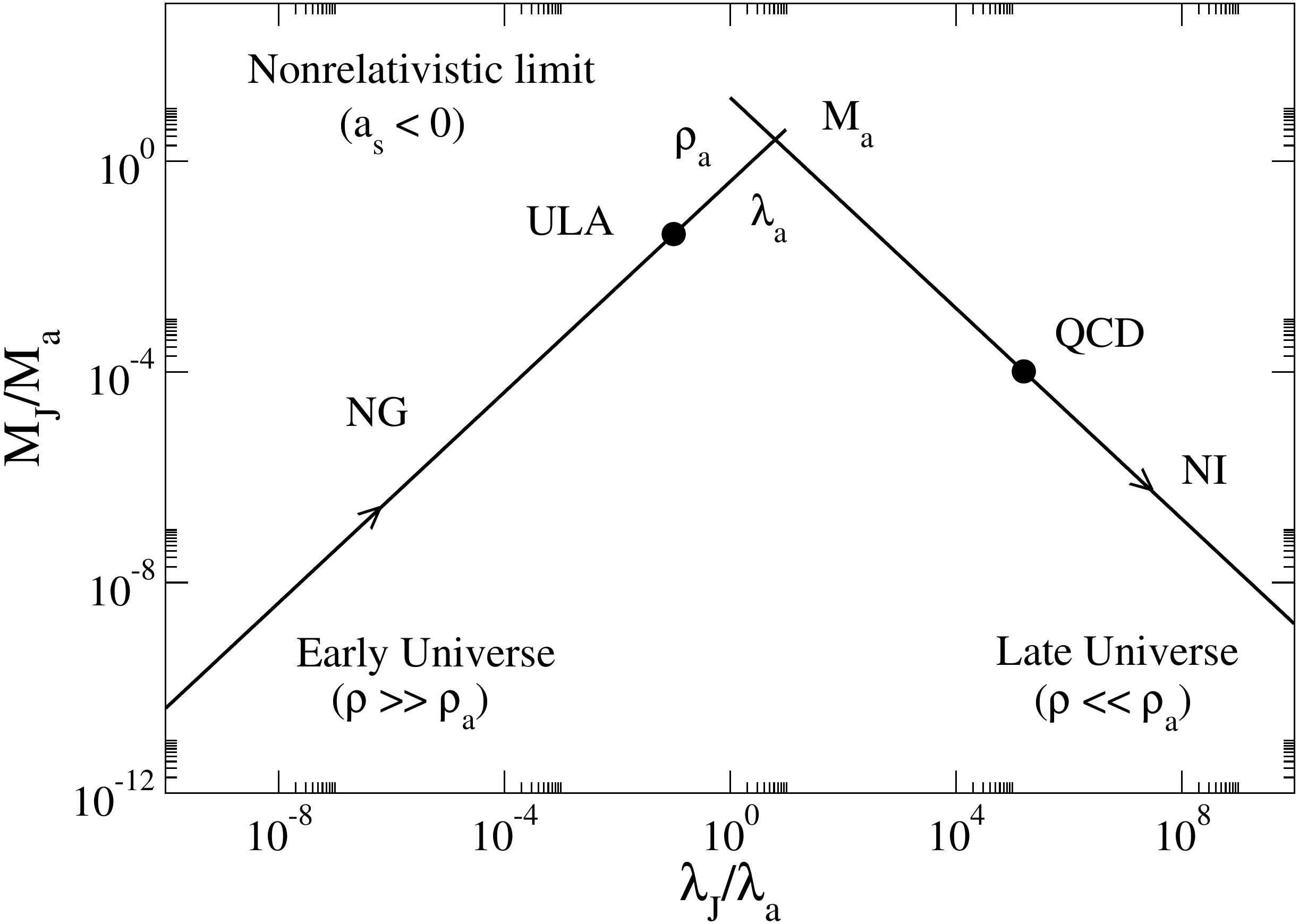}} 
\caption{Jeans mass-radius relation $M_J(\lambda_J)$  parametrized by the
density of the Universe $\rho$ in the nonrelativistic
limit ($a_s<0$). The arrows indicate the cosmic
evolution of the
Universe from the nongravitational (NG) regime at high densities [see Eq.
(\ref{wqo})] to the
noninteracting (NI) regime at low densities [see Eq. (\ref{alb1})]. This
curve can be compared with
the mass-radius relation $M(R)$ (parametrized by the central density
$\rho_0$) of nonrelativistic self-gravitating
BECs with an attractive self-interaction (see Fig. 6 of \cite{prd2}).
The bullets correspond to QCD axions and ULAs at the epoch of
radiation-matter equality considered in
Secs.  \ref{sec_appqqcd} and  \ref{sec_forg}.}
\label{nonrelativisticNEG}
\end{figure}

The Jeans mass-radius relation $M_J(\lambda_J)$ presents a maximum $M_a$ at
$\lambda_a$, corresponding to a density $\rho_a$ (see Fig.
\ref{nonrelativisticNEG}).  The
transition density $\rho_a$ where $\lambda_J\sim\lambda_a$ and $M_J\sim
M_a$ separates the nongravitational regime from the noninteracting regime.

\subsection{Transition scales in the ultrarelativistic limit}
\label{sec_tsur}

We consider a SF in the ultrarelativistic limit. 

\subsubsection{$a_s>0$}
\label{sec_tsura}

When $a_s>0$ the following results can be
deduced from Eqs. (\ref{urni1}) and (\ref{urtf1})  using the
scales of Appendix
\ref{sec_is}.

The TF era (Sec. \ref{sec_urtf}) corresponds to:

(i) $\nu\gg 1$.

(ii)  $\rho\gg \rho_{R}$ (high densities/early
Universe).

(iii) $\lambda_J\sim\lambda_H\ll\lambda_R$ and $M_J\sim M_H \ll
M_R$.

The noninteracting era  (Sec. \ref{sec_urni}) corresponds to:

(i) $\nu\ll 1$.

(ii)  $\rho\ll \rho_{R}$ (low densities/late
Universe).

(iii) $\lambda_J\sim \lambda_H \gg\lambda_R$ and $M_J\sim M_H \gg
M_R$.

The Jeans mass-radius relation $M_J(\lambda_J)$ increases
monotonically.  The
transition density $\rho_R$ where $\lambda_J\sim\lambda_R$ and $M_J\sim
M_R$ separates the TF regime from the noninteracting regime.

\subsubsection{$a_s<0$}
\label{sec_tsurb}

When $a_s<0$, the density  in the ultrarelativistic limit has the
fixed value $\rho_i$ given by Eq. (\ref{badlu}). The following results can be
deduced from Eqs. (\ref{str1}) and (\ref{urng1}) using the
scales of Appendixes  \ref{sec_cs} and \ref{sec_effS}.

The noninteracting limit (Sec. \ref{sec_urni}) corresponds to:

(i) $\sigma\ll 1$.

(ii)   $|a_s|\ll r_S$ 

(iii) $\lambda_J\sim \lambda_H\ll\lambda_C$ and $M_J\sim M_H\ll
M_C$.

The nongravitational limit  (Sec. \ref{sec_urng}) corresponds to:

(i) $\sigma\gg 1$.

(ii)   $|a_s|\gg r_S$ 

(iii) $\lambda_J\sim\lambda_C\ll \lambda_H$ and $M_J\ll
M_C\ll M_H$.

The Jeans mass-radius relation $M_J(\lambda_J)$ presents a maximum
$M_C$ at $\lambda_C$ corresponding to $|a_s|\sim r_S$. This transition value 
separates the noninteracting regime from the nongravitational regime.

\subsection{Transition scales for fermionic dark matter}
\label{sec_fermit}

We consider a gas of fermions.  The following results can be
deduced from Eqs.  (\ref{fermi11}) and (\ref{fermi1}),
or by comparing Eqs. (\ref{fermi13r}) and (\ref{fermi2}),
using the
scales of Appendix \ref{sec_chas}.

\subsubsection{General relativistic treatment}
\label{sec_fermita}

The ultrarelativistic era (Sec. \ref{sec_fermiur}) corresponds to:

(i) $\mu\gg 1$.

(ii) $\rho\gg \rho_F$ (high densities/early Universe).

(iii)  $\lambda_J\sim \lambda_H\ll\lambda_F$ and $M_J\sim
M_H\ll M_F$.

The nonrelativistic era (Sec. \ref{sec_ferminr}) corresponds to:

(i) $\mu\ll 1$.
 
(ii) $\rho\ll \rho_F$ (low densities/late Universe). 

(iii) $\lambda_F\ll\lambda_J\ll\lambda_H$ and $M_J\ll M_F\ll M_H$.

\begin{figure}[h]
\scalebox{0.33}{\includegraphics{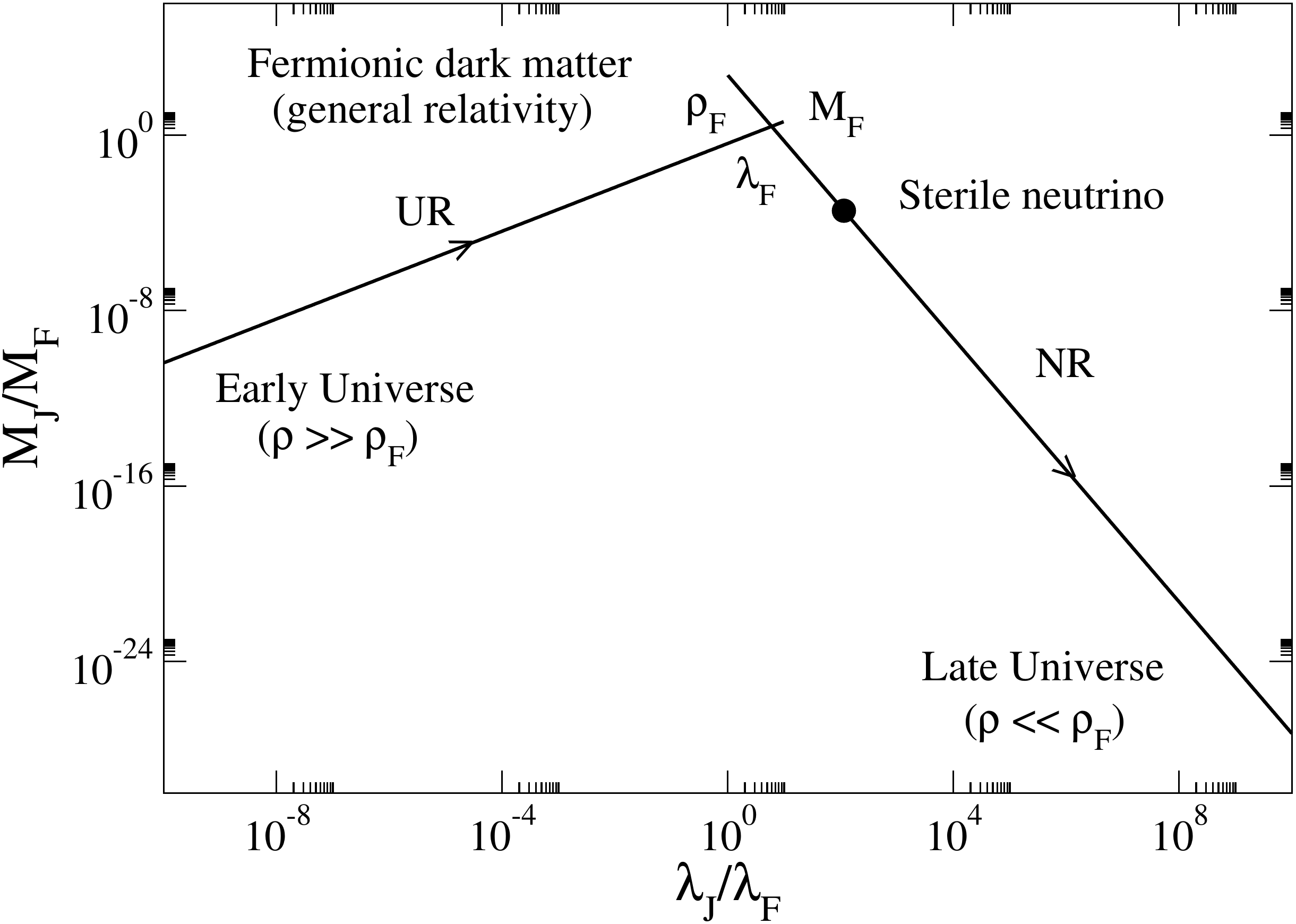}} 
\caption{Jeans mass-radius relation $M_J(\lambda_J)$  parametrized by the
density of the Universe $\rho$ for fermionic DM treated in the framework of
general relativity. The arrows indicate the cosmic
evolution of the
Universe from the ultrarelativistic (UR) regime at high densities [see Eqs.
(\ref{fermi18})
and (\ref{fermi18b})] to the
nonrelativistic (NR) regime at low densities [see Eqs. (\ref{fermi10})
and (\ref{fermi10sam})]. This curve can be compared with
the mass-radius relation $M(R)$ (parametrized by the central energy density
$\epsilon_0$) of self-gravitating fermions in the
context of general relativity (see Fig. 7
of \cite{ns}). The bullet corresponds to fermions (sterile neutrinos) at
the epoch of
radiation-matter equality considered in
Sec.  \ref{sec_ferminr}.}
\label{fermi}
\end{figure}

The Jeans mass-radius relation $M_J(\lambda_J)$ presents a maximum
$M_F$ at $\lambda_F$ corresponding to a density $\rho_F$ (see Fig. \ref{fermi}).
The transition 
density $\rho_F$ where $\lambda_J\sim \lambda_F$ and $M_J\sim M_F$
separates the ultrarelativistic regime from the nonrelativistic regime.
 This is also the density below which the
Jeans length is smaller than the Hubble length (horizon), allowing the
formation of the large-scale structures of the universe.

\subsubsection{Newtonian treatment}
\label{sec_fermitb}

The ultrarelativistic era (Sec. \ref{sec_fermiur}) corresponds to:

(i) $\mu\gg 1$.

(ii) $\rho\gg \rho_F$ (high densities/early Universe).

(iii)  $\lambda_J\ll\lambda_F$ and $M_J\sim  M_F$.

The nonrelativistic era (Sec. \ref{sec_ferminr}) corresponds to:

(i) $\mu\ll 1$.
 
(ii) $\rho\ll \rho_F$ (low densities/late Universe). 

(iii) $\lambda_F\ll\lambda_J\ll\lambda_H$ and $M_J\ll M_F\ll M_H$.

\begin{figure}[h]
\scalebox{0.33}{\includegraphics{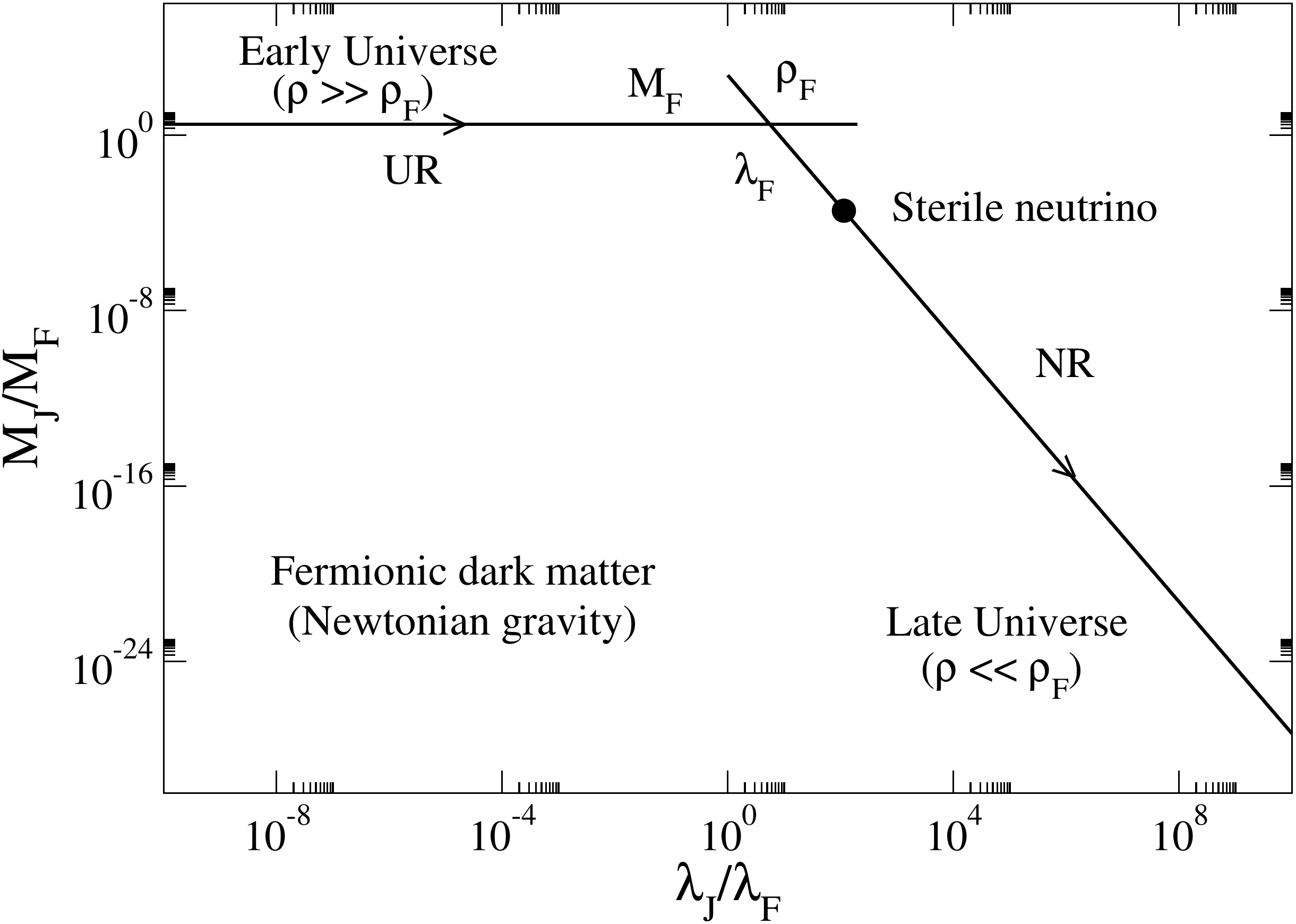}} 
\caption{Jeans mass-radius relation $M_J(\lambda_J)$  parametrized by the
density of the Universe $\rho$ for fermionic DM treated in the framework of
Newtonian gravity. The arrows indicate the cosmic
evolution of the
Universe from the ultrarelativistic (UR) regime at high densities [see Eq.
(\ref{mac5})] to the
nonrelativistic (NR) regime at low densities [see Eqs. (\ref{fermi10})
and (\ref{fermi10sam})]. This curve can be compared with
the mass-radius relation $M(R)$ (parametrized by the central density
$\rho_0$) of self-gravitating fermions in the
context of Newtonian gravity (see Fig. 2 of \cite{chandramr}).
The bullet corresponds to fermions (sterile neutrinos) at
the epoch of
radiation-matter equality considered in
Sec.  \ref{sec_ferminr}.}
\label{fermiN}
\end{figure}

The Jeans mass-radius relation $M_J(\lambda_J)$ decreases monotonically (see
Fig. \ref{fermiN})
starting from the maximum value $M_F$ at $\lambda_J\rightarrow 0$ corresponding
to a density $\rho\rightarrow +\infty$ (ultrarelativistic regime).

\section{About the (non)importance of the self-interaction}
\label{sec_imp}

In this Appendix, we obtain precise criteria determining the importance, or
nonimportance, of
the self-interaction of the bosons in the linear and nonlinear regimes of
structure formation.

\subsection{The linear regime of structure formation\\
 (Jeans problem)}
\label{sec_impl}

We consider the nonrelativistic regime corresponding to the matter
era (see Sec. \ref{sec_app}).\footnote{In the ultrarelativistic regime (see Sec.
\ref{sec_anrad}), a repulsive self-interaction ($a_s>0$) is negligible
for the Jeans problem at an epoch where the density is $\rho$ if $\nu=2
\rho/3\rho_R\ll 1$ (see Appendixes \ref{sec_dv2}, \ref{sec_is} and
\ref{sec_tsur}) i.e. $a_s/m^3\ll c^2/\rho\hbar^2$. In the opposite case, we are
in the TF limit. For an attractive self-interaction ($a_s<0$), the
ultrarelativistic regime corresponds to a density
$\rho_i=m^3c^2/12\pi|a_s|\hbar^2$ (see Sec. \ref{sec_jia}) and the
self-interaction is negligible for the Jeans problem if $\sigma=3|a_s|/2r_S\ll
1$ (see Appendixes \ref{sec_effS} and
\ref{sec_tsur}) i.e. $|a_s|\ll r_S=2Gm/c^2$. In the opposite
case, we are in the nongravitational limit.} As
shown in Appendixes \ref{sec_dv1} and \ref{sec_tsnr}, in the linear regime of
structure formation (Jeans problem), the noninteracting limit is valid when
$\alpha=(\rho/\rho_a)^{1/2}\ll 1$. Inversely, when $\alpha\gg 1$ we are in
the TF limit (for $a_s>0$) or in the nongravitational limit (for $a_s<0$).
Using Eq. (\ref{dv4}) the validity of the noninteracting limit can be written as
\begin{equation}
\frac{|a_s|}{m^2}\ll \left (\frac{G}{\rho \hbar^2}\right )^{1/2}.
\label{imp1}
\end{equation}
This criterion may also be expressed in terms of the dimensionless
self-interaction constant $\lambda=8\pi a_s m c/\hbar$ or in terms of the
axion decay constant $f=(\hbar c^3 m/32\pi |a_s|)^{1/2}$ (see, e.g.,
\cite{phi6}).

For an illustration, let us consider an ULA of mass $m=2.92\times 10^{-22}\,
{\rm eV}/c^2$ (see Appendix D of \cite{suarezchavanis3}) and let us
apply the criterion (\ref{imp1}) at the epoch of radiation-matter
equality ($\rho_{\rm eq}=8.77\times 10^{-14}\, {\rm
g}/{\rm m}^3$) which marks the beginning of structure formation. We
then find that, concerning the linear Jeans problem, the self-interaction of the
bosons can be neglected provided that $|a_s|\ll 10^{-63}\, {\rm fm}$ or,
equivalently, $|\lambda|\ll 10^{-91}$ or $f\gg 10^{15}\, {\rm GeV}$. As we have
seen in Secs. \ref{sec_urtf} and \ref{sec_urng}, these conditions are not always
satisfied for ULAs. This implies that the self-interaction of ULAs usually has
to be taken into account even if it looks very small at first sight.

\subsection{The nonlinear regime of structure formation (DM halos $=$
equilibrium states)}
\label{sec_impnl}

Let us briefly recall the argument, given in
Appendix L of \cite{phi6}, concerning the validity, or invalidity, of the
noninteracting
limit in the nonlinear regime of structure formation, when we treat DM halos as
equilibrium states of the GPP equations (we restrict ourselves to the
nonrelativistic regime which is fully relevant for DM halos).

Let us consider the most compact halo that we know  and let us assume
that it corresponds to the ground state ($T=0$) of a self-gravitating BEC (see
Appendix D of \cite{suarezchavanis3}). To be specific, we identify this halo
with
Fornax which has a mass $M_{\rm ground}\sim 10^{8}\, M_{\odot}$ and a radius 
$R_{\rm ground}\sim 1\, {\rm kpc}$. Our argument also applies to the
solitonic core of large DM halos which may also correspond to the
ground state of the GPP equations. As shown in \cite{prd2}, in the nonlinear
regime of structure formation, the noninteracting limit is valid when $M_{\rm
ground}\ll {\hbar}/{\sqrt{Gm|a_s|}}$. For bosons with a repulsive
self-interaction
($a_s>0$),  when  $M_{\rm ground}\gg {\hbar}/\sqrt{Gm a_s}$ we are in the TF
limit. For bosons with an attractive self-interaction ($a_s<0$), there is no
equilibrium state when $M>M_{\rm max}=1.012 \, {\hbar}/\sqrt{Gm|a_s|}$. The
validity of the noninteracting limit can be written as
\begin{equation}
m |a_s|\ll \frac{\hbar^2}{GM_{\rm ground}^2}.
\label{imp2}
\end{equation}
This criterion can be expressed in terms of the
dimensionless self-interaction constant alone as
\begin{equation}
|\lambda|\ll \left (\frac{M_P}{M_{\rm ground}}\right )^2\sim 10^{-92}.
\label{imp3}
\end{equation}
Therefore, in
order to be able to
neglect the self-interaction of the bosons, $|\lambda|$ has to be small with
respect to $10^{-92}$ (!). This striking condition was
stressed in Appendix A.3 of \cite{prd2}. This condition is not satisfied for QCD
axions, nor for ULAs in general. This implies that the self-interaction of
QCD axions and, usually, ULAs has to be taken into account in the
nonlinear regime even if it looks very small at first sight \cite{phi6}.

Interestingly, the condition of Eq. (\ref{imp3}) is the same condition as the
one obtained in
the linear regime (see the numerical application in Appendix \ref{sec_impl}). We
note that, contrary to the condition obtained in Appendix \ref{sec_impl}, the
criterion of Eq. (\ref{imp3}) is independent of the mass of the boson. If we
assume $m=2.92\times 10^{-22}\, {\rm eV}/c^2$ then we get the same conditions
$|a_s|\ll 10^{-63}\, {\rm fm}$ or $f\gg 10^{15}\, {\rm GeV}$ as those obtained
in Appendix \ref{sec_impl}. Alternatively, if we argue that the criteria
(\ref{imp1}) and (\ref{imp2}) are both marginally satisfied (which needs not be
the case in reality) we find that 
\begin{equation}
m\sim \frac{\hbar \rho_{\rm eq}^{1/6}}{G^{1/2}M_{\rm ground}^{2/3}}\sim
10^{-22}\, {\rm eV/c^2},
\label{imp4}
\end{equation}
\begin{equation}
|a_s|\sim \frac{\hbar}{G^{1/2}M_{\rm ground}^{4/3}\rho_{\rm eq}^{1/6}}\sim
10^{-63}\, {\rm fm},
\label{imp5}
\end{equation}
which provide curious relations between the ULA mass and their scattering
length, the density of the Universe at the epoch of radiation-matter equality,
and the mass of ultracompact halos.

\end{document}